\documentclass[10pt]{article}
\usepackage[english]{babel}
\usepackage{babel}
\usepackage{graphicx, epsfig}
\usepackage{amsmath}
\usepackage{amssymb}
\usepackage[T1]{fontenc}
\usepackage[latin9]{inputenc}
\usepackage[letterpaper]{geometry}
\def\LL{{\cal L}}
\def\FF{{\cal F}} 
\def\MM{{\cal M}}

\def\SS{{\cal S}}

\def\be{\begin{equation}}
\def\ee{\end{equation}}
\def\bea{\begin{eqnarray}}
\def\eea{\end{eqnarray}}
\def\ba{\begin{array}}
\def\ea{\end{array}}
\newcommand{\lsim}{\,\raise 0.4ex\hbox{$<$}\kern -0.8em\lower 0.62ex\hbox{$\sim$}\,}
\newcommand{\gsim}{\,\raise 0.4ex\hbox{$>$}\kern -0.7em\lower 0.62ex\hbox{$\sim$}\,}

\def\de{\mathrm{DE}}
\def\dde{{\rm DE}}
\newcommand{\ca}{{c_a^2}}
\newcommand{\cs}{{\hat{c}_s^2}}
\newcommand{\ax}{a_\times}
\newcommand{\brho}{\bar{\rho}}
\newcommand{\dep}{\delta p}
\newcommand{\der}{\delta\!\rho}
\newcommand{\hdep}{\delta \hat{p}}
\newcommand{\hder}{\delta\!\hat{\rho}}
\newcommand{\bap}{\bar p}
\newcommand{\HH}{{\mathcal H}}

\newcommand{\p}{\phi}
\newcommand{\rmd}{{\rm d}}
\newcommand{\ltr}{\left(t\right)}
\newcommand{\lzr}{\left(z\right)}
\def\chk{\checkmark}
\newcommand{\etal}[1]{et al. }

\newcommand{\lyxaddress}[1]{
\par {\raggedright #1
\vspace{1.4em}
\noindent\par}
}

\makeatother

\begin{document}

\title{ DARK ENERGY IN PRACTICE}

\maketitle

\author{
\begin{center}
Domenico Sapone\footnote{domenico.sapone@uam.es}
\end{center}}

\lyxaddress{
\begin{center}
Departamento de F\' isica Te\'orica and Instituto de F\'isica Te\'orica, \\ 
Universidad Aut\'onoma de Madrid IFT-UAM/CSIC,\\ 
$28049$ Cantoblanco, Madrid, Spain.
\end{center}}

\begin{abstract}

In this paper we review a part of the approaches that have been considered to explain the 
extraordinary discovery of the late time acceleration of the Universe. 
We discuss the arguments that have led physicists and astronomers to accept dark energy as 
the current preferable candidate to explain the acceleration. 
We highlight the problems and the attempts to overcome the difficulties related to such a 
component. 
We also consider alternative theories capable of explaining the acceleration of the Universe, 
such as modification of gravity. 
We compare the two approaches and point out the observational consequences, reaching the 
sad but foresightful conclusion that we will not be able to distinguish between a Universe 
filled by dark energy or a Universe where gravity is different from General Relativity.
We review the present observations and discuss the future experiments that 
will help us to learn more about our Universe. 
This is not intended to be a complete list of all the dark energy models but 
this paper should be seen as a review on the phenomena responsible for the acceleration. 

Moreover, in a landscape of hardly compelling theories, it is an important task 
to build simple measurable parameters useful for future experiments that will 
help us to understand more about the evolution of the Universe.

\end{abstract}

\tableofcontents{}

\section{Introduction}

Modern cosmology was born in the 1920s as one of the first 
applications of General Relativity for studying the properties
of the {\em cosmos} on large scales. If the traditional astrophysics and astronomy 
were studying single objects (e.g. stars) or clusters of such objects (e.g. galaxies), 
characterizing the local properties of the Universe, with a more sophisticated instrumentation 
it was possible to observe deeper in the sky and at larger scales.
More generally, if the global properties of the cosmos are to be described, we refer to the 
Universe as the first object under investigation by cosmology.

Modern cosmology has a precise birth-date: the discovery of Cepheids and 
ordinary stars in {\em Nebulae} by E.~P.~Hubble. After he recognized that most 
of the nebulae were galaxies, Hubble also pointed out that they were 
moving away from each other. 
Such expansion, currently named the Hubble flow, has been confirmed over the years by the 
whole set of data we now possess. Even if Hubble's intuition was extraordinary, his 
data were too noisy. At the distances where he pretended to see an average 
expansion, the motion of galaxies is still dominated by their peculiar motions. 
Therefore local measurements of moving objects are not accurate to prove the total 
expansion of the Universe. It is at scales larger than the galactic scale in which 
the dynamic evolution is mostly due to pure gravity. Galaxies, therefore, are the 
occupants of the Universe whose rules are set by general relativity and their 
average distances are increasing with time.

An expanding Universe filled only by ordinary matter (e.g. matter whose stars are formed) 
will at some point slow down due to gravitational attraction. In 1998, astronomers, 
observing supernovae, realized that their apparent luminosities were reduced as an 
effect of more red-shifting; astronomers pointed out the Universe is not only in 
phase of expansion but this expansion is accelerating. Although the supernova 
observations, see Refs. \cite{sn1} and \cite{sn2} has led to the general 
acceptance that the expansion of the Universe is accelerating, we are still far 
from a consensus on what is responsible for the acceleration. Although there is a 
name for the phenomenon, dark energy, there is as of yet no convincing physical explanation.

Nonetheless, other recent observations have put serious doubts on the completeness 
and understanding of the cosmological models, the theoretical and experimental 
considerations are:
\begin{itemize}
\item Primordial nucleosynthesis: predicts the baryonic abundance which is in 
agreement with the observations but it is not enough to produce the observed 
gravitational effects. 
This has led people to believe that a new form of matter, which is not 
(or weakly) coupled to the electro-magnetic field, has to be added in the 
Universe, known as {\em Dark Matter}.
\item Inflation: (needed to solve the curvature and horizon problems) requires 
an extension of standard model of particle physics.
\item Supernovae type Ia (SNIa): show the Universe is in a phase of accelerated expansion.
\item Large Scale Structure (LSS): provides another test for the existence of dark energy.
It also shows there is a lack of matter. Therefore, this is also another evidence of dark matter. 
\item Cosmic Microwave Background Anisotropies (CMB): suggest the total energy density 
in the Universe has to be equal to a particular critical value.
\end{itemize}

The result of these observations has built a scenario in which the Universe is in a phase 
of accelerated expansion (SNIa) and its matter density (LSS) is not sufficient to set the 
curvature equal to zero (CMB).

This review is organized as follow: we first introduce the basic equations 
for studying the dynamic of the Universe, section \ref{basic1} and \ref{basic2}; 
in section \ref{hubble-flow} we report in more detail the Hubble's flow 
and definition of cosmic distances.

The observational evidence that has led cosmologists to accept dark energy are 
reported in section \ref{firstobs}; in section \ref{DEmodels} we list the most studied 
models for dark energy whereas in section \ref{MGmodels} are shown some of 
the models of modified gravity that are able to lead the Universe to a 
phase of accelerated expansion.
In section \ref{comparisons} we compare the two different approaches, 
i.e. dark energy (or equivalently to add a new form of matter in the Universe) 
and modified gravity (or revising the laws of gravity).

In section \ref{tests} we present a list of possible test that a model 
has to pass in order to be considered a good {\em dark energy} candidate.

In section \ref{phenomenological} we relate theory to observations 
and we show how phenomenological parameters can be helpful to distinguish, 
say, a cosmological constant from a dynamical dark energy or the latter 
from some form of modified gravity. We end with the new observational tests 
in section \ref{newtests}.

\section{Basic equations}\label{basic1}

General Relativity theory has led to an increasing development in the understanding of 
our Universe. The most interesting aspect of this theory is that the dynamics of every 
object can be expressed by a single tensorial equation:
\be
G_{\mu\nu} = \frac{8\pi G}{c^4} T_{\mu\nu} \label{Einstein}
\ee
known as Einstein equation. Here $G_{\mu\nu}$ represents Einstein tensor, which 
defines the geometry of the Universe, $G$ is Newton's constant, $T_{\mu\nu}$ denotes 
the energy-momentum tensor, which describes the matter and the energy content of the 
Universe and $c$ is the speed of light. In other words, Einstein equation relates 
the geometry of the Universe to its content.

Einstein equations are in general complicated non-linear equations and  it is 
impossible to solve them analytically and only few exact, analytical solutions are 
known, involving strong symmetry constraints;
however under the conditions of homogeneity and isotropy, which is approximately true at 
large scales, the most general line of 
element for an expanding Universe is the Friedmann-Lema\^itre-Robertson-Walker metric (FLRW):
\be
\rmd s^{2}=-\rmd t^{2}+a^{2}(t)\left[\frac{\rmd r^2}{1-Kr^2}+r^2\left(\rmd\theta^2+\sin^2\theta \rmd\phi^2\right)\right]\label{metric1}
\ee
where $a\left(t\right)$ is the scale factor and represents the evolution in time of 
all the physical scales (every scale $\ell_{0}$ will change in time as 
$\ell\left(t\right)=a\left(t\right)\ell_{0}$); $K$ is the curvature parameter 
(it can be $-1$, $1$ or $0$ meaning an open, closed or flat Universe respectively); 
$t$ is the cosmic time and the coordinates $r$, $\theta$ and $\phi$ are spherical 
comoving coordinates.

In the FLRW spacetime the energy-momentum tensor takes the simple perfect fluid form:
\be
T^{\mu\nu}=\left(\rho+p\right)u^{\mu}u^{\nu}+p~g^{\mu\nu} \label{em1}
\ee
where $u^{\mu}$ is the four velocity of the fluid and $\rho$ and $p$ are the energy 
density and the pressure of a fluid, respectively.

Einstein equation gives rise then to the Friedmann equations:
\bea
H^{2}  +\frac{K}{a^{2}}&=& \frac{8\pi G}{3}\rho  \label{first-Fried}\\
\left( \frac{\ddot{a}}{a} \right) &=& -\frac{4 \pi G}{3} \left( \rho+3p \right) 
\label{second-Fried}
\eea
where the dot denotes the derivative with respect to the cosmic time $t$ and 
$H=\dot{a}/a$ is called the Hubble parameter which describes the expansion rate 
of the Universe. The energy density $\rho$ should be thought as a sum running 
over all the forms of matter that fill the Universe, the same discussion has to 
be made for the term $\left( \rho+3p \right)$.

In both the Friedmann equations the left hand side is given by the left hand side 
of Einstein equation, in other terms the geometry of a homogeneous and isotropic 
fluid is fully characterized by the scale factor $a\left(t\right)$, whereas the 
right hand side of Friedmann equations come directly from the energy momentum 
tensor $T^{\mu\nu}$, Eq.~(\ref{em1}), hence geometry is related to matter.

The Einstein tensor satisfies Bianchi's identity: 
\be
G^{\mu}_{\nu;\mu}=0
\ee
where $_{;\mu}$ represents the covariant derivative; from Eq.~(\ref{Einstein}) 
follows automatically:
\be
T^{\mu}_{\nu;\mu}=0
\ee
which gives the continuity equation of a fluid in the FLRW background Universe:
\be
\dot{\rho}+3H\left(\rho+p\right)=0.\label{continuity}
\ee
Let us now rewrite Eq.~(\ref{first-Fried}) as:
\be
\Omega\left(t\right)-1=\frac{K}{\left(aH\right)^2}
\ee
where $\Omega\left(t\right)=\rho\ltr/\rho_{c}\ltr$ is the dimensionless density 
parameter and $\rho_c\ltr=3H^2/8\pi G$ is the critical energy density parameter.
Then, the matter content determines the spatial geometry of the Universe, i.e.:
\bea
\Omega >1 &\rightarrow& K = +1 \rightarrow {\bf closed}~{\rm Universe}\\
\Omega = 1 &\rightarrow& K = 0 \rightarrow {\bf flat}~{\rm Universe} \\
\Omega < 1 &\rightarrow& K = -1\rightarrow {\bf open}~{\rm Universe}.
\eea
Recent measurements of Cosmic Microwave Background Radiation (CMBR) have shown that 
the Universe is spatially flat ($\Omega = 1$ or $K = 0$) to within a few percent, 
see Ref.~\cite{cmb-5y} . This is also a result that comes naturally from 
inflation in the early Universe, see Ref.~\cite{liddle-lyth}. So we assume 
hereinafter, if not else specified, that  $K$ is equal to $0$.

\section{The evolution of the Universe with a barotropic perfect fluid}\label{basic2}

Let us now consider the case in which the Universe is filled only by one single perfect 
barotropic\footnote{We define a fluid to be barotropic if the pressure $p$ 
depends strictly only on the energy density $\rho$: $p=p\left(\rho\right)$} 
fluid with an equation of state parameter:
\be
w = \frac{p}{\rho}.
\ee
If $w$ is constant one can analytically find the evolution for $\rho$ and the scale 
factor $a\ltr$ by solving the Eqs.~(\ref{first-Fried}) and (\ref{continuity}):
\bea
&&\rho\propto a^{-3\left(1+w\right)}\\ \label{rho-t}
&& a\ltr\propto \left(t-t_{0}\right)^{\frac{2}{3\left(1+w\right)}} \label{a-t}
\eea
where $t_{0}$ is a constant. An analytical expression for the Hubble parameter can be found:
\be
H\ltr =\frac{2}{3\left(1+w\right)\left(t-t_{0}\right)}\label{H-t}.
\ee
The above solutions, Eqs.~(\ref{rho-t}) and (\ref{H-t}), are valid for all the value 
of $w$ except for $w=-1$. When the parameter of equation of state $w$ is $-1$, i.e. 
$p=-\rho$ it follows from Eq.~(\ref{rho-t}) that the energy density $\rho$ is constant; 
this corresponds to the cosmological constant, denoted by $\Lambda$. In this case the 
Hubble parameter $H$ is also constant and the scale factor evolves as:
\be
a\ltr = \exp\left(Ht\right) \label{a-t-exp}
\ee 
which corresponds to the de-Sitter Universe. It is clear from Eq.~(\ref{a-t-exp}) 
that a cosmological constant cannot be an appropriate candidate for inflation in the early 
Universe because otherwise inflation would not have ended. In turn a cosmological constant 
can be a good candidate for the dark energy today; for the moment we assume the 
cosmological constant to be the only model for dark energy.

The parameter of equation of state $w$ fully characterizes the fluid at background 
level, for instance:
\bea
Rad.&:&w=\frac{1}{3},~a\ltr\propto \left(t-t_{0}\right)^{1/2},~\rho\propto a^{-4}\\
Dust&:&w=0,~a\ltr\propto \left(t-t_{0}\right)^{2/3},~\rho\propto a^{-3}\\
\Lambda&:&w=-1,~a\ltr\propto \exp\left(Ht\right),~\rho=const.
\eea
Both cases, radiation and dust, correspond to a decelerated Universe; this can be 
easily seen if we insert the corresponding value of the parameter of equation of 
state in the Eq.~(\ref{second-Fried}), which gives $\ddot{a}<0$. In order to have 
cosmic acceleration, we need $\ddot{a}$ to be positive, in other words we require: 
\be
w<-1/3,~p<-\rho/3.\label{min-condit}
\ee
This is clearly an upper bound for the parameter of equation of state; we can say 
that: the more the $w$ is negative the bigger $\ddot{a}$ is.

In order to explain the late phase acceleration in the Universe we require an exotic 
form of energy with negative pressure\footnote{The energy density is assumed to be 
positive.} satisfying Eq.~(\ref{min-condit}).

This concept is completely new and it comes directly from Einstein equations 
because of relativistic effects (absent in Newtonian gravity). 
To show this we can consider a homogeneous sphere 
with radius $r$ and energy density $\rho$ in which a point particle with mass $m$ 
is moving. In Newtonian gravity, the equation of motion for such a particle is given by:
\be
m\ddot{r}=-\frac{Gm}{r^2}\left(\frac{4\pi r^3\rho}{3}\right)
\ee
leading to:
\be
\frac{\ddot{r}}{r}= -\frac{4\pi G}{3}\rho
\ee
which looks like Eq.~(\ref{second-Fried}) except for the missing pressure term. 
Then, we need a large negative pressure, to counter-balance the gravitational 
attraction, in order to accelerate the Universe. As shown before, Newtonian gravity 
can only decelerate the expansion of the Universe.

\section{Hubble's flow and Cosmic Distances}\label{hubble-flow}

\subsection{Hubble's flow}

As mentioned before, in 1929 E.~P.~Hubble, see Ref.~\cite{Hubble}, 
observed that the spectral lines of the chemical elements in the galaxies were 
shifted from their normal position, they manifested a shift towards the red part 
of the electro-magnetic spectrum (hence the name redshift), more precisely towards 
lower frequencies with respect to those obtained in the laboratories. 
Such a characteristic, analogous to the Doppler effect\footnote{But it is not. 
Firstly because, at very large distances we would have had objects moving faster 
than the speed of light. Secondly, in General Relativity the presence of matter 
induces a gravitational redshift.}, is expressed in terms of parameter $z$ given by:
\be
z=\frac{\lambda_{0}-\lambda_{e}}{\lambda_{e}}\label{redshift1}
\ee
where $\lambda_0$ and $\lambda_e$ are the observed and emitted wavelengths, 
respectively. He concluded that objects were moving away from us and he also 
derived, for small redshift, a direct relation between the redshift and 
the distance of an object: 
\be
z = H_{0}D.
\ee
The redshift increases with distance or equivalently as we go 
back to the past the redshift increases.

Here $H_0$ is the present value of the Hubble parameter called Hubble constant, 
which is usually written as:
\be
H_0 = 100~h~{\rm Km~sec^{-1}Mpc^{-1}}
\ee
where $h$ describes the uncertainty of the Hubble parameter; measurements constraint 
this value to be, see Ref.~\cite{HST}:
\be
h=0.72\pm 0.08.
\ee
Moreover, an object at the distance $D$ will show a {\em recession} velocity 
proportional to its distance; to show this, let us consider two objects separated 
by a comoving distance $r_0$, than their physical distance will be $R\ltr=a\ltr r_0$, 
if we derive the latter we find:
\be
V_{r}=\dot{a}r_0=\frac{\dot{a}}{a} R\ltr =HR\ltr
\ee
thus the relative velocity of the two objects is independent on the position of the observer.

\subsection{Cosmic distances}

An important concept related to observational tools in an expanding background 
is associated with the definition of distance.
In fact most of the evidence in dark energy comes from distance measurements. 
Therefore, in order to discuss the observational constraints, we need to introduce 
the cosmic distances in the FLRW Universe. Let us rewrite the metric (\ref{metric1}) 
in the following form:
\be
\rmd s^2=-\rmd t^2+a\ltr\left[\rmd\chi^2+\left(S_{K}\left(\chi\right)\right)^2\rmd\Omega\right]\label{metric2}
\ee
where $\rmd\Omega = \rmd\theta^2+\sin\theta^2\rmd\phi^2$ and where:
\be
S_{K}\left(\chi\right)=\begin{cases}
\sin\left(\chi\right) & K = +1\\
\chi & K = 0\\
\sinh\left(\chi\right) & K = -1
\end{cases}
\ee

\subsubsection{Comoving Distance}

Let us consider the metric given by Eq.~(\ref{metric2}), we can evaluate the physical 
distance along a null geodesic simply setting: $\rmd s^2=-c^2\rmd t^2+a^2\ltr \rmd\chi^2=0$; 
that all light rays travel along null geodesic. Let us suppose that light is 
emitted by a source at redshift $z$ (corresponding to $\chi_e$), at time $t_e$ 
and reaches the observer situated at $z=0$ (corresponding to $\chi = 0$) 
at time $t_0$; the comoving distance can be evaluated simply by integrating the 
null geodesic equation:
\be 
D_{c}=-\int_{t_0}^{t_e}{\frac{c}{a\ltr}\rmd t}=\frac{c}{a_0}\int_{0}^{z}{\frac{\rmd z'}{H\left(z'\right)}}.
\ee

\subsubsection{Luminosity Distance}

Another way of defining a distance is through the luminosity of the objects. The 
luminosity distance plays an important role in observational cosmology.
Let us consider a source with an intrinsic luminosity $\LL_{s}$ at some distance $d_{s}$, 
the flux we observe will be $\FF$, which is defined as amount of energy per unit 
surface, that is $\FF = \LL_0/\SS$, the area of a sphere at $z=0$ in a curved 
space time is $\SS = 4\pi \left(a_0S_{K}\left(\chi\right)\right)^2$.
The luminosity distance can be written as:
\be
D_{L}^2=\left(a_0S_{K}\left(\chi\right)\right)^2\frac{\LL_s}{\LL_0}\label{Dl1}.
\ee
The luminosity of an object is defined as the amount of energy emitted per unit time. 
If we denote by $\Delta E_e$ the energy of light emitted by an object in the time 
interval $\Delta t_e$ with absolute luminosity $\LL_s$ and by $\Delta E_0$ 
the energy that reaches the observer, we have:
\bea
\LL_e &=&\frac{\Delta E_e}{\Delta t_e}\\
\LL_0 &=&\frac{\Delta E_0}{\Delta t_0}.
\eea
Since the energy of the photon is inversely proportional to its wavelength 
$\lambda$ we have that:
\bea
\frac{\Delta E_e}{\Delta E_0}=\frac{\lambda_0}{\lambda_e}=(1+z)
\eea
where we have used Eq.~(\ref{redshift1}).

Similar considerations can be made for the time interval, so we have that 
$\Delta t_e/\Delta t_0=\lambda_e/\lambda_0 = (1+z)$. Hence we find:
\be
\frac{\LL_s}{\LL_0} = \frac{\Delta E_e}{\Delta E_0}\frac{\Delta t_0}{\Delta t_e} = \left(1+z\right)^2.\label{ratio-L}
\ee
Inserting Eq.~(\ref{ratio-L}) into Eq.~(\ref{Dl1}) the luminosity distance reads:
\be
D_L = a_0S_{K}\left(\chi\right)\left(1+z\right)\label{D_L}.
\ee

\subsubsection{Angular Diameter Distance}

The angular diameter distance is the distance under which an object of actual size 
$\Delta x$ is seen under an angle $\Delta\theta$, that is:
\be
D_{A} = \frac{\Delta x}{\Delta\theta}.
\ee
The size $\Delta x$ of a source lying on the surface of a sphere or radius $\chi$ 
at the time $t_1$ with the observer at the center is given, in FLRW metric, by:
\be
\Delta x = a\left(t_1\right)S_{K}\left(\chi\right)\Delta\theta.
\ee
The angular diameter distance will become:
\be
D_{A}=a\left(t_1\right)S_{K}\left(\chi\right)=\frac{a_0}{1+z}S_{K}\left(\chi\right)\label{D_A}
\ee
where we have used the relation $z+1=a_0/a\left(t_1\right)$. Comparing the last 
one with Eq.~(\ref{D_L}) we find the relation, called distance duality:
\be
D_L=\left(1+z\right)^2D_A. \label{distance-duality}
\ee
The distance duality relation holds for any metric (not just FLRW) in which photons travel 
on null geodesic and it depends crucially on photon conservation\footnote{For instance if 
photons are scattered from dust or free electrons, the luminosity distance $D_L$ will be 
reduced but the angular diameter distance is basically unaffected, as a consequence the 
relation (\ref{distance-duality}) does no longer hold.}, 
see Refs.~\cite{Etherington,BassKunz}.

\section{Observational evidence for dark energy}\label{firstobs}

The existence of dark energy is supported by many observations. It was in 1998 
when two groups, see Refs. \cite{sn1} and \cite{sn2}, firstly and independently reported 
that the Universe is a phase of accelerated expansion by observing the luminosity 
of Supernovae Type Ia (SN Ia) which are huge explosions of some stars. 
This event is extremely luminous and can be used as a test in observational cosmology. 
The Supernovae are classified according to their spectral characteristics. The most useful 
in cosmology are those called Type Ia because they are poor in hydrogen\footnote{It is the
 most abundant element in the Universe.} and the mechanism generating this explosion is 
assumed to be known. 
The explosion of Type Ia occurs when the mass of a white dwarf in a binary system 
(usually its companion is a red giant star) exceeds the Chandrasekhar  
limit by absorbing the gas from the other star. 
The general belief is that SN Ia are formed in the same way in all the Universe, which 
means they have a common absolute luminosity. Hence SN Ia are considered to 
be {\em standard candles} by which the luminosity distance can be determined. 
In particular, the light curve of a SN Ia 
is correlated with its peak luminosity to a precision of about $7\%$, in other words, 
the brighter the SN the longer it takes to vanish, see Ref.~\cite{sahni}.

Other observations, apart from SN Ia, support the evidence for dark energy and these are: 
\begin{itemize}
\item Age of the Universe compared to the oldest star.
\item Cosmic Microwave Background (CMB).
\item Baryon Acoustic Oscillation (BAO).
\item Large Scale Structure (LSS).
\end{itemize}

\subsection{Constraints from SN Ia}

In astronomy the magnitude is often used as quantity to categorize sources; also in this 
case we have the absolute magnitude $\MM$ which is related to the logarithm of luminosity and 
the apparent magnitude ${\it m}$ which is related to the flux. Absolute and apparent 
magnitude are related to the luminosity distance via the relation:
\be
{\it m}-\MM=5\log\left(\frac{D_L}{{\rm Mpc}}\right) + 25. \label{magnitude-rel}
\ee 
Since the absolute magnitude $\MM$ is known {\em a priori} (under the assumption 
of standard candle) then the luminosity distance can be derived from 
Eq.~(\ref{magnitude-rel}) once the relative magnitude ${\it m}$ is measured. 
The redshift of the corresponding source can be obtained simply by measuring 
the frequency of the photons.

Let us consider now, as an example, a FLRW flat background ($S_K\left(\chi\right)=\chi$), 
the luminosity distance Eq.~(\ref{D_L}) reads:
\be
D_L\lzr = c\frac{1+z}{H_{0}}\int_{0}^{z}{\frac{\rmd z}{E\lzr}}
\ee
where $E\lzr=H\lzr/H_0$. The Hubble parameter can be expressed in terms 
of luminosity distance:
\be
H\lzr = c\left[\frac{d}{\rmd z}\left(\frac{D_L\lzr}{1+z}\right)\right]^{-1}.
\ee
Once the luminosity distance is measured by observations then we can determine the expansion 
rate $H\lzr$ and this is crucial to understand our Universe because we know the Hubble 
parameter depends on all the components present in the Universe. 
The energy density in Eq.~(\ref{first-Fried}) has to be thought as the sum 
over all the components, namely relativistic matter, non-relativistic matter, 
cosmological constant (and eventually other stuff):
\be
\rho = \sum_{i}{\rho_{0,i}\left(\frac{a}{a_0}\right)^{-3\left(1+w\right)}}=\sum_{i}{\rho_{0,i}\left(1+z\right)^{3\left(1+w\right)}}.\label{densities-sum}
\ee
Using Eq.~(\ref{densities-sum}) the Hubble parameter reads:
\be
H^2\lzr=H_0^2\sum_{i}{\Omega_{0,i}\left(1+z\right)^{3\left(1+w\right)}}\label{eq:hubble-sum}
\ee
where $\Omega_{0,i}=\rho_{0,i}/\rho_{0,c}=3H_0^2\rho_{0,i}/8\pi G$, where we have used the 
critical energy density defined previously.

\begin{figure}[pb]
\centerline{\epsfig{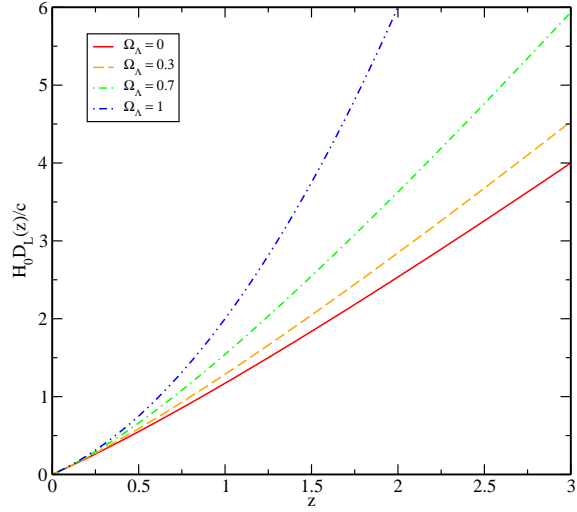}}
\vspace*{8pt}
\caption{The luminosity distance $D_L$ for a flat cosmology with two components, 
matter $w_m=0$ and a cosmological constant $w_\Lambda=-1$ for different values 
of $\Omega_\Lambda$. \label{fig:DL-plot}}
\end{figure}

\begin{figure}[pb]
\centerline{\epsfig{file=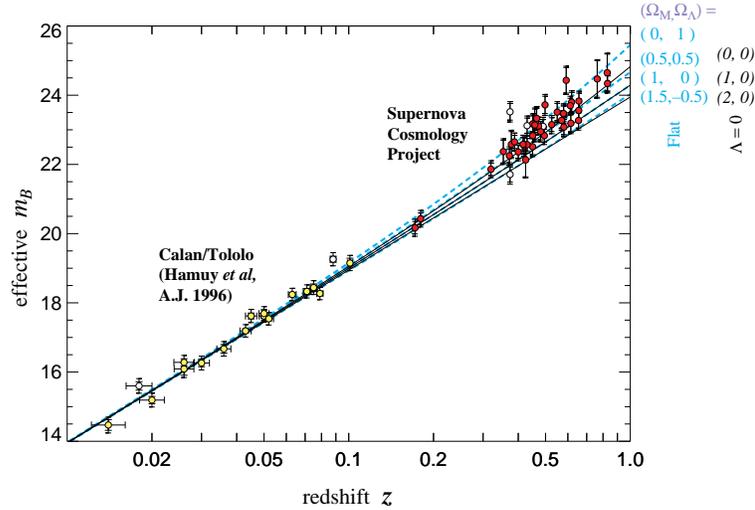,width=10.0cm}}
\vspace*{8pt}
\caption{The effective apparent magnitude $m_B$ for 42 high-redshift SN Ia from the 
Supernova Cosmology Project, and 18 low-redshift SN Ia from the Calan/Tololo 
Supernova Survey, after correcting both sets for the SN~Ia lightcurve width-luminosity
relation. The inner error bars show the uncertainty due to measurement
errors, while the outer error bars show the total uncertainty when
the intrinsic luminosity dispersion, 0.17 mag, of lightcurve-width-corrected
Type Ia supernovae is added in quadrature.
The horizontal error bars represent the assigned
peculiar velocity uncertainty of 300 km s$^{-1}$.
The solid curves are the theoretical $m_B^{\rm effective}(z)$
for a range of cosmological models with zero cosmological constant:
$(\Omega_{\rm M},\Omega_\Lambda) = (0,0)$ on top, $(1,0)$ in the middle
and (2,0) on bottom.  The dashed curves are for a range of flat
cosmological models: $(\Omega_{\rm M},\Omega_\Lambda) = (0,1)$
on top, $(0.5,0.5)$ second from top, $(1,0)$ third from the top,
and (1.5,-0.5) on the bottom. From Ref.~2.
 \label{fig:perlmutter98}}
\end{figure}

As an example, let us consider a Universe filled only by non-relativistic matter $w=0$ 
and cosmological constant $w=-1$ (relativistic matter plays a small role in the late time 
and can be neglected), the Hubble parameter and the luminosity distance become:
\bea
H^2\lzr &=& H_0^2\left[\Omega_{m,0}\left(1+z\right)^{3}+\Omega_{\Lambda,0}\right]\\
D_L\lzr &=& \frac{c}{H_0}\left(1+z\right)\int_{0}^{z}{\frac{\rmd z'}{\sqrt{\Omega_{m,0}\left(1+z\right)^{3}+\Omega_{\Lambda,0}}}}.\label{DL-plot}
\eea
In Fig.~(\ref{fig:DL-plot}) we plot the luminosity distance Eq.~(\ref{DL-plot}). 
We remind the reader that we are working in a flat FLRW background which 
implies that $\Omega_{m,0}+\Omega_{\Lambda,0}=1$. 
As we can see the luminosity distance increases with the value of the cosmological constant. 
Fig.~(\ref{fig:perlmutter98}) illustrates the observational values of the luminosity distance 
versus redshift $z$ together with the theoretical curves derived from Eq.~(\ref{DL-plot}). 
The plot clearly shows that a Universe with only matter ($\Omega_m = 1$) does 
not fit to the data. 
Perlmutter {\it et al.} showed that cosmological constant is present at the $99\%$ confidence 
level, giving the value of the matter energy density $\Omega_{m_0} = 0.28^{+0.09}_{-0.08}$ 
($1\sigma$ statistical) in the flat Universe with a cosmological constant which 
has to be roughly the $70\%$ of the total energy density of the Universe. 
Since at low redshift the models are indistinguishable and moreover observational 
data suffers from statistical and systematic errors, one cannot conclude that 
the Universe is accelerating from a small data set. Recent experiments 
as the Super Nova Legacy Survey (SNLS), Hubble Space Telescope (HST) and 
Equation of State: SupErNovae trace Cosmic Expansion (ESSENCE), 
see Refs.~\cite{SNLS,HST1,ESSENCE} respectively, have measured 
the luminosity distances of SN Ia at higher redshift giving tighter constraints 
on the dark energy parameters and confirming the trend that the Universe is 
in a phase of accelerated expansion.

\begin{figure}[pb]
\centerline{\epsfig{file=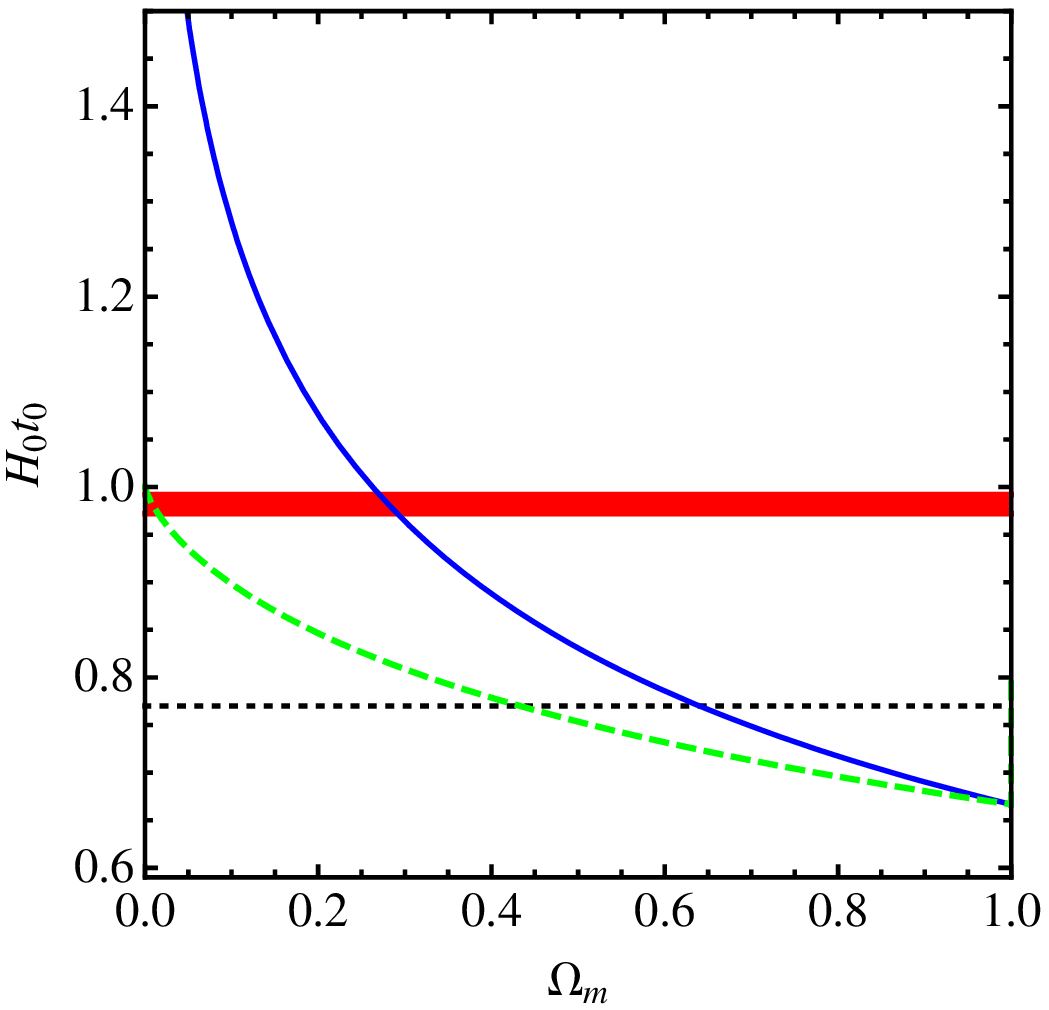,width=5.7cm}}
\vspace*{8pt}
\caption{The cosmic age $t_0$ in units $H_0^{-1}$ versus $\Omega_{m_0}$. The blue solid 
curve describes a Universe with the cosmological constant, the green dashed curve 
corresponds to an open Universe without a cosmological constant, the black dotted line 
gives the lower bound of the cosmic age from the globular cluster. The red stripe gives 
the constrains on the age of the Universe from WMAP 5-year (assuming the $\Lambda$CDM). 
\label{fig:age}}
\end{figure}

\subsection{Age of the Universe}

The inverse of the Hubble parameter gives roughly the age of the Universe. 
Once an expression for the Hubble parameter is taken then one can find the 
corresponding age of the Universe and compare it with the oldest stars which 
gives a lower age boundary. 
Let us consider the Hubble parameter given in Eq.~(\ref{eq:hubble-sum}) in a 
FLRW Universe and taking into account all the component, i.e. relativistic 
and non-relativistic matter and dark energy:
\be
H\lzr = H_0\left[\Omega_{r_0}\left(1+z\right)^4+\Omega_{m_0}\left(1+z\right)^3+\Omega_{DE_0}\left(1+z\right)^{3\left(1+w_{DE}\right)}\right]^{1/2}
\ee
where we have not made any assumption on the nature of the dark energy yet. 
Using the relation ${\rm d}t = -{\rm d}z/\left[\left(1+z\right)H\right]$ 
we can find the age of the Universe to be:
\be
t_{0} = \int_{0}^{\infty}{\frac{\rmd z}{\left(1+z\right)H\lzr}}\label{eq:age-general}
\ee
The integral in Eq.~(\ref{eq:age-general}) is too complicated to solve analytically 
and we are forced to make some assumptions. The contribution of the radiation term in 
Eq.~(\ref{eq:age-general}) can be neglected since the radiation dominated era 
is much shorter than the total age of the Universe, or to put it differently, 
the integral in Eq.~(\ref{eq:age-general}) is not hardly affected for redshift $z>1000$.
Moreover, $\Omega_{r_0}$ is of the order of $10^{-4}$, then it is a good approximation to 
neglect radiation. Let us now consider the case of a cosmological constant ($w=-1$), 
then the age of the Universe is:
\be
t_{0} = \frac{1}{H_0}\int_{0}^{\infty}{\frac{\rmd z}{\left(1+z\right)\left[\Omega_{m_0}\left(1+z\right)^3+\Omega_{DE_0}\right]^{1/2}}}.\label{eq:age-flrw}
\ee
integrating the last one, we have:
\be
t_{0} = \frac{1}{3H_0}\frac{1}{\sqrt{1-\Omega_{m_0}}}\log\left[\frac{1+\sqrt{1-\Omega_{m_0}}}{1-\sqrt{1-\Omega_{m_0}}}\right]
\ee
where we used $\Omega_{m_0}+\Omega_{DE_0}=1$. In the limit $\Omega_{DE_0}\rightarrow0$ we find:
\be
t_{0} = \frac{2}{3H_0}.\label{age-matter}
\ee
A Universe without cosmological constant would have an age within the range 
$8.2~{\rm Gyr}<t_{0}<10.2~{\rm Gyr}$, where we have used $h=0.72$. 
However, different groups have estimated the age of the globular cluster, 
see Refs.~\cite{Carretta,Jimenez,Hansen}, 
to be within the range of $12.7~{\rm Gyr}<t_{0}<13.5~{\rm Gyr}$. 
In most cases the globular clusters seem to be older than $11~{\rm Gyr}$ 
(upper bound given by Eq.~(\ref{age-matter})), being inconsistent with 
the age of a matter dominated Universe. The age of the Universe can be increased by taking 
into account the dark energy with an equation of state close to $-1$ as shown 
in Fig.~(\ref{fig:age}). 

Also in an open Universe $\Omega_{K_0}>0$ it is possible to increase the cosmic age 
but it will be still insufficient to bring the cosmic age up to $13{\rm Gyr}$. As an example, 
let us consider a Universe with $\Omega_{m_0}+\Omega_{K_0}=1$ and zero cosmological constant, 
then the age of the Universe will be:
\be
t_0 = \frac{1}{H_0}\frac{1}{1-\Omega_{m_0}}\left[1+\frac{\Omega_{m_0}}{2\sqrt{1-\Omega_{m_0}}}\log\left(\frac{1-\sqrt{1-\Omega_{m_0}}}{1+\sqrt{1-\Omega_{m_0}}}\right)\right]
\ee 
when $\Omega_{m_0}=1$ we recover Eq.~(\ref{age-matter}); in the limit 
of $\Omega_{K_0}=1$ then the age of the Universe is $t_0=1/H_0=13{\rm Gyr}$. 
However the curvature $\Omega_{K_0}$ has been constrained  by WMAP 
measurements, see Ref.~\cite{cmb-5y}, to be much smaller than 
unity. An open Universe without dark energy is inconsistent with the 
age of the oldest stars present in the Universe. The flat Universe 
with the cosmological constant is consistent with WMAP constraints for a 
value of matter density $0.271<\Omega_{m_0}<0.289$, see Fig.~(\ref{fig:age}).

\subsection{Cosmic microwave background}

During the last decade an avalanche of experiments, 
together with the most recent WMAP satellite, have measured the small and 
large angular temperature fluctuations of the Cosmic Microwave Background Radiation. 
Such measurements have detected a series of acoustic peaks in the anisotropy power 
spectrum confirming the earlier predictions about the evolution of pressure 
waves in the primordial photon-baryon plasma. 
The specific features of such peaks are sensitive to the value of the 
cosmological parameters, in particular to $\Omega_{tot}$, $\Omega_b$ 
and the scalar spectral index $n_{s}$.

\begin{figure}
\begin{centering}
\epsfig{figure=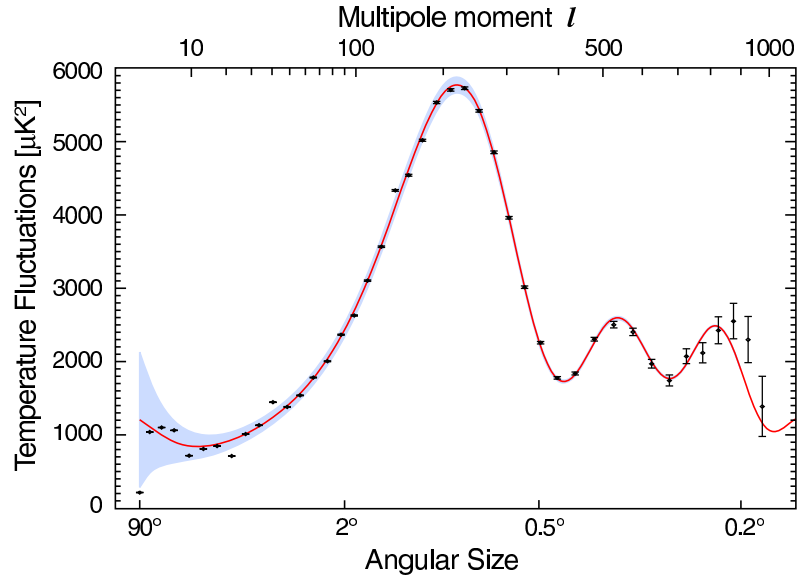,
  width=4.3in}
\caption{The CMB power spectrum $\ell\left(1+\ell\right)C_{\ell}/2\pi$ versus the 
multiple $\ell$ and the angular size $\theta$ The curve represent the theoretical 
prediction of the power spectrum whereas the black points are the WMAP 5-year data. 
Taken from the webpage of WMAP: http://map.gsfc.nasa.gov/ \label{fig:CMB}}
\end{centering}
\end{figure}

So the CMB power spectrum provides information on combinations of fundamental 
cosmological parameters. 
The physical processes responsible for the anisotropies are well understood and 
allows a complete prediction on the shape of the CMB power spectrum for a 
given cosmological model, see Refs.~\cite{bond-efst1,bond-efst2,peebles-bao,silk-bao}. 

Generally, before recombination, the baryons in the Universe were locked to the photons 
(via Thompson scattering) of the CMB, and the photon pressure interacting against 
gravitational instability, due to matter, produced sound waves in the plasma. 
After recombination, the baryons and the photons separated, but the effects of the 
acoustic oscillations remained imprinted in the spatial structures of the baryons 
and eventually on the dark matter. The physical length scale of the acoustic 
oscillations depends on the sound horizon of the Universe at the epoch of recombination, 
where the sound horizon is the comoving distance that a sound wave can travel before 
recombination and depends only on the baryon and matter densities. 
The relative heights of the acoustic peaks in the CMB anisotropy power spectrum 
measure these densities with excellent accuracy.

The CMB anisotropies can be thought of as fluctuations in temperature $\delta T/T$ around 
the mean black body temperature $T\simeq 2.726 {\rm K}$ of the cosmological radiation. 
During the radiation dominated era the equation describing the effective temperature 
fluctuations $\delta T$ of the CMB has the form of a harmonic oscillator sourced by a 
function of the gravitational potential ($\psi$), the amplitude of these oscillations are 
modulated by the photon-baryon sound speed $c^{2}_{s} = 1/3(1+3\rho_{b}/\rho_{r})$. 
Therefore when photons decouple from baryons their energy carry an imprint 
of such oscillations. 
The characteristic frequencies of these oscillations are fixed by the size of the sound 
horizon at the decoupling $\lambda_{sh}=\int{c_{s}d\tau}$. 
Therefore we have a series of compressions and rarefactions at scales 
$k_{m}\lambda_{sh}= m\pi$. 
Today such scales appear at angles that are multiple integers of the angular size of 
the sound speed horizon at the decoupling $\theta_{hs}=\lambda_{sh}/D_{K}(\tau_{lss})$, 
where $D_{K}(\tau_{lss})$ is the distance to the last scattering surface for a 
spacetime with curvature $K$. 
As a consequence, the position of the {\em acoustic} peaks in the power spectrum 
depends on the geometry of the Universe. 
For a flat cosmology the peaks will appear at the multipoles $\ell_{m} = m\ell_{sh}$. 
However the acoustic oscillations are perturbed by the evolution of the 
gravitational potential which shift the position of the peaks by an amount 
that depends on the cosmological parameters that are relevant before recombination. 
Then, it is easy to understand that the position of the peaks and their amplitudes 
give information on cosmological parameters.

In Fig.~(\ref{fig:CMB}) we show the predicted CMB temperature anisotropies 
$\ell\left(1+\ell\right)C_{\ell}/2\pi$ versus the multiple $\ell$ together with 
the WMAP 5-year data, see Ref.~\cite{cmb-5y}.

\subsection{Baryon acoustic oscillations}\label{sec-standard-ruler}

The acoustic peaks in the CMB power spectrum are predicted to be present also 
in the clusters of galaxies as a series of weak fluctuations (as 
a function of scale), usually called {\em wiggles}, see 
Ref.~\cite{meikins,springel,seo-eisenstein-bao,white-bao,eisensteinetal}.

\begin{figure}
\begin{centering}
\epsfig{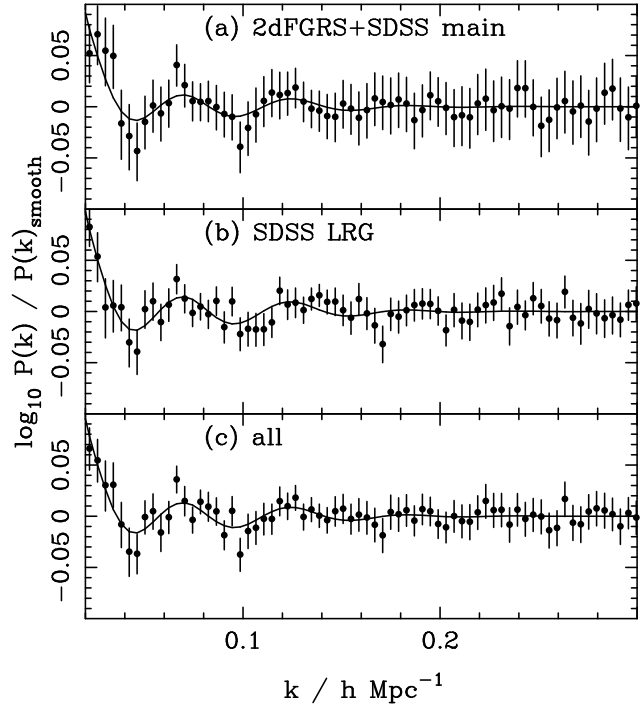}
\caption{Ratio of the full observed power spectra to smoothed spectra to show 
the oscillating BAO component.  
The power spectra are calculated from (a) the combined SDSS and $2$dFGRS 
main galaxies, (b) the SDSS DR$5$ LRG sample 
and (c) the combined samples (solid symbols with $1\sigma$)
The solid line in each panel is the fit to the data. From Refs.~\cite{w-percival-bao.}.}
\label{fig:bao-07}
\end{centering}
\end{figure}

Measuring the acoustic peaks in the galaxy power spectrum at high precision, 
matching those already measured in the CMB power spectrum, 
would provide a spectacular confirmation of the standard cosmological model 
in which mass over densities grow from the initial fluctuations present in the CMB. 
In Fig.~(\ref{fig:bao-07}) are plotted the power spectra divided by a smooth 
power spectrum (without baryon oscillations).

The position of the peaks and troughs in Fourier space can be calculated 
form linear physics and they are considered as a {\em standard ruler}, 
see Refs.~\cite{blake-glaze,seo-eisenstein-bao-ruler}. 
Therefore an analysis of the galaxy power spectrum in a redshift survey 
containing acoustic peaks can be used to measure cosmological parameters. 
BAO have been detected at low redshift in the $2$dFGRS and SDSS galaxy sample, 
see Refs.~\cite{coleetal,sdss_bao,huetsi06}.

What is a standard ruler?

The comoving size of an object at some redshift along the line of 
sight ($\lambda_{\parallel}$) and the transverse ($\lambda_{\perp}$) 
direction are related to the observed sizes $\Delta z$ and 
$\Delta \theta $ of the object by the Hubble 
parameter $H$ and angular diameter distance $D_{A}$. When the true scales, 
$\lambda_{\parallel}$ and $\lambda_{\perp}$, are known, measurements of 
the observed dimensions, $\Delta z$ (deep in space) and $\Delta \theta$ 
(wide in angle), give estimates of the Hubble parameter and the 
angular diameter distance: these are the {\em standard rulers}.

In turns the Hubble parameter and the angular diameter distance (integral over the 
redshift of the inverse of Hubble parameter) are related to the matter content of 
the Universe, and information on the cosmological parameters can be extracted.

\subsection{Large scale structures}

The conventional paradigm for the formation of structures in the Universe is based 
on the growth of small perturbations due to the gravitational instabilities. 
In this picture, some mechanism is required to generate small perturbations in 
energy density in the very early phase of the Universe. The central quantity 
that it is needed to describe the growth of structures is the density contrast, 
defined as:
\be 
\delta\left(t,\vec{x}\right)=\frac{\rho\left(t,\vec{x}\right)-\bar{\rho}\left(t\right)}{\bar{\rho}\left(t\right)}
\label{eq:density-contrast-gen}
\ee 
which describes the change in energy density $\rho\left(t,\vec{x}\right)$ 
compared to the background $\bar{\rho}\left(t\right)$.

Since one is often interested in 
statistical behaviour of structures in the Universe, it is conventional to 
assume that $\delta$ and other related quantities are elements of an ensemble. 
The most accredited model of structure formation suggests that the initial 
density perturbations in the early Universe can be represented as a Gaussian 
random variable and a given initial power spectrum. The last quantity is defined 
through the relation: 
\be
P\left(k\right)(2\pi)^3\delta_{D}\left(\vec{k}-\vec{k'}\right)=\langle\delta(\vec{k})\delta^{*}(\vec{k'})\rangle
\ee 
where $\delta(k)$ is the Fourier transform of $\delta\left(t,\vec{x}\right)$ and 
$\langle...\rangle$ means averaging over the ensemble. Though gravitational 
clustering will make the density contrast non Gaussian at late times, the power 
spectrum (and the correlation function) remains the most important quantity that 
needs to be studied in structure formation. 

When the density contrast is small, its evolution can be studied by linear 
perturbations theory and each Fourier mode $\delta_{k}\left(t\right)$ will grow independently. 

\begin{figure}[pb]
\centerline{\epsfig{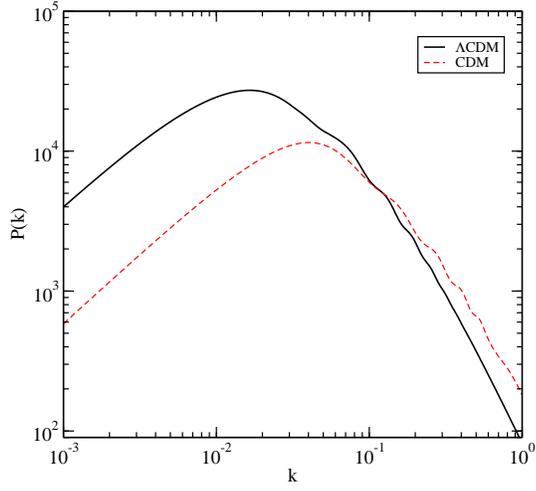}}
\vspace*{8pt}
\caption{The matter power spectrum in the flat $\Lambda{\rm CDM}$ model with 
$\Omega_{m_0}=0.289$  and in the ${\rm CDM}$ model with $\Omega_{m_0}=1$. 
The power spectra are evaluated with CAMB code. \label{fig:matterps}}
\end{figure}

We can say that the large scale structures of the Universe start to grow after 
matter-radiation equality ($t_{eq}$ or equivalently $a_{eq}$) 
when the pressure support due to photons becomes smaller than the gravitational 
attraction due to the non relativistic matter component. Since non-relativistic 
matter has a negligible pressure relative to its energy density, the gravitational 
attraction becomes stronger and objects in the Universe start to form. 
The matter-radiation equality fixes the position of the peak in the matter 
power spectrum, see Fig.~(\ref{fig:matterps}); the wavenumber $k_{eq}$ characterizes 
the border of the "large scale" and "small scale" modes and it corresponds to the 
one that entered the Hubble radius at the radiation-matter equality, i.e. 
$k_{eq}=a\left(a_{eq}\right)H\left(a_{eq}\right)$. 

The observations of large scale structures such as galaxy clustering provides 
another independent observational test for the existence of dark energy. 
In Fig.~(\ref{fig:matterps}) we plot the matter power spectrum for two different 
flat cosmologies: $\Lambda{\rm CDM}$ where $\Omega_{m_0} = 0.3$ and for ${\rm CDM}$ 
where $\Omega_{m_0}=1$, evaluated with CAMB code, see Ref.~\cite{camb}

For instance the peak position of the matter power spectrum $P_{m}\left(k\right)$ 
is shifted toward larger scales (smaller $k$) in the presence of the dark energy. 
Hence the scale of the peak position can be used as a probe of the dark energy.

\begin{figure}[pb]
\centerline{\epsfig{file=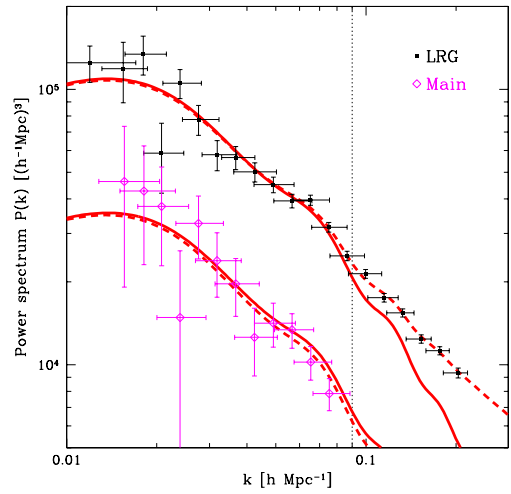,width=7.0cm}}
\vspace*{8pt}
\caption{The measured power spectra with error bars for the full luminous red 
galaxies (LRG) and the main galaxy sample of the 2dF survey. The solid curves 
show the theoretical prediction for the $\Lambda{\rm CDM}$ in the linear 
perturbation theory. The dashed curves include the non-linear correction 
to the matter spectrum. From Ref.~\cite{SDSSgal}.}
\label{fig:tegmarkps}
\end{figure}

In Fig.~(\ref{fig:tegmarkps}) are shown the galaxy power spectra for the Luminous 
Red Galaxy (LRG) and main galaxy samples of the SDSS, see Ref.~\cite{SDSSgal}. 
As it can be seen from the figure the position of the peaks are  around 
$0.01h{\rm Mpc}<k<0.02h{\rm Mpc}$ confirming that $\Lambda{\rm CDM}$ is favored 
over the ${\rm CDM}$ model. It is worth remembering that the galaxy power spectra 
alone do not provide tight bounds on the dark energy density parameter $\Omega_{DE_0}$ 
but the key point is that the observations of large scale structure are consistent 
with the existence of the dark energy, see Ref.~\cite{lss}.

\section{A bit of dark energy in theory}\label{DEmodels}

\subsection{Cosmological constant $\Lambda$}

The cosmological constant $\Lambda$ is the simplest candidate for the 
dark energy because it is characterized by a constant energy density in 
both time and space and its parameter of equation of state $w$ is 
constant and equal to $-1$. It comes also as quite natural 
term to be added in the Lagrangian, it appears in fact as a constant and 
linear term to $R$ (Ricci scalar).

From the point of view of particle physics the cosmological constant is considered to 
be the energy density of the vacuum.

As previously said the Einstein tensor $G_{\mu\nu}$ satisfies the Bianchi 
Identities $\nabla_{\mu}G^{\mu\nu}=0$ and the energy conservation is automatically 
verified $\nabla_{\mu}T^{\mu\nu}=0$. Since the metric tensor $g_{\mu\nu}$ is constant 
with respect the covariant derivative there is a freedom to add terms 
in the Einstein equation proportional to $g_{\mu\nu}$ without violating the total invariance. 
Then the Einstein equation given in Eq.~(\ref{Einstein}) can be written as:
\be
R_{\mu}-\frac{1}{2}g_{\mu\nu}R+\Lambda g_{\mu\nu} = 8\pi GT_{\mu\nu}\label{metric-lam}
\ee
where $\Lambda$ is a constant. In a FLRW metric the modified Einstein equation gives:
\bea
&&H^{2} = \frac{8\pi G}{3}\rho -\frac{K}{a^2}+\frac{\Lambda}{3} \label{dot-aoa}\\
&&\frac{\ddot{a}}{a}= -\frac{4\pi G}{3}\left(\rho+3p\right)+\frac{\Lambda}{3}.\label{ddot-a}
\eea
It is clear to see from Eq.~(\ref{ddot-a}) that $\Lambda$ contributes positively 
acting, at the background level, it acts as a repulsive force against gravity.

\subsection{The first problem: fine tuning}

Let us consider again the Hubble parameter:
\be
H^2= \frac{8\pi G}{3}\rho -\frac{K}{a^2}+\frac{\Lambda}{3}.
\ee
In order to have a phase of accelerated expansion today we require 
cosmological constant $\Lambda$ to be of the order of the square of the 
Hubble parameter today, that is:
\be
\Lambda \approx H_{0}^2 =(2.13h\times10^{-42}{\rm GeV})^2.
\ee
In terms of energy density this corresponds to:
\be
\rho_{\Lambda} = \frac{\Lambda m_{{\rm pl}}^2}{8\pi}\approx 10^{-47}{\rm GeV}^4
\ee
where $m_{\rm pl}= G^{-1}=1.22\times10^{19}{\rm GeV}$ is the Planck mass.
As previously said, the cosmological constant is associated with the 
vacuum energy of an empty space. The vacuum energy density is evaluated by 
the sum of the zero point energies of a quantum fields of mass $m$:
\be
\rho_{vac}= \int_{0}^{\infty}{\frac{\rmd^{3} k}{\left(2\pi\right)}^3\frac{1}{2}\sqrt{k^2+m^2}}
\ee
where $k$ is the momentum. Of course such an integral shows a ultraviolet 
divergence: $\rho_{vac}\approx k^{4}$. Our belief is that field theory is valid 
up to a certain maximum value of frequency $k_{max}$, moreover we expect 
General Relativity to hold just before the Planck scale: 
$m_{\rm pl}=1.22\times10^{19}{\rm GeV}$. Hence if we choose $k_{max}$ to be 
the Planck scale then the vacuum energy density will be:
\be
\rho_{vac}\approx \frac{k_{\rm pl}^4}{16\pi^2}\simeq10^{74}{\rm GeV}^4.
\ee
The ratio between the theoretical and the observed value of the cosmological constant is:
\be
\frac{\rho_{vac,th}}{\rho_{vac,ob}}=\frac{10^{74}{\rm GeV}^4}{10^{-47}{\rm GeV}^4}=10^{121}.
\ee
The theoretical value then is about $121$ order of magnitude larger than the observed value.

This ratio can be reduced if we assume a different cut-off scale $k_{max}$. 
We expect that in supersymmetry theories for a certain number of bosonic 
degrees of freedom there will be exactly the same number of fermionic degrees 
of freedom so that the total contribution to the zero point energy will vanish. 
If the supersymmetry is broken, as it certainly 
is nowadays because we do not live in a supersymmetric Universe, then 
the vacuum energy will be non-zero. It is known that supersymmetry is broken 
at scale: $M_{{\rm SUSY}}\approx 10^{3}{\rm GeV}$ which will reduce the ratio 
approximately by $60$ orders of magnitude but it is still not able to bring 
it down to the observed one. 
There have been a huge number of attempts in order to solve the fine tuning 
problem involving, QCD, supersymmetry, string theory, etc. a partial list 
is Refs. \cite{kachru}-\cite{Kane:2003qh}.
Anyway, none of them seems still convincing and moreover 
self satisfactory.

\subsection{The second problem: the coincidence}

The second problem of the dark energy is related again to its value today. 
Aside from its origin and the fine tuning shown in the previous subsection, 
its value today is comparable with the present matter energy density, 
in principle these two quantities should be unrelated. 
In other words, the value $\Omega_{\Lambda_0}$ is comparable 
(to a factor two or three) to $\Omega_{m_0}$, for no obvious reason. 

\begin{figure}[pb]
\centerline{\epsfig{file=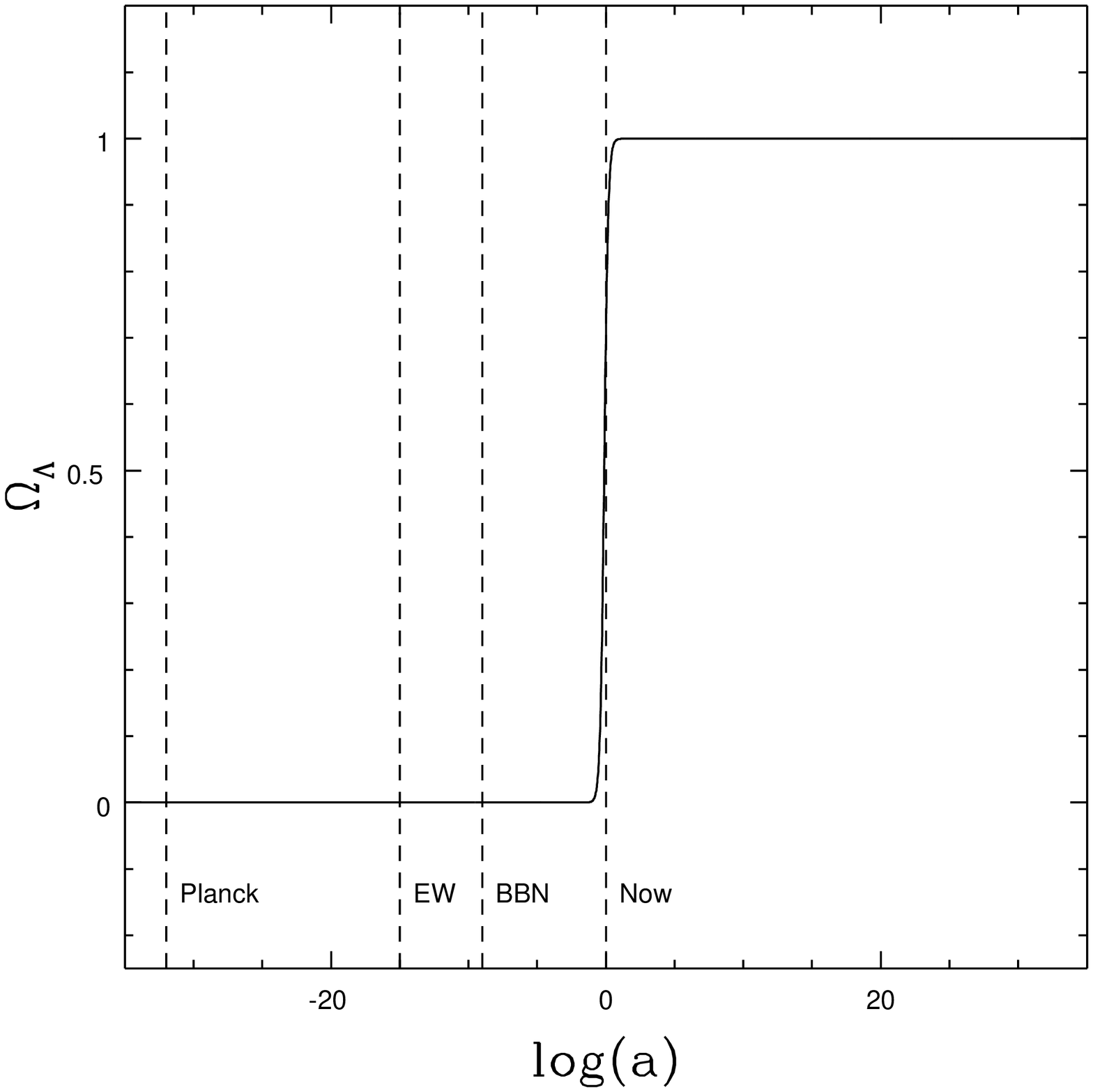,width=7.0cm}}
\vspace*{8pt}
\caption{Plot of the $\Omega_{\Lambda}$ versus $\log a$ assuming flat space 
with $\Omega_{\Lambda_0}=0.7$. There are also indicated the Planck era, the 
electroweak phase transition (EW) and the Big Bang Nucleosynthesis (BBN).
From Ref.~\cite{carroll}}
\label{fig:coincidence}
\end{figure}

Let us consider the energy density for a general fluid, that is:
\be
\rho\ltr=\rho_{0}a^{-3\left(1+w\right)}=\rho_{0}\left(1+z\right)^{3\left(1+w\right)}.
\ee
The time at which the matter energy density ($w_m=0$) coincides with the 
cosmological constant ($w_\Lambda = -1$) is given by:
\be
z_{c} = \left(\frac{\Omega_{\Lambda}}{\Omega_{m}}\right)^{1/3}-1
\ee
and for $\Omega_{\Lambda} \approx 0.7$ and $\Omega_{m} \approx 0.3$ 
(best fit values) then $z_{c}\approx 0.3$ (remember that $z=0$ means today). 
In Fig.~(\ref{fig:coincidence}) is shown $\Omega_{\Lambda}$ as a function 
of the scale factor $a$ for the $\Lambda{\rm CDM}$ model. 
As it can be seen, at early time the cosmological 
constant would be negligible, while at later times the density of matter 
will be zero and the Universe will be empty.Strangely (or hopefully) 
we are living in that short era, cosmologically speaking, when both 
dark matter and dark energy are comparable in magnitude. 
This is the {\em coincidence problem}.

This is not a characteristic of the cosmological constant $\Lambda$ only 
but of all the dark energy models.

As we will show in the next section there have also been here attempts to 
solve, at least partially, the coincidence problem.

\subsection{Scalar field models of dark energy}

The cosmological constant considered so far corresponds to a perfect fluid 
with an equation of state parameter $w=-1$ constant in time and space. 
This seems to be a good candidate because, as we have seen so far, 
observations give a value of $w$ close to $-1$. However, all the 
observations say little about the variation on time of such a parameter, 
so we cannot take for granted either that the cosmological constant 
is the {\em true} model or that the parameter of equation of state 
has always been constant. We can think to extend our horizon 
and to consider models in which the parameter of equation of state 
is a function of time.

The scalar fields is not a new concept to physics, it comes quite naturally in particle 
physics and it can be also a good candidate for dark energy.
So far an ample of dark energy models have been proposed: quintessence, K-essence, 
phantom, tachyonic, coupled dark matter and dark energy models, just to name the few. 

Here we are not interested on reporting the details of all the models 
present in the literature or all the details of a particular model. 
We are more interested on the impact on phenomenological cosmology or,  
so to say, we are more interested on the question: what should we do once 
we have a new model of dark energy?
We briefly discuss the different main models while giving a list 
of papers where more details can be found.

\subsubsection{Quintessence}

Caldwell {\it et al.}, see Ref.~\cite{caldwell}, named quintessence a canonical scalar 
field $\phi$ with a potential $V\left(\phi\right)$ responsible for the late time acceleration. 
The quintessence model is described by the action:
\be
S=\int{\rmd^{4}x\sqrt{-g}\left[\frac{1}{2k^2}R+\LL_{\phi}\right]}+S_M \label{action}
\ee
where the $R$ is the Ricci scalar, $k^2=8\pi G$, $S_M$ is the matter action\footnote{We take 
into account all the forms of matter since it is present in the Universe as well.} 
and $\LL_{\phi}$ is the scalar field Lagrangian, defined as:
\be
\LL_{\phi}= -\frac{1}{2}g^{\mu\nu}\partial_{\mu}\partial_{\nu}\phi-V\left(\phi\right).
\ee
The energy-momentum tensor of quintessence is:
\bea
T_{\mu\nu}^{(\phi)} &=&-\frac{2}{\sqrt{-g}}\frac{\delta\left(\sqrt{-g}\LL_{\phi}\right)}{\delta g^{\mu\nu}}\\
&=&\partial_{\mu}\partial_{\nu}\phi-g_{\mu\nu}\left[\frac{1}{2}g^{\alpha\beta}\partial_{\alpha}\phi\partial_{\beta}\phi+V\left(\phi\right)\right]
\eea
see Ref.~\cite{kolbturner} for more details.

The energy density and the pressure are, in a FLRW background:
\bea
\rho_{\phi} &=& -T_{0}^{0~(\phi)} = \frac{1}{2}\dot{\phi}^2+V\left(\phi\right)\\
p_{\phi} &=&\frac{1}{3}T_{i}^{i~(\phi)} = \frac{1}{2}\dot{\phi}^2-V\left(\phi\right)
\eea
and the parameter of equation of state is:
\be
w_{\phi} = \frac{p_{\phi}}{\rho_{\phi}}= \frac{\dot{\phi}^2+2V\left(\phi\right)}{\dot{\phi}^2-2V\left(\phi\right)}.
\ee
The scalar field also satisfy the continuity equation:
\be
\dot{\rho}_{\phi}+3H\left(\rho_{\phi}+p_{\rho}\right) = 0. \label{eq:continuity-phi}
\ee
We can also evaluate the Klein-Gordon equation just varying the action given in 
Eq.~(\ref{action}) with respect to $\phi$, that is:
\be
\ddot{\phi}+3H\dot{\phi}+\frac{\rmd V}{\rmd \phi}=0.
\ee
The Hubble parameter and  its first derivative are:
\bea
H^2 &=& \frac{8\pi G}{3}\left[\rho_{M}+\frac{1}{2}\dot{\phi}^2+V\left(\phi\right)\right] \\ \label{hubble-phi}
\dot{H} &=& -4\pi G\left[\rho_M+p_M+\dot{\phi}^2\right] \label{dothubble-phi}
\eea
where we used the continuity equation for  the matter component and the 
Klein-Gordon equation to evaluate Eq.~(\ref{dothubble-phi}).

The acceleration parameter will be given by:
\be
\frac{\ddot{a}}{a}= -\frac{4\pi G}{3}\left[\left(\rho_M+3p_M\right)+\left(\dot{\phi}^2-V\left(\phi\right)\right)\right]
\ee
and in order to have a phase of accelerated expansion today\footnote{We have shown before as 
relativistic and non-relativistic matter are unable to accelerate the Universe.} 
we need $\dot{\phi}^2<V\left(\phi\right)$. This means that we require a flat potential 
to accelerate the expansion. 

We can make a comparison with the inflation (where a scalar field is used in order 
to have an inflationary period) where two parameters are used, 
see Ref.~\cite{bassetwands}:
\be
\epsilon = \frac{1}{16\pi G}\left(\frac{1}{V}\frac{\rmd V}{\rmd \phi}\right)^2,~~~\eta =\frac{1}{8\pi G}\frac{1}{V}\frac{\rmd^2V}{\rmd\phi^2}\label{slow-roll-conditions}.
\ee
Inflation occurs when $\epsilon\ll1$ and $\left|\eta\right|\ll1$, these are called 
slow-roll conditions. If these conditions are satisfied we have:
\be
\dot{\phi}^2\ll V\left(\phi\right),~~|\ddot{\phi}|\ll |3H\dot{\phi}|. \label{slowrollcondi}
\ee
We can write the equation of state parameter as:
\be
1+w_{\phi} = \frac{\dot{\phi}^2}{\frac{1}{2}\dot{\phi}^2+V\left(\phi\right)} = \frac{16\pi G\epsilon V\left(\phi\right)}{9H^2\left(1+\xi\right)^2\rho_{\phi}}
\ee
where $\xi = \ddot{\phi}/3H\dot{\phi}$. The above equation also shows that the 
equation of state parameter for the quintessence can be bigger or at most equal 
to $-1$ (for positive values of potentials). In the slow roll limit, 
Eq.~(\ref{slowrollcondi}), we have $\rho_{\phi}\approx V\left(\phi\right)$ 
and $\xi\ll 1$ implying $1+w_{\phi} = 2\epsilon/3$. 
The deviation of $w_{\phi}$ from $-1$ is determined by the slow roll parameter $\epsilon$.

There is now the problem in finding the proper potential which is able to lead to an 
accelerated expansion. For this purpose it is interesting to consider, as an example, 
a scalar field which gives a power law expansion: 
\be
a\ltr\propto t^{p}
\ee 
and in order to have an accelerated expansion we need $p>1$. Let us now 
consider Eq.~(\ref{dothubble-phi}) with only a scalar field:
\be
\dot{H} = -4\pi G\dot{\phi}^2
\ee
where we have neglected matter for a sake of simplicity. From the last equation we 
can obtain a solution for the scalar field:
\be
\phi = \int{\left(-\frac{\dot{H}}{4\pi G}\right)^{1/2}\rmd t}\label{phi-1}
\ee
and from Eq.~(\ref{dothubble-phi}) we have the potential:
\be
V\left(\phi\right) = \frac{3H^2}{8\pi G}\left(1+\frac{\dot{H}}{3H^2}\right). \label{V-phi-1}
\ee
Solving Eq.~({\ref{phi-1}}) and (\ref{V-phi-1}) we have:
\be
\phi = \sqrt{\frac{p}{4\pi G}}\ln t,~~V\left(\phi\right)=V_0\exp\left(-\sqrt{\frac{16\pi G}{p}\phi}\right)
\ee
where $V_0 = p(3p-1)/8\pi G$.

The exponential potentials can give rise to the accelerated expansion, moreover 
they have cosmological scaling solution in which the energy density of the field 
$\rho_{\phi}$ follows the energy density of matter $\rho_{m}$.

Of course all the potential that are not steeper than the exponential potential can 
lead to an accelerated expansion; for instance a class of potentials is:
\bea
V\left(\phi\right) &=& M^{4+p}\phi^{-p}~~(p>0) \\
V\left(\phi\right) &=& M^{4+p}\phi^{-p}\exp(\alpha\phi^2/m_{\rm pl}^2)
\eea
where $\alpha$ is a positive number 
and $M$ is a constant; for the first class of models the potentials do not possess 
any minimum and  the field just roll down to its potential towards infinity; 
whereas for the second class of models the potentials have a minimum 
at which the field is trapped corresponding to $w_{\phi}=-1$, 
see Ref.~\cite{ratrapeebles}.

In general it is complicated to construct a quintessence model because it is difficult to 
find a corresponding model in particle physics due to the energy scale which is too 
low. As we said, in order to lead to the acceleration today we require a rather flat 
potential to satisfy the condition $\epsilon\ll 1$ and $|\eta |\ll1$ (slow roll 
parameters given in Eq.~(\ref{slow-roll-conditions})). We can also give an upper 
bound to the mass field of the quintessence starting from the slow roll condition:
\be
\left|\frac{\rm m_{pl}^2}{8\pi}\frac{1}{V}\frac{\rmd^2 V}{\rmd \phi^2}\right|\ll1.
\ee 
The mass of the field is simply given by second derivative of the potential 
$m_\phi=\rmd^2 V/\rmd \phi^2$, hence the mass field of the quintessence today 
has to satisfy the condition:
\be
\left|m_{\phi}\right|\lsim \sqrt{\frac{8\pi V_0}{m_{\rm pl}^2}}H\simeq H_0\approx 10^{-33}{\rm eV}
\ee
which means that $m_\phi$ has to be extremely small in order to have an accelerated 
phase today. Nevertheless another problem arises because such a light scalar field 
could in principle (and easily) couple with ordinary matter; but such a 
coupling could lead to a time dependence on the constants of nature. 
However, there are tight constraints on the time variation of the 
constants of nature and, clearly, any viable model must also satisfy these constraints, 
see Ref.~\cite{carroll1}.

We should also have in mind that the quintessence models have been proposed 
to explain (or alleviate) the coincidence problem; this is because scalar field models 
have global attractor solutions which sub-dominantly track the dominant 
component of the cosmological fluid; however, it is still not clear which has to 
be the mechanism to lead quintessence to dominate at late times and 
fine-tuning the initial conditions is required. 

The key problem here is to find a physically motivated and self-consistent quintessence 
model able to account for a late time acceleration in the Universe, a partial list of 
quintessence models motivated by particle physics is given in 
Refs. \cite{binetruy}-\cite{chibasiino}.

\subsubsection{K-essence}

Quintessence is based on scalar fields which possess flat potentials 
to lead to a late time acceleration of the Universe. 

However, it is also possible that the acceleration of the Universe arises from a 
modification of the kinetic term of the scalar field. Originally this class of 
model was introduced to drive inflation at high energies, the so called K-inflation, 
see Refs. \cite{armendariz1,garrida}; recent works extended this 
model and it was realized that it could have been also applied to the late time Universe, 
giving the name: K-essence, see Refs. \cite{armendariz2,armendariz3}.

Moreover, K-essence models were introduced as a good dark energy candidates because within 
a huge range of initial conditions the field automatically goes to a period in 
which $w\simeq -1$. In this scenario both initial conditions and coincidence problems 
are almost solved; however, other instabilities arise.

The K-essense is characterized by a non-canonical kinematic term; the most general 
action for such model is
\be
S=\int{\rmd^4x\sqrt{-g}P\left(\phi,X\right)} \label{action-k-essence}
\ee
where the Lagrangian density $P\left(\phi,X\right)$ corresponds to the pressure 
$p_{\phi}$; here $\phi$ is the scalar field and $X=-\frac{1}{2}\left(\nabla\phi\right)^2$.

From the action Eq.~(\ref{action-k-essence}) we can derive the energy momentum tensor:
\be
T_{\mu\nu}^{(\phi)} = -\frac{2}{\sqrt{-g}}\frac{\delta\left(\sqrt{-g}P\right)}{\delta g^{\mu\nu}} = P_{,X}\partial_{\mu}\phi\partial_{\nu}\phi+g_{\mu\nu}P
\ee
where the suffix $",X"$ represents the partial derivative with respect to $X$. 
The energy momentum tensor of the K-essence is the same as that of the perfect 
fluid $T_{\mu\nu}=\left(\rho+p\right)u_{\mu}u_{\nu}+p~g_{\mu\nu}$ with velocity 
$u_{\mu} = \partial_{\mu}\phi/\sqrt{2X}$ and pressure and energy density:
\bea
p_{\phi} &=& P\left(\phi,X\right) \\
\rho_{\phi} &=&2X\frac{\partial P\left(\phi,X\right)}{\partial X}-P\left(\phi,X\right).
\eea 
The equation of state of the K-essence is:
\be
w_{\phi} = \frac{p_{\phi}}{\rho_{\phi}}= \frac{P}{2XP_{,X}-P}\label{w-k-essence}
\ee
where we omitted the $\phi$ and $X$ dependence in $P$ for simplicity. Moreover 
the K-essence field satisfies the continuity equation:
\be
\dot{\rho}_{\phi}+3H\left(\rho_{\phi}+p_{\phi}\right)=0.
\ee
We can evaluate the propagation speed of sound $c_{s}^2$, which is defined as:
\be
c_{s}^2=\frac{\partial p_{\phi}/\partial X}{\partial\rho_{\phi}/\partial X}=\frac{P_{,X}}{P_{,X}+2XP_{,XX}}.
\ee
Equation (\ref{w-k-essence}) shows that the $P$ plays a crucial role in 
determining the equation of state of the scalar field $\phi$. Usually K-essence 
models have a Lagrangian of the form $P\left(\phi,X\right)=f\left(\phi\right)L\left(X\right)$ 
but the choice is basically arbitrary.

Moreover, the idea is that the field $\phi$ is considered as a low 
energy effective degree of freedom of some fundamental theory; such a model 
should respect some basic properties, for instance Lorentz invariance and causality. 
It has been shown that K-essence field during some epoch propagates with a speed 
larger than the speed of light, breaking clearly causality, see Ref.~\cite{camille}.
 
At last, it might well be that the linear kinetic energy in $X$ has a negative sign 
leading to quantum instabilities, such a field is called phantom or ghost scalar field.
A partial list of K-essence models can be found in 
Refs.~\cite{barger-marfatia}-\cite{chibascherrer}.

\subsubsection{Phantom}

All the models considered so far have a characteristic equation of state parameter 
whose value cannot be less than $-1$. However, recent observations indicate that 
the value of $w_{DE}$ has to be around $-1$ and moreover, they 
allow the possibility of a $w_{DE}<-1$. 
If we require that the dark energy equation of state parameter to be less than $-1$ 
then from Eq.~(\ref{w-k-essence}) we have:
\be
P_{,X}<0 \label{phantom-condition}.
\ee
The simplest model that satisfies the Eq.~(\ref{phantom-condition}) is a scalar 
field with a negative kinetic energy:
\be
P\left(\phi, X\right) = -X-V\left(\phi\right)
\ee
where $V\left(\phi\right)$ is the field potential. This particular scalar field 
is called phantom or ghost scalar field; its energy density and pressure are:
\be
\rho_{\phi}=-\frac{1}{2}\dot{\phi}^2+V\left(\phi\right),~~~p_{\phi}=-\frac{1}{2}\dot{\phi}^2-V\left(\phi\right).
\ee
The parameter of equation of state is:
\be
w_{\phi} = \frac{\dot{\phi}^2+2V\left(\phi\right)}{\dot{\phi}^2-2V\left(\phi\right)}
\ee
and for $\dot{\phi}^2/2<V\left(\phi\right)$ we have $w_{\phi}<-1$. 

However, there is also the problem of finding a self consistent and physically 
motivated scalar potential. In top of that, if the equation of state parameter is always 
less than $-1$ this will lead to a big-rip singularity. For instance, 
if the field potential has the form
\be
V\left(\phi\right)=V_{0}e^{-\mu\phi}
\ee
where $\mu$ is a constant then the equation of state will be always $w_{\phi}<-1$. 
The big-rip singularity can be avoided if other field potentials are considered, for instance:
\bea
V\left(\phi\right) &=& V_{0}e^{-\phi^2/\sigma^2} \\
V\left(\phi\right) &=& V_{0}\left[\cosh\left(\beta\phi m_1{pl}\right)\right]^{-1}\\
\eea
and these are called bell type potentials. In this case the equation of state starts from a 
value less than $-1$ but approaches to $w_{\phi}=-1$; this is because the phantom 
field climbs up the potential, due to its negative kinetic energy, but after a 
certain period of time the field settles at the maximum of the potential.

Even if the dynamics of a phantom field seem acceptable at the classical level 
unfortunately it suffers from ultra violet quantum instabilities, this is because 
the energy density of a phantom field is not bound from below, which leads to 
an unstable vacuum state, see Refs.~\cite{Carrollwless} and \cite{saridakis}. In order to avoid 
such a problem we need to have a phantom field that is weakly couple to any 
other normal field; but even if we assume that the phantom field is not coupled to 
any form of matter, they couple to gravitons which mediate the process:
\be
{\rm Vacuum} \rightarrow 2\phi+2\gamma
\ee
$\phi$ being a ghost and $\gamma$ a photon. In order not to have an overproduction 
of photons we required a Lorentz invariance breaking with a cut-off at the 
scale of few ${\rm MeV}$, such a breaking still does not have a natural 
interpretation in the standard model, see Ref.~\cite{Cline}.

A Lagrangian with a negative kinetic term of the form $P = -X+V(\phi)$ not 
always leads to quantum instabilities. If higher order kinetic terms are added in the 
Lagrangian, i.e. $P = -X+X^2$, instabilities 
can be avoided. However, it is difficult to use such a Lagrangian as a 
candidate for dark energy itself, this comes directly from the small value of 
the energy density of the scalar field which gives $X\gg X^2$, though the 
stability cannot always be guaranteed. 
See Refs.~\cite{nojiriodint-phontom}-\cite{dutta-scherrer-phantom} for a list of papers 
on the phantom model.

\subsection{Dark energy as a fluid}

\subsubsection{Perfect fluid}

So far we have considered only scalar fields as a candidate for dark energy; 
an entire class of models exists involving perfect fluids. 
Here we consider a single fluid model with a general energy momentum 
tensor\footnote{We apologize for being sometimes too repetitive.}:
\be
T_{\mu\nu}=\left(\rho+p\right)u_{\mu}u_{\nu}+pg_{\mu\nu}+\left[q_{\mu}u_{\nu}+q_{\nu}u_{\mu}+\pi_{\mu\nu}\right]
\label{eq:general-EM}
\ee
where, beside the familiar symbols $\rho$, $p$ and $u_{\mu}$ for the energy density, 
pressure and four-velocity, 
we introduced the heat flux vector $q_{\mu}$ and the viscous shear tensor $\pi_{\mu\nu}$. 
The terms appearing in the square brackets in the Eq.~(\ref{eq:general-EM}) are important 
only when the internal energy of the fluid is comparable to the total energy 
density. For a perfect fluid they are all zero. 

The covariant derivative $T_{\nu;\mu}^{\mu}=0$ gives the conservation equation:
\be
\dot{\rho}+3H\left(\rho+p\right)=0
\ee
valid at background level. From the above equation we can derive the expression 
for the energy density as a function of time under the assumption that the 
pressure is a function of the energy density $p = w\rho$. 
For the sake of generality, we consider $w$ to be a function of time, 
the energy density scales as:
\be
\rho= \rho_0\exp\left[-3\int_{a_0=1}^{a}{\frac{1+w(a')}{a'}\rmd a'}\right]
\label{eq:w-general}
\ee
where $\rho_0$ is the energy density of the fluid today.

Eq.~(\ref{eq:w-general}) can be written in a more familiar way:
\be
\rho=\rho_0a^{-3\left(1+\hat{w}\right)}
\ee
where 
\be
\hat{w}\left(a\right)=\frac{1}{\ln a}\int_{1}^{a}{\frac{w\left(a'\right)}{a'}\rmd a'}
\label{eq:what}.
\ee
At background level a perfect fluid is characterized by its equation of 
state parameter $w$. 
There is now the problem to parameterise $w\left(a\right)$. As realized 
from the previous discussions, there are different possibilities: 
according to the preferred model the equation of state parameter 
can be a constant (as it was for the cosmological constant) or it can 
be a function of time (as it was for the scalar field).
There is not a unique general expression for the dark energy parameter 
of equation of state if we consider the dark energy as being a fluid.
The easiest thing to do is to Taylor expand $w\left(a\right)$ around $a_0= 1$; we have, 
see Refs.~\cite{chevpol,linder2003} (usually called CPL paramatererization):
\be
w\left(a\right) = w_0 + w_{a}\left(1-a\right) = w_{0} + w_{a}\frac{z}{1+z}\label{w-linder}
\ee
where $w_0$ is the value of the parameter of equation of state today and $w_a$ 
gives the variation on time of $w\left(a\right)$. The values of the 
parameters are usually assumed to be $w_0\sim -1$ because observations 
tell us that the value of the parameter of equation of state today
has to be close to $-1$ and $w_a\sim 0$ or small enough because we do not 
have any strong evidence that the dark energy varies so much over time.
Moreover, the expression given in Eq.~(\ref{w-linder}) is very practical because it 
also avoids the complication of having unrealistic behavior. e.g., $w\ll -1$.

\subsubsection{Chaplygin gas}

Another model involving a fluid is known as Chaplygin gas. Its pressure takes 
a simple form:
\be
p = -\frac{A}{\rho}\label{pressure-chaplygin}
\ee
where $A$ is a positive constant. The continuity equation Eq.~(\ref{continuity}) gives:
\be
\rho = \sqrt{A+\frac{B}{a^6}}
\ee
where $B$ is also a constant. The expression of the energy density for the Chaplygin gas 
has two interesting asymptotic behaviors:
\bea
\rho &\sim& a^{-3}~~{\rm when}~~a\ll \left(\frac{B}{A}\right)^{1/6}\\
\rho &\sim& -p\sim \sqrt{A} ~~{\rm when}~a\gg \left(\frac{B}{A}\right)^{1/6}.
\eea
The Chaplygin gas behaves as pressureless dust at early times (when $a$ is small) whereas 
at late time (when $a\sim 1$) it behaves as dark energy with $w\sim-1$ leading to an 
accelerated expansion; hence the Chaplygin gas can be also a good candidate for unified dark 
energy and dark matter models. However, observations exclude the Chaplygin gas being 
a good candidate for dark energy. This is because the sound speed (the velocity with which 
perturbation propagates in the fluid) of such a model takes:
\be
c_s^2= \frac{\partial p}{\partial \rho} = \frac{A}{\rho^2}=\frac{Aa^6}{B+Aa^6}=\begin{cases}
c_s^2\sim 0 & a\ll1\\
c_s^2\simeq 1 & a\sim1\end{cases}
\ee
hence it behaves as pressureless dust in the Matter Dominated Era ($a\ll 1$) and it 
approaches the speed of light at late times ($a\sim1$)\footnote{The importance of 
the sound speed will be clear when we discuss the perturbation of a fluid.}.  
A sound speed of the order of the speed of light has implications on the growth of 
perturbations. A non zero sound speed gives a non zero Jeans  
scale (characteristic scale for the growth of perturbations); the higher the sound speed 
the larger is the Jeans scale. All the perturbation smaller of such characteristic scale 
cannot grow, in particular they should manifest an oscillation behavior. The fact that 
the latter are not observed in the matter power spectrum has led people to rule out 
the Chaplygin gas as a competitive candidate. It was also found that combining the 
results from CMB and SNIa, the Chaplygin gas is ruled out at more than the $99\%$ 
confidence level, see Ref. \cite{AFBC}

This situation can be partially alleviated if we consider a generalized Chaplygin gas; 
in this case the pressure becomes:
\be
p = -\frac{A}{\rho^\alpha}
\ee
with $0<\alpha<1$. However, $\alpha$ has been constrained to be $\left|\alpha\right|<10^{-5}$ at 95\% 
confidence level, see Ref.~\cite{sandvik}. Then it seems that $\Lambda$CDM model is 
the prefered model.
A partial list of generalized Chaplygin gas models is 
Refs. \cite{fabris1}-\cite{jlu}.

\subsubsection{Coupled dark energy}

As we said there is still the problem of explaining why the value of the 
dark energy density today is comparable with the present matter energy density. 
This fact may suggests that the dark energy is coupled to dark matter. 
The problem is finding a physically motivated coupling term. Let us recall 
the Einstein equation:
\be
G_{\mu\nu} = 8\pi GT_{\mu\nu}\label{Einstein-coupled},
\ee
we know that from the Bianchi Identity, the Einstein tensor $G_{\mu\nu}$ is conserved 
which guarantees the conservation of the energy momentum tensor:
\be
\nabla_\mu T_{\mu\nu}=0\label{conservation-coupled}.
\ee
However, the energy momentum tensor that appears on the right hand side of 
Eq.~(\ref{Einstein-coupled}) has to be considered as the sum of the 
energy momentum tensors of all the species present in  the Universe, 
$T_{\mu\nu}=\sum_{i} T^{(i)}_{\mu\nu}$. Usually it is taken for granted that all the 
species are decoupled (but always coupled through gravity)
so that the Eq.~(\ref{conservation-coupled}) holds independently for all 
the species: $\nabla_\mu T^{(i)}_{\mu\nu}=0$. However interactions 
between different fluids, say matter and dark energy, might be likely 
to happen. Hence Eq.~(\ref{conservation-coupled}) (which is always valid) 
is the starting point for finding the coupled model. 

Several coupling forms between dark energy and dark matter have been discussed in literature. 
Amendola proposed, in Ref.~\cite{amendola-coupled}, a quintessence scenario coupled with 
dark matter as a direct consequence of non minimally coupled theories; the coupling term has 
the form $Q\rho_{m}\dot{\phi}$.

Another interaction, which is based on the fluid description, is $H\rho_{m}\delta$, 
where $\delta$ is a dimensionless coupling, see Ref.~\cite{majsapam}, \cite{fuzfa}, 
\cite{dalal}, \cite{zimdhal}, \cite{campoherrera}, \cite{caldera}.

As an example, let us consider the dark matter and the dark energy, coupled together, 
with an interaction of the form:
\bea
&&\dot{\rho}_{m}+3H\rho_{m} = H\rho_{m}\delta \label{m-coupled} \\ 
&&\dot{\rho}_{de}+3H\left(\rho_{de}+p_{de}\right) = -H\rho_{m}\delta \label{de-coupled}.
\eea
The coupling $\delta$ can either be constant or time varying

\begin{itemize}
\item Constant $\delta$ 

The Eqs.~(\ref{m-coupled}) and (\ref{de-coupled}) can be solved easily, giving:
\bea
\rho_{m}(a)&=&\rho_{m_0}a^{-3+\delta} \\
\rho_{de}(a)&=&\rho_{de_0}a^{-3(1+w_{de})}+\frac{\rho_{m_0}\delta}{\delta+3w_{de}}\left[a^{-3(1+w_{de})}-a^{-3+\delta}\right]
\eea
where we assumed $w_{de}$ to be constant. The Hubble parameter then reads:
\be
\frac{H^2}{H_{0}^2} = \frac{\Omega_{m_0}}{\delta+3w_{de}}\left[3w_{de}a^{-3+\delta}+\delta a^{-3\left(1+w\right)}\right]+\Omega_{de_0}a^{-3(1+w_{de})}.
\ee
Observational constraints on the coupling constant $\delta$ are plotted 
in Fig.~(\ref{plot-del-const}), combining SNIa, CMB and BAO.

\begin{figure}
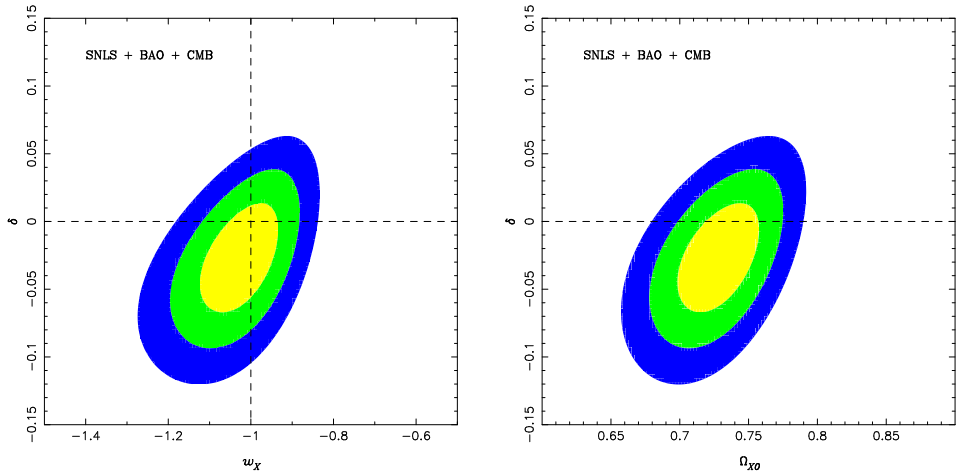

\begin{center}
\includegraphics[width=60mm]{conall.eps}\hspace{5mm}
\includegraphics[width=60mm]{conall2.eps}
\caption{Probability contours at $1\sigma$, $2\sigma$ and $3\sigma$ respectively combining 
SNLS, BAO and CMB data.
The left panel shows observational contours
in the ($w_X, \delta$) plane marginalized over
$\Omega_{X0}$, whereas the right panel shows contours
in the ($\Omega_{X0}, \delta$)
plane marginalized over $w_X$.
The best-fit model parameters correspond to
$\delta=-0.03$, $w_X=-1.02$ and
$\Omega_{X0}=0.73$. Form Ref.~\cite{guota}}
\label{plot-del-const}
\end{center}
\end{figure}

\item Time varying $\delta$ 

In order to have an expression for the Hubble parameter we need to find a parameterisation 
of $\delta(a)$. Generally speaking, we are faced with two possibilities: either we write 
down a parameterisation for $\delta(a)$ and then find from this the relation between $\rho_{m}(a)$ 
and $\rho_{de}(a)$, or first give the latter and then derive a function $\delta(a)$. 
Here we show this second approach.
The basic relation we start from is that the fluid densities scale according to the 
following relation, see Ref.~\cite{dalal}
\be
\frac{\rho _{de}(a)}{\rho _{m}(a)}=\frac{\rho_{de_0}}{\rho_{m_0}}a^{\xi }\label{rap}
\ee
where $\xi $ is a constant parameter. Moreover, we approximate $w_{de}$ as a constant: 
it is clear however that a more complete analysis should allow for a time-dependent equation of state.

The relation given in Eq.~(\ref{rap}) has two useful properties: \emph{a}) it includes 
all scaling solutions (defined as those with $\rho \sim a^{m}$) and \emph{b}) the functions 
$\rho _{m}(a)$ and $\rho _{de}(a)$ can be calculated analytically. Since for uncoupled dark energy 
models with constant equation of state one has $\rho _{m}\sim a^{-3}$ and 
$\rho _{de}\sim a^{-3(1+w_{de})}$ it appears that the relation Eq.~(\ref{rap}) reduces to 
this case for $\xi =-3w_{de}$. Conversely, if $\xi \not =-3w_{de}$ then the matter density 
deviates from the $a^{-3}$ law.

To find a parameterisation of $\delta(a)$ we can derive with respect to time Eq.~(\ref{rap}) 
and substituting Eq.~(\ref{m-coupled}) and Eq.~(\ref{de-coupled}), we find then:
\be
\delta\left(a\right) = -\Omega_{de}\left(a\right)\left(3w_{de}+\xi\right) \label{delta-coupled}
\ee
where $\Omega_{de}(a)=\rho_{de}(a)/(\rho_{de}(a)+\rho_{m}(a))=\Omega_{de_0}/(\Omega_{de_0}+\Omega_{m_0}a^{-\xi})$, 
hence we can write Eq.~(\ref{delta-coupled}) as:
\be
\delta(a)=\frac{\delta_0}{\Omega_{de_0}+\Omega_{m_0}a^{-\xi}}
\ee
where $\delta_0 = -\Omega_{de_0}\left(3w_{de}+\xi\right)$ is the present value of the coupling.

Using Eq.~(\ref{m-coupled}) and (\ref{de-coupled}) with the Eq.~(\ref{rap}) we can find the 
conservation equation for the effective fluid $\rho_{tot}=\rho_{m}+\rho_{de}$, that is:
\be
\dot{\rho}_{tot}= -3H\left[1+w\left(1+\frac{\rho_{m_0}}{\rho_{de_0}}a^{-\xi}\right)^{-1}\right]\rho_{tot}
\ee
integrating the latter we find the Hubble parameter:
\be
H^2 = H_0^2a^{-3}\left[\Omega_{m_0}+\Omega_{de_0}a^{\xi}\right]^{-3w_{de}/\xi}.
\ee
There are three parameters $\{w_0, \xi, \Omega_{de_0}\}$ to constrain, we can 
use $\delta_0$ instead of $\xi$ since they are related via 
$\delta_0 = -\Omega_{de_0}\left(3w_{de}+\xi\right)$. Data analysis using SNIa, 
CMB and BAO gives the bounds $-0.4<\delta<0.1$, $-1.18<w_{de}<-0.91$ and 
$0.69<\Omega_{de_0}<0.77$ at the $95\%$ confidence level, see Ref. \cite{guota}.

\end{itemize}

\section{Dark energy: modification of gravity}\label{MGmodels}

Because of the unfeasibility of solving the theoretical problems related to dark energy, 
cosmologists have started to look at other models. An alternative approach postulates that 
General Relativity is only accurate on small scales and has to be modified on cosmological scales. 
We have seen that the contribution of the matter content is given by the energy momentum tensor 
on the right hand side of the Einstein equation, whereas the left hand side represents the geometry. 
In order to have a late time accelerated phase we included an extra component, called dark energy, 
in the Universe which is meant to add an extra term on the right hand side of Einstein equation.
However, an effective accelerated phase can be also reproduced by modifying the geometry of the 
Universe or by simply change the laws of gravity. To understand that, let us assume that the 
Newtonian gravity is described by:
\be
F=G\frac{m~M}{r^2}e^{-r/r_0}\label{eq-newt-ex}
\ee
where $m$ and $M$ represent the masses of two objects, $r$ is their distance and $r_0$ 
is some typical fixed distance. For small value of $r$ Eq.~(\ref{eq-newt-ex}) becomes similar 
to the traditional Newtonian force, i.e. $F\simeq G\frac{m~M}{r^2}$ and we are tempting 
to say that the last one represents the correct theory but if we look a little bit further, 
say at distances comparable to $r_0$ then the gravitational attraction  becomes weaker as 
a consequence we are led to say that there has been an extra force which has counterbalanced 
the gravitational attraction. This is what happens in the modified gravity models. 

This review is not meant to explain in detail the different models existing nowadays 
but for the sake of completeness we will briefly list the most studied modified gravity 
models, pointing out the main characteristics useful for the phenomenological cosmology.

\subsection{$f\left(R\right)$ gravity}

The Einstein-Hilbert action
\be
S=\int{\Big[\frac{1}{2k}R+\LL_{M}}\Big]\sqrt{-g}\rmd^4x
\ee
is the base of General Relativity, by simply varying the latest with respect to 
the metric we can obtain the Einstein equations seen before. 

The $f\left(R\right)$ gravity intends to modify the action itself; in principle 
there is no {\em a priori} reason to consider a gravitational Lagrangian as a linear 
function of the Ricci scalar $R$. Let us consider an action of the form, 
see Refs.~\cite{capoz,carroll-d}:
\be
S=\int{f\left(R\right)\sqrt{-g}\rmd^4x}\label{f-R-action}
\ee
where $R$, in the Einstein-Hilbert action, has been replaced by a generic 
function $f\left(R\right)$ of $R$. 
The field equation can be obtained by varying the action (\ref{f-R-action}) 
with respect to the metric $g_{\mu\nu}$:
\be
f'(R)R_{\mu\nu}-\frac{1}{2}f(R)g_{\mu\nu}-\Big[\nabla_\mu\nabla_\nu-g_{\mu\nu}\square\Big]f'(R)=8\pi GT_{\mu\nu}\label{eq:field-fR}
\ee
where $\square=g_{\mu\nu}\partial^{\mu}\partial^{\nu}$ is the d'Alembert operator, 
the prime refers to the derivative with respect the Ricci scalar $R$, i.e. 
$f'(R) = \partial f(R)/\partial R$, and $T_{\mu\nu}$ is the matter energy 
momentum tensor. Taking the trace of the Eq.~(\ref{eq:field-fR}) we have:
\be
3\square f'(R)+Rf(R)-2f(R)=8\pi GT \label{eq:trace-field-fR}
\ee
being $T=-\rho+3p$ the trace of the energy momentum tensor. 

Working in the flat FLRW metric, Eqs.~(\ref{eq:field-fR}) and (\ref{eq:trace-field-fR}) read:
\bea
H^2 &=& \frac{8\pi G}{3f'(R)}\Big\{\rho +\frac{1}{2}\left[Rf'(R)-f(R)\right]-3H\dot{R}f''(R)\Big\} \label{hubble-fR}\\
2\dot{H}+3H^2 &=& -\frac{8\pi G}{f'(R)}\Big\{ p +\dot{R}^2+2H\dot{R}f''(R)-\ddot{R}f''(R)+\\
&+& \frac{1}{2}\left(f(R)-Rf(R)\right)\Big\}.
\eea
As we said $f(R)$ theories are interesting because they lead to accelerated expansion 
without the need for dark energy. To see this we can define an effective energy density and pressure as:
\bea
\rho_{eff} &=& \frac{1}{2f'(R)}\Big[Rf'(R)-f(R)\Big]-\frac{3}{f'(R)}H\dot{R}f''(R) \\
p_{eff} &=&\frac{1}{f'(R)}\Big\{\dot{R}^2f'''(R)+2H\dot{R}f''(R)+\ddot{R}f''(R)+\frac{1}{2}\left[f(R)-Rf'(R)\right]\Big\}
\eea
where $\rho_{eff}$ has to be positive to guarantee a positive Hubble parameter 
(one way to see this is to set in Eq.~(\ref{hubble-fR}) $\rho=0$, hence, 
in vacuum the curvature correction can be viewed as an effective fluid).

The effective equation of state parameter is:
\be
w_{eff} = \frac{\dot{R}^2f'''(R)+2H\dot{R}f''(R)+\ddot{R}f''(R)+\frac{1}{2}\left[f(R)-Rf'(R)\right]}{\frac{1}{2}\Big[Rf'(R)-f(R)\Big]-3H\dot{R}f''(R)}.
\ee
The effective parameter of equation of state depends on the function $f(R)$; 
let us give two examples: first, we consider a $f(R)\propto R^n$ and the scale 
factor is assumed to be a generic power law $a(t) = (t/t_0)^\alpha$, we obtain, see Ref.~\cite{capoz}:
\be
w_{eff}=-\frac{6n^2-7n^2-1}{6n^2-9n+3}
\ee
for $n\neq 1$ and $\alpha$ is given by:
\be
\alpha = \frac{-2n^2+3n-1}{n-2}.
\ee
According to the value of $n$ a desired value for $w_{eff}$ can be obtained. 
For instance if $n=2$ then $w_{eff}=-1$ and $\alpha=\infty$ as desired. 
However, this model was first considered as candidate of inflation where 
$f\left(R\right)=R+\alpha R^2$ which leads to an accelerated epoch but it ends 
as soon as $\alpha R^2<R$, see Ref.~\cite{starobinsky}.

The second example uses the function $f(R)= R-\mu^{2\left(n+1\right)}/R^n$, where $\mu$ 
is a generic parameter, see Ref.~\cite{capoz}; in this case, 
assuming the scale factor to be a power law, we have:
\be
w_{eff}=-1+\frac{2\left(n+2\right)}{3\left(2n+1\right)\left(n+1\right)}.
\ee
For instance, if $n=1$ then we have $w_{eff}=-2/3$, see Ref.~\cite{carroll-d}.
However, also this class of models are not good candidates for the late 
time acceleration for several reason: first of all they are excluded by solar system 
constraints (this is because $f''<0$), see Ref.~\cite{chiba,olmo1,navarro,chibaerick,nojiriodint}; 
then, in the Ref.~\cite{amenpolar,amenpolar1} was shown that this class of models 
lead to a different growth of the scale factor during matter dominated era, 
hence inconsistent with observations. 

It is also worth mentioning that some of the problems listed above can be solved if the 
Palatini formalism of $f\left(R\right)$ gravity is used; in such a formalism 
the independent variable is not anymore the metric tensor $g_{\mu\nu}$ but the 
connection $\Gamma_{\mu\nu}^{\tau}$. As a consequence the Ricci tensor calculated 
in the Palatini formalism $R_{\mu\nu}\left(\Gamma\right)$ will differ from the 
one calculated in terms of the metric tensor $R_{\mu\nu}\left(g\right)$.
Even though all the stabilities problems are solved at the background level, there 
is still the problem of the evolution of matter density perturbations which may 
manifest a violent growth or a series of damped oscillations being incompatible with 
observations, see Ref.~\cite{tsujuddin}.
 
Good review of $f(R)$ theory can be found in Ref.~\cite{sotiriou} and 
\cite{felicetsujikawa}.

\subsection{Brane world: DGP}

It might also well be that we live in an Universe with more than four dimensions 
(three spatial dimensions and one temporal dimension); although the other extra dimensions 
are compactified on some manifold in order to obtain 4-dimensional gravity theories. 
One of the best studied examples is the Dvali-Gabadadze-Parroti brane-world model (DGP), 
see Ref.~\cite{dgp}. In this model the 4-dimensional brane is embedded in a 5-dimensional 
Minkowski bulk with an infinitely large extra dimension. Matter is confined in the 
4-dimensional brane and only gravity can propagate in the 5-dimensional bulk. Gravity 
leaks off the 4-dimensional brane into the 5-dimensional bulk only on large scales, 
say $\lambda>r_{c}$. On small scales, $\lambda<r_{c}$, gravity is effectively bound to 
the brane and 4-dimensional dynamic is recovered with very good approximation. Then the 
gravity leakage leads to the observed late-time acceleration of the expansion of the 
Universe. In other words the acceleration is not due to the presence of a dark energy 
component but rather to the weakening of gravity.

The action, for this model, is:
\be
S=-\frac{M_{5}^2}{2}\int{\rmd^5y\sqrt{-g}R_{5}}-\frac{M_{pl}^2}{2}\int{\rmd^4x\sqrt{-h}R_{4}}+\int{\rmd^4\sqrt{-h}\LL_{m}}+S_{\rm GHY}\label{eq:action-DGP}
\ee
where $g_{\alpha\beta}$ is the metric in the bulk and $h_{\mu\nu}$ is the induced metric 
on the brane; $\LL_m$ is the matter Lagrangian confined, as we said, on the brane. The 
term $S_{\rm GHY}$ is the Gibbons-Hawking-York\footnote{The necessity of such a term was 
first pointed out by York and later it was refined by Gibbons and Hawking.} boundary term 
which needs to be added all the time the space time manifold has a 
boundary\footnote{Adding this extra term the variational principle is well-defined.}.

The characteristic scale we said before is given by the ratio of the Planck mass $M_{\rm pl}$ 
and its counter part $M_5$ in the 5-dimensional bulk, that is:
\be
r_{c}=\frac{M_{\rm pl}^2}{2M_5}.
\ee
The action (\ref{eq:action-DGP}) gives the following modified Hubble equation:
\be
H^2-\frac{\epsilon}{r_c}H=\frac{8\pi G}{3}\rho\label{eq-hubble-DGP-g}
\ee
where $\epsilon=\pm 1$, this because the DGP model gives two different solutions, we will 
consider here only the self-accelerating one: $\epsilon = 1$; $\rho$ is the total energy 
density of the fluid confined on the brane and it satisfies the standard conservation 
equation\footnote{Matter is confined on the brane and hence there is no flow of matter outside the brane.}:
\be
\dot{\rho}+3H\left(\rho+p\right)=0\label{eq:conserv-dgp}.
\ee
Eq.~(\ref{eq-hubble-DGP-g}) can be written, assuming only matter present on the brane 
and zero curvature, as:
\be
H=H_0\Big[\sqrt{\Omega_{r_c}}+\sqrt{\Omega_{r_c}+\Omega_{m}a^{-3}}\Big]\label{eq-hubble-DGP-2}
\ee
where $\Omega_m$ is the matter density parameter and
\be
\Omega_{r_c}=\frac{1}{4r_c^2H_0^2}.
\ee
Setting $z=0$ in Eq.~(\ref{eq-hubble-DGP-2}) we get:
\be
\Omega_{r_c}=\frac{\left(1-\Omega_m\right)^2}{4}.
\ee
Comparing Eq.~(\ref{eq-hubble-DGP-g}) to the normal Friedmann equation with an additional 
dark energy component, we see that we can move the crossover term to the right hand side 
and think of it as a dark energy contribution with $\rho_\de = 3H/(8\pi G r_c)$.
Looking at the conservation equation (\ref{eq:conserv-dgp}) we find that it is solved if the effective dark energy 
has an equation of state given by
\be
1+w_\de = -\frac{\dot{H}}{3H^2} .
\ee
However, things are not as nice as they look. In DGP model there has to be a sort of 
fine-tuning as well\footnote{Problem that we tried to solve in the $\Lambda$CDM scenario.}: 
we need to quantify the cross-over scale $r_c$. If we consider the equation Eq.~(\ref{eq-hubble-DGP-g}) 
we realize that in a CDM domination situation characterized by $\rho\propto a^{-3}$ the Universe 
will approach the de Sitter solution:
\be
H\rightarrow H_{\infty}=\frac{1}{r_c}.
\ee
This means we can have an accelerated phase at late time without invoking dark energy but 
the price to pay, because we require the accelerated phase to happen now, is that the 
cross-over scale has to correspond to the present Hubble radius today, i.e. $r_c\sim H_0$. 
With this fine tuning the DGP model seems to fit the SNe observations; 
however, when we add other observations such as BAO and CMB shift the DGP 
model is ruled out, see Ref.~\cite{martmajer}.

However, the DGP model itself suffers from several stability problems (namely ghost), 
see Ref.~\cite{kkoyama}-\cite{dgpr}; moreover it is also about to be 
excluded from observations, especially from solar system tests, see Ref.~\cite{fanget} 
and \cite{lombriser}. 
These difficulties have led people to extend the DGP model to higher dimensions, called 
cascading DGP, see Ref.~\cite{cascadingdgp1} and \cite{cascadingdgp2}. 
Recently, people have developed a covariant cascading gravity in $6{\it D}$ brane-world model 
to study the cosmological solutions, see Ref.~\cite{cascadingcosmology}. 

Another way to keep the DGP model ghost-free is to assume a light scalar field which is 
coupled to metric tensor; furthermore, it is fundamental to have only second order time derivatives 
in the field equations otherwise this would rise to extra degree of freedom which are likely 
to give rise to a ghost state. However, these theories might have superluminal excitations, 
see Ref.~\cite{nicolis} and \cite{silvakoyama}.

A review of brane-world cosmology can be found in Ref.~\cite{wands}.

\subsection{Neither of the two}

Instead of modifying the right hand side of Einstein equation (adding dark energy) or the 
left hand side (modifying gravity) to explain the observed late time acceleration, a 
third possibility is to drop the assumption that the Universe is homogeneous an large scales. 

The assumption that the Universe is homogeneous and isotropic is known to be violated 
at late times when structures formation starts to be non-linear. Therefore, it is 
important first to quantify the effect of the breakdown of linearity before invoking new physics. 

The main idea comes from the fact that the Friedmann equation is derived 
from the Einstein equation $G_{\mu\nu}=8\pi GT_{\mu\nu}$ which is non-linear in the metric $g_{\mu\nu}$
(the Einstein tensor $G_{\mu\nu}$ contains quadratic and quartic terms in the metric). 
The Universe expansion and the scale factor are supposed to describe the average dynamics of the 
Universe, hence they should be evaluated from the Einstein equation averaged over large scales:
\be
\langle G_{\mu\nu}\rangle=8\pi G\langle T_{\mu\nu}\rangle\label{eq:backreaction}.
\ee
However, the approach of the FLRW model is to define first the average 
metric $\langle g_{\mu\nu}\rangle={\rm diag}(-a^2,a^2,a^2,a^2)$, an average density and pressure 
and then to compute the Einstein equation, that is:
\be
G_{\mu\nu}\left(\langle g_{\mu\nu}\rangle\right) = 8\pi G T_{\mu\nu}\left(\langle\rho\rangle,\langle p\rangle\right)
\label{eq:usual-Einstein}.
\ee 
This equation leads to the Friedmann equation, but clearly Eq.~(\ref{eq:backreaction}) and Eq.~(\ref{eq:usual-Einstein}) 
do not give the same results because of the non linear structure of $G_{\mu\nu}$. Moreover 
$G_{\mu\nu}$ contains time derivatives and spatial derivatives of $g_{\mu\nu}$; hence, it is possible that 
$g_{\mu\nu}$ is very close to $\langle g_{\mu\nu}\rangle$ while at the same time $G_{\mu\nu}$ and 
$\langle G_{\mu\nu}\rangle$ may be significantly different.
Fortunately, this effect can be neglected before the time of equality of matter and 
radiation, because it is clear that the Universe is nearly homogeneous before that time, but 
departure from homogeneity should be taken into account when structures start to form.
The majority of cosmologists believe that the standard approach provides a good approximation, relying 
especially on the N-body simulations. A few experts challenge this picture and try to compute the proper 
averaging, see Refs.~\cite{buchert,rasanen}.

This idea is of course the most appealing and economical one that one might think of, 
but there are doubts that it could work. The problem is that technically it is 
impossible to solve analytically the Einstein equation (if special symmetries are considered) 
and moreover N-body simulation is still based on Newtonian gravity. 

However, the idea of considering the effect of inhomogeneities has led people to 
consider a different metric instead of the usual FLRW one. One example is the 
Lemaitre-Tolman-Bondi (LTB) model. In this model the Milky Way has to be near the center of 
a large underdense region, called void, in order to be consistent with SN Ia observations and 
CMB, see Ref.~\cite{garciahoeg1} and \cite{garciahoeg2}. 
Of course in this case we need to violate the Copernican principle requiring that 
we live a privileged location, see Refs~\cite{alnes,enqvist}.

\section{Dark energy or modify gravity?}\label{comparisons}

There is now strong observational evidence that the expansion of the Universe
is accelerating. The standard explanation invokes an unknown ``dark energy''
component. But such scenarios are faced with serious theoretical problems, which has
led to increased interest in models where instead General Relativity is
modified in a way that leads to the observed accelerated expansion. The
question then arises whether the two scenarios can be distinguished. Here
we show that this may not be so easy. We demonstrate explicitly that a
generalized dark energy model can match the 
Dvali-Gabadadze-Porrati model and reproduce the $3+1$ dimensional metric 
perturbations. Cosmological observations are then unable 
to distinguish the two cases.

\subsection{Background quantities: is this all we need?}

It is straightforward to explain the observed accelerated expansion of the 
late-time Universe, within the framework of FLRW cosmology by simply 
introducing a cosmological constant or a more general (dynamic) dark energy component.

An alternative approach postulates that General Relativity is only accurate up to 
certain scales and has to be modified on cosmological distances. This in turn
leads to the observed late-time acceleration of the expansion of the Universe, 
see Ref.~\cite{CD,BDEL,RM}.

One important question is whether a scenario where the gravity is modified 
can be distinguished from one invoking an invisible dark energy component. 
It is well known that {\em any} expansion history (as parameterised by the 
Hubble parameter $H(t)$) can be generated by choosing a suitable 
equation of state for the dark energy
(parameterised by the equation of state parameter $w=p/\rho$ of the dark
energy):
\be
w(z)=\frac{\frac{2}{3}\left(1+z\right)H(z)H'(z)-H(z)^2}{H(z)^2-H_0^2\Omega_{m}\left(1+z\right)^{3}}
\ee
where the prime refers to the derivative with respect to the redshift.
Let us illustrate this explicitly for the DGP model, for which the Hubble
parameter evolves as Eq.~(\ref{eq-hubble-DGP-g}); then the 
effective dark energy equation of state is given by:
\be
w(z) =-1 +\frac{1}{3}\left(1+z\right)\frac{H'(z)}{H(z)} .
\ee
Consequently, it is impossible to rule out dark energy based on measurements of
the cosmic expansion history (e.g. SN-Ia data).

Lately there have been claims that if we are able to measure the growth rate of structures 
then we can rule out different models. This because different theories predict different 
growth of structure, then fixing the parameter of equation of state $w$ (effective or not) 
we can predict the evolution of the growth rate; if we observe a different growth then the 
theory is wrong and can be ruled out.

\subsection{The role of perturbations}\label{pert-section}

Here we want to evaluate the growth of matter perturbations in two different scenarios: 
\begin{itemize}
\item General relativity and dark energy 
\item Modified gravity (without dark energy)
\end{itemize}
in order to verify the previous statement that at perturbations level different 
models, with the same equation of state parameter, can be distinguished.

In order to compare models of dark energy with observations like the Cosmic 
Microwave Background (CMB) and large scale structure (LSS), 
it is important to study the evolution of density perturbations in a 
Universe containing dark energy. 

As mentioned, the conventional paradigm for the formation of structures in 
the Universe is based on the growth of small perturbations due to the 
gravitational instabilities. 
The central quantity is the density contrast $\delta\left(t,\vec{x}\right)$, 
defined in Eq.~(\ref{eq:density-contrast-gen}).

The growth of perturbations depends on the characteristics of the fluid itself, 
namely pressure, pressure perturbation and perhaps anisotropic stress. 
For ordinary matter this quantities are known, so the growth of the density 
contrast can be easily evaluated. However, for dark energy these parameters are 
{\em a priori} unknown and need to be measured. Usually it is taken for granted 
that dark energy does not cluster on small scales because it is assumed that 
its sound speed is too high to allow perturbations to gravitationally 
collapse. For some specific models, i.e. scalar field models, 
this is a good approximation but for a general dark energy fluid this may 
not be the case and perturbations in energy density have to be studied. 

The dynamic evolution of dark energy can affect a number of observables, including the 
spectrum and the growth of large scale structures, weak gravitational lensing, 
SNIa apparent luminosity and CMB anisotropies. One particular manifestation 
occurs in the late time Integrated 
Sachs-Wolfe (ISW) effect, which measures the evolution of the gravitational potential as the 
Universe enters a phase of dark energy domination. This effect is only significant on large 
scales (low multiples), since small scale fluctuations in the gravitational potential smooth 
out along the line of sight. And it is only significant at late time since potentials evolve 
the same as the background during matter domination. The ISW effect has been detected in 
cross-correlations between CMB temperature anisotropies and surveys of large scale structure, 
see Ref.~\cite{bedo}.

It is imperative to deeply study the perturbations equations in order to compare 
different models with observations, as they clearly depend on them.

Here we provide the perturbation equations in a dark energy dominated Universe 
starting with a barotropic fluid and we then extend to a more general fluid.
We will concentrate then only to dark energy perturbations. 
However, also at perturbations level, dark energy suffers of serious stability 
problems. We face these difficulties and we show the attempts made 
in order to encompass these stabilities. 

For simplicity, we consider a flat  Universe containing only (cold dark) 
matter and dark energy, so that the Hubble parameter is given by
\be
H^2 = \left(\frac{1}{a}\frac{\rmd a}{\rmd t}\right)^2 = H_{0}^{2}\left[\Omega_m a^{-3}+\left(1-\Omega_m \right)a^{-3\left(1+\hat{w}\left(a\right)\right)} \right]
\ee
where $\hat{w}(a)$ is given in Eq.~(\ref{eq:what}).

We will consider linear perturbations about a spatially-flat background
model, defined by the line of element:
\be
\rmd s^{2} = a^{2} \left[ -\left( 1+2A\right) \rmd\eta^{2}+2B_{i}\rmd{\eta}\rmd x^{i}+\left( \left( 1+2H_{L}\right) \delta_{ij}+2H_{Tij} \right) \rmd x_{i}\rmd x^{j} \right]
\label{pert_0_ds}
\ee
where $A$ is the scalar potential; $B_{i}$ a vector shift; $H_{L}$ is the
scalar perturbation to the spatial curvature; $H_{T}^{ij}$ is the trace-free
distortion to the spatial metric, see Ref.~\cite{hu_lect} for more details.

We will assume that the Universe is filled with perfect fluids only,
so that the energy momentum tensor takes the simple form:
\be
T^{\mu\nu}=\left( \rho+p\right) u^{\mu}u^{\nu} +p~g^{\mu\nu}
\label{EMT}
\ee
where $\rho$ and $p$ are the density and the pressure of the 
fluid respectively and $u^{\mu}$ is the four-velocity.

The components of the perturbed energy momentum tensor can be written as:
\bea
T_{0}^{0} &=& - \left( \bar\rho + \delta\rho \right) \\
T_{j}^{0} &=& \left( \bar\rho + \bar{p} \right) \left( v_{j} - B_{j} \right) \\
T_{0}^{i} &=& \left( \bar\rho + \bar p \right) v^{i} \\
T_{j}^{i} &=& \left( \bar{p} + \delta{p} \right) \delta_{j}^{i} + \bar{p}~\Pi_{j}^{i}.
\eea
Here $\bar\rho$ and $\bar p$ are the energy density and pressure of the
homogeneous and isotropic background Universe,
$\der$ is the density perturbation, $\dep$ is the pressure perturbation,
$v^{i}$ is the vector velocity and $\Pi_{j}^{i}$ is the anisotropic stress
perturbation tensor which represents the traceless component of the $T_{j}^{i}$. 
So far the treatment of the matter and metric is fully general and applies 
to any form of matter and metric. 

As it was at background level we have to choose now a preferred metric. 
Here, we want to investigate only the scalar modes of the perturbation equations.
We choose the Newtonian gauge (also known as the longitudinal gauge) which is
very simple for scalar perturbations because they are characterized by two
scalar potentials $\psi$ and $\phi$; the metric Eq.~(\ref{pert_0_ds}) becomes:
\be
\rmd s^{2} = a^{2} \left[ -\left( 1+2\psi \right) \rmd\eta^{2} + \left( 1-2\phi\right) \rmd x_{i}\rmd x^{i} \right]
\label{pert_newton_ds}
\ee
where we have set the shift vector $B_{i}=0$ and $H_{T}^{ij}=0$. The advantage of using the
Newtonian gauge is that the metric tensor $g_{\mu\nu}$ is diagonal and
this simplifies the calculations. Moreover, as we will see later, it is very useful to use 
the Newtonian gauge for the observational tests.

In the conformal Newtonian gauge, the first-order perturbed Einstein
equations give, see Ref.~\cite{mabe} for more details:
\bea
k^2\phi + 3\frac{\dot{a}}{a} \left( \dot{\phi} + \frac{\dot{a}}{a}\psi\right) &=& 4\pi G a^2 \delta T^0{}_{\!0} \,,\label{ein-cona}\\
k^2 \left( \dot{\phi} + \frac{\dot{a}}{a}\psi \right)&=& 4\pi G a^2 (\bar{\rho}+\bar{P}) \theta\,,\label{ein-conb}\\
\ddot{\phi} + \frac{\dot{a}}{a} (\dot{\psi}+2\dot{\phi})+\left(2\frac{\ddot{a}}{a} - \frac{\dot{a}^2}{a^2}\right)\psi+ \frac{k^2}{3} (\phi-\psi)
&=& \frac{4\pi}{3} G a^2 \delta T^i{}_{\!i}\,,\label{ein-conc}\\
k^2(\phi-\psi) &=& 12\pi G a^2 (\bar{\rho}+\bar{P})\sigma\,,\label{ein-cond}
\eea
where $\sigma$ is related to $\Pi_{j}^{i}$ through:
\be
\left(\bar\rho+\bar{p}\right)\sigma = -\left(\hat{k}_i\hat{k}_j-\frac{1}{3}\delta_{ij}\right)\Pi_{j}^{i}.
\ee
The energy-momentum tensor components in the Newtonian gauge become:
\bea
T_{0}^{0} &=& -\left( \brho + \delta\rho \right) \label{00_EM}\\
ik_i T_{0}^{i} &=& -ik_i T_{i}^{0} = \left(\brho + \bap \right) \theta  \label{0i_EM}\\
T_{j}^{i} &=& \left( \bar p + \delta p \right) \delta_{j}^{i} +\bar{p}\Pi_{j}^{i} \label{ij_EM}
\eea
where we have defined the variable $\theta=ik_j v^j$ which represents the divergence
of the velocity field.

The perturbations equations are obtained taking the covariant derivative of the perturbed 
energy momentum tensor, i.e. $T_{\nu;\mu}^{\mu}=0$. We have:
\bea
\dot\delta &=& -\left( 1+w \right) \left( \theta - 3\dot\phi \right)
-3\frac{\dot a}{a} \left( \frac{\delta p}{\bar\rho} - w\delta \right)~~~~~~~~~~~~~~~~~~~{\rm for}~~~~\nu=0 \label{d_pert}\\
\dot\theta &=& -\frac{\dot a}{a} \left( 1-3w \right) \theta -
\frac{\dot{w}}{1+w}\theta +k^{2}\frac{\delta{p}/\bar\rho}{1+w} + k^{2}\psi - k^2\sigma~~~~~{\rm for}~~~~\nu=i \label{t_pert}.
\eea
The equations above are valid for any fluid; for non-relativistic matter 
(for instance for cold dark matter) we need simply to 
set $w=0$, $\delta p=0$ and $\sigma = 0$.

\bea
\dot\delta &=& - \theta + 3\dot\phi \\
\dot\theta &=& -\frac{\dot a}{a}\theta + k^{2}\psi.
\eea
The above equations depend on the evolution of the gravitational potentials 
which depend on the evolution of the perturbations of all the species 
in the Universe (if the cluster at all!) just because the right hand 
side of Eqs.~(\ref{ein-cona})-(\ref{ein-cond}) have to be considered as 
the sum over all the species. 
As we mentioned at background level, any fluid is characterized by its 
own equation of state parameter $w$ but at first order perturbation level 
we need to specify also $\delta p$ and $\sigma$ in order to close the system. 

Here we want to focus on dark energy perturbations because, as we will see, they play an 
important role on the evolution of matter perturbations, hence on the evolution of 
large scale structures.

The equation of state parameter can be inferred by measuring the 
Hubble expansion whereas $\delta p$ and $\sigma$ depend on the intrinsic characteristic 
of the fluid and the choice of these two functions 
is almost arbitrary because we have not been able to detect dark energy in the 
laboratories yet.

However, particular attention has to be paid to the pressure perturbation $\delta p$
because it is responsible for the growth of perturbations.

The pressure perturbation of a particular component is related to its 
velocity dispersion $v$; if the pressure distribution of this component
is higher than the gravitational attraction then the density perturbations can 
not growth. Only if we lower the pressure perturbation then the amplitude of the 
density contrast can increase. 
This is what happens for (dark) matter: where the pressure perturbation is basically 
zero (it is often said that matter is pressureless); in the last case the gravitational 
attraction is always bigger than the pressure support and the perturbations can 
grow at any time and any scale.

The velocity dispersion (or equivalently pressure perturbation)  
will define a characteristic scale called the Jeans scale: $\lambda_{J}=1/k_{J}$. 
In brief, all the perturbations smaller than this typical scale 
cannot grow just because the pressure support is bigger than the 
gravitational attraction; if a particular perturbation is bigger 
than the Jeans scale then it is free to grow because it does not 
feel the pressure support inside this region. 
Hence we need to characterize the pressure perturbation for any clustering 
fluid present in the Universe. 

However, the discussion is here a bit more complicated since the Jeans scale 
is in comoving coordinates and hence it grows on time; it may happen that 
at some time $t_1$ a perturbation with a scale $\lambda$ (or equivalently 
$k\sim 1/\lambda$, anyway it is a fixed number) was bigger than the Jeans 
scale but at some $t_2>t_1$ the Jeans scale has become bigger than $\lambda$ 
and this particular perturbation it is said that it has entered inside the sound 
horizon and the perturbation at this corresponding scale has stopped growing. 

Usually all the fluids in the Universe are considered to be 
adiabatic (the definition of adiabatic fluid will be clear in the next section), so 
the pressure perturbation takes a simple form:
\be
\dep = c_{a}^{2} \der =\left(w - \frac{\dot{w}}{3 \HH (1+w)} \right) \der
\label{eq:ad-pert-p}
\ee
where we have introduced a new quantity $c_a^2=\dot{\bar{p}}/\dot{\bar{\rho}}$, 
called adiabatic sound speed.
For regular matter and radiation the adiabatic sound speed is constant: $c_{a}^{2}=0$ and 
$c_{a}^{2}=1/3$, respectively. 
We also pretend dark energy to be adiabatic and hence to have the same expression 
for the pressure perturbation but soon we realize that if we assume the equation 
of state parameter for dark energy to be constant and negative (negative 
in order to lead to an acceleration phase) 
we automatically have a negative sound speed with a consequently exponential 
growth of the perturbations. Moreover, if $w$ is a time dependent function close to $-1$, 
the second term in Eq.~(\ref{eq:ad-pert-p}) will diverge.

The latter ones force us to go beyond the assumption of adiabatic fluid. In the next section 
we will also consider a different attempt in order to stabilize the pressure perturbation.

We should also consider the anisotropic stress; it will be a bit bias 
to assume it to be zero. The choice of $\sigma$ is even more difficult 
as it might rise from a pure geometrical configuration or from a real internal 
degree of freedom. A familiar class of models represents the anisotropic stress 
as a viscosity term of the density perturbations, hence it should be clearly connected 
to the the velocity perturbations and its first derivative

Later we will see the importance of the anisotropic stress and 
how it may influence the growth of matter inhomogeneities.

\subsubsection{How to deal with crossing models}

We saw before that in the Newtonian gauge we have three parameters to measure 
in order to close the system and to solve the equations of perturbations: $w$, $\delta p$ 
and $\sigma$. 

Current limits on the equation of state parameter $w=p/\rho$ of dark energy
seem to indicate that $p\approx -\rho$, see Refs.~\cite{cora,wlim}, 
sometimes even that $p<-\rho$, Ref.~\cite{phantom}, often called
{\em phantom energy}, see Ref. \cite{caldwell1}. Although there is
no problem considering $w<-1$ for the background evolution, there
are apparent divergencies appearing in the perturbations 
when a model tries to cross the ``phantom divide'' $w=-1$, see Ref.~\cite{crossing}. 
Even though this region may be unphysical at the
quantum level, Ref.~\cite{cht,Cline:2003gs}, it is still important to be able to probe
it, not least to test for alternative theories of gravity or
higher dimensional models which can give rise to an effective
phantom energy, see Refs.~\cite{para,nojiri,vli,vik2,kosh}.

However, at the level of cosmological first order perturbation 
theory, there is no fundamental limitation that prevents an effective
fluid from crossing the phantom divide.

As $w \rightarrow -1$ the terms in Eqs.~(\ref{d_pert}) and (\ref{t_pert}) 
containing $1/(1+w)$ will generally
diverge. This can be avoided by replacing $\theta$ with a new variable
$V$ defined via $V=\left( 1+w \right) \theta$. This corresponds to rewriting
the $0i$ component of the energy momentum tensor as 
$ik_j T_{0}^{j}=\brho V$ which avoids problems if $T_{0}^{j}\neq0$ when
$\bap=-\brho$. Replacing the time derivatives by a derivative with respect
to the scale factor $a$ (denoted by a prime), we obtain:
\bea
\delta' &=& 3(1+w) \psi' - \frac{V}{Ha^2} 
- 3 \frac{1}{a}\left(\frac{\dep}{\bar\rho}-w \delta \right) \label{eq:delta} \\
V' &=& -(1-3w) \frac{V}{a}+ \frac{k^2}{H a^2} \frac{\dep}{\bar\rho}
+(1+w) \frac{k^2}{Ha^2}\psi  \label{eq:v}
\eea
where we are assuming (at the moment) the anisotropic stress $\sigma$ to be zero, 
implying $\phi=\psi$, see Eq.~(\ref{ein-cond}).

In this form everything looks perfectly finite even at $w=-1$.
But we still need to give an expression
for the pressure perturbations.

In addition to the fluid perturbation equations we need to add
the equation for the gravitational potential $\psi$. If there are several fluids
present, then the evolution of each of them will be governed by their own set
of equations for their matter variables $\{\delta_i,V_i\}$, linked by a
common $\psi$ (which receives contributions from all the fluids)\footnote{We assume 
that there are no couplings beyond gravity between the fluids.}. 

The evolution of a fluid is therefore governed by Eqs.~(\ref{eq:delta}) and 
(\ref{eq:v}), supplemented by a prescription for the internal physics (given
by $\dep$) and the external physics through $\psi$ and $\psi'$, by Eq.~(\ref{ein-cona}).

We start with a barotropic fluid highlighting the problems; we extend the discussion to a more 
general case trying to overcome these difficulties.

\begin{itemize}
\item {\bf Barotropic fluid}

We define a fluid to be barotropic if the pressure $p$ depends only strictly
on the energy density $\rho$: $p=p(\rho)$. These fluids 
have only adiabatic perturbations, so that they are often called adiabatic.
We can write their pressure as
\be
p(\rho) = p(\bar{\rho}+\delta\rho) 
= p(\bar{\rho}) + \left.\frac{dp}{d\rho}\right|_{\bar{\rho}} \delta\rho
+ O\left((\delta\rho)^2\right).
\label{eq:baro_exp}
\ee
Here $p(\brho) = \bap$ is the pressure of the isotropic and homogeneous
part of the fluid. Introducing
$N\equiv \ln a$ as a new time variable, we can rewrite the background 
energy conservation equation, $\dot{\brho} = -3 \HH (\brho+\bap)$ in terms of $w$,
\be
\frac{dw}{dN} = \frac{d\brho}{dN} \frac{dw}{d\brho}
= -3 (1+w) \brho  \frac{dw}{d\brho} .
\label{eq:w_baro}
\ee
We see that the rate of change of $w$ slows down as $w\rightarrow -1$, and
$w=-1$ is not reached in finite time except if $dw/d\brho$ diverges
(or $\brho$, but that would lead to a singular cosmology), see Ref.~\cite{vik1}.
The physical reason is that we demand
$p$ to be a unique function of $\rho$, but at $w=-1$ we find that $\dot{\brho}=0$.
If the fluid crosses $w=-1$ the energy density $\rho$ will first decrease
and then increase again, while the pressure $p$ will monotonically
decrease (at least near the crossing), Fig.~(\ref{fig-1a}). It is therefore impossible 
to maintain a one-to-one relationship between $p$ and $\rho$ (our starting point), 
see Fig.~(\ref{fig-1b}) (notice that the maximum of $p$ and the minimum
of $\rho$ do not coincide).

Let us have a closer look at the perturbations. The second term
in the expansion Eq.~(\ref{eq:baro_exp}) can be re-written as
\be
\left.\frac{dp}{d\rho}\right|_{\bar{\rho}}
= \frac{\dot{\bar{p}}}{\dot{\bar{\rho}}} = w - \frac{\dot{w}}{3 \HH (1+w)} 
\equiv c_a^2
\ee
where we used the equation of state and the conservation
equation for the dark energy density in the background.
We notice that the adiabatic sound speed $c_a^2$ will necessarily diverge
for any fluid where $w$ crosses $-1$. 

\begin{figure}
\begin{centering}
\epsfig{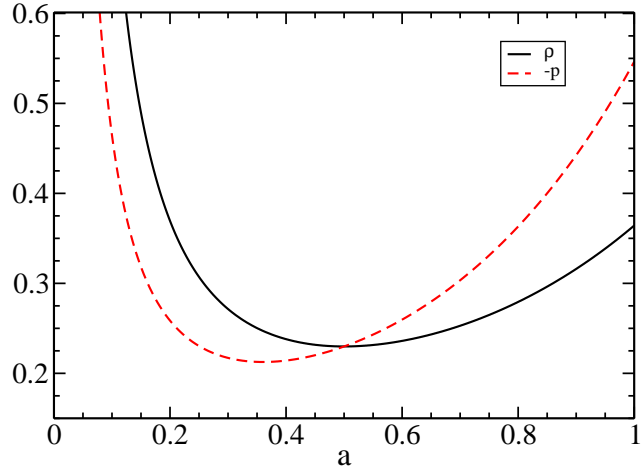}
\caption{Energy density (black solid line) and pressure (red dashed line) as
  functions of the scale factor. 
  $w$ crosses $-1$ at $\ax=0.5$, so that $\rho$ is minimal at this point
  while $p$ decreases monotonically there (but has a maximum earlier). 
  }
\label{fig-1a}
\end{centering}
\end{figure}

\begin{figure}
\begin{centering}
\epsfig{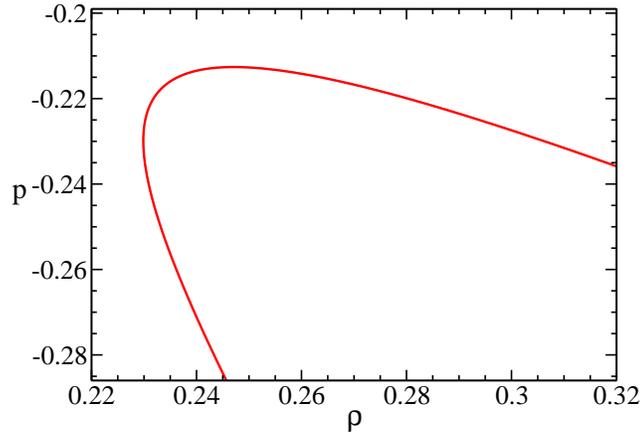}
\caption{The pressure as function of the energy
  density. The graph
  of $p(\rho)$ shows that $p$ is not a single valued function of the energy
  density, and that there is a point with an infinite slope (corresponding to
  the minimum of $\rho$ and the divergence of $\ca$). The point where $p(\rho)$
  has a zero slope corresponds to the maximum of $p$ and a vanishing $\ca$.
  }
\label{fig-1b}
\end{centering}
\end{figure}

For a perfect barotropic fluid the adiabatic sound speed $\ca$ turns out to
be the physical propagation speed of perturbations. It should therefore never 
be larger than the speed of light, otherwise the theory becomes acausal, 
see Refs.~\cite{ah,bcd}. 
The condition $\ca\le1$ for $(1+w)>0$ (our point of departure) is equivalent to
\be
-\frac{dw}{dN} \le 3 (1+w) (1-w) .
\ee 
Therefore, no barotropic fluid can pass through $w=\pm 1$ from above without
violating causality. Even worse, the pressure perturbation defined in Eq.~(\ref{eq:ad-pert-p}) 
will necessarily diverge if $w$ crosses $-1$ and $\der\neq0$.

\item {\bf Non-adiabatic fluids}

The discussion of the barotropic fluid shows that we have to violate
the constraint that $p$ be a function of $\rho$ alone. At the level
of first order perturbation theory, this amounts
to changing the prescription for $\dep$ which now becomes an arbitrary
function of $k$ and $t$. This problem is conceptually similar to
choosing the background pressure $\bap(t)$, where the conventional
solution is to compare the pressure with the energy density by
setting $\bap(t) = w(t) \brho(t)$. In this way we avoid having to deal
with a dimensionfull quantity and can instead set $w$, which has no units
(up to a factor of the speed of light) and so is generically of order
unity and often has a simple form.

It certainly makes sense to try a similar approach for the pressure
perturbation. However, there are two relevant variables that we could
compare $\dep$ to, the fluid velocity $V$ and the perturbation in the
energy density $\der$. Clearly it would be counterproductive to replace
a single free function by two free functions, and it would lead to
degeneracies between the two. Another problem is that the perturbation
variables depend on the gauge choice. But in this case the two problems
cancel each other, leading to a simpler solution: going to the rest-frame
of the fluid both fixes the gauge and renders the fluid velocity
physically irrelevant, so that we can now write, see Ref. \cite{bedo}:
\be
\hdep = \cs \hder , \label{eq:csrest}
\ee
where a hat denotes quantities in the rest-frame. The physical
interpretation is that $\cs(k,t)$ is the speed with which fluctuations
in the fluid propagate, i.e. the sound speed. 

The rest-frame is chosen
so that the energy-momentum tensor looks diagonal to an observer in
this frame. 

Let us calculate the pressure perturbation defined by
Eq.~(\ref{eq:csrest}) in the conformal Newtonian frame, following Ref. \cite{bedo}:
breaking the single link between $\rho$ and $p$ amounts to the introduction
of entropy perturbations. A gauge invariant entropy perturbation variable 
was found to be $\Gamma \equiv \frac{\delta{p}}{\bar{p}}-\frac{c_a^2}{w}
\frac{\der}{\brho}$, see Ref. \cite{KS,bedo}. By using
\be
w \Gamma = \frac{\dot{p}}{\rho} \left(\frac{\dep}{\dot{p}}-
\frac{\der}{\dot{\rho}}\right)
\ee
and the expression for the gauge transformation of $\der$, Ref. \cite{KS},
\be
\hder = \der + 3 \HH \brho \frac{V}{k^2}
\ee
we find that the pressure perturbation is given by
\be
\dep = \cs \der + 3 \HH \left(\cs - \ca\right) \brho \frac{V}{k^2} .
\label{eq:dp_rest}
\ee
As $\ca \rightarrow \infty$ at the crossing, it is impossible that all
other variables stay finite except if $V\rightarrow 0$ fast enough;
and this is not in general the case, except if
$\cs\rightarrow0$ or $w'\rightarrow0$ at crossing.

To this end, we consider again the structure of the perturbation
equations near $w=-1$. Inserting the expression for $\dep$ into our
system of perturbation equations we find
\bea
\delta' &=& 3(1+w) \psi' - \left(\frac{k^2}{Ha^2}+3H(\cs-\ca) \right) \frac{V}{k^2}-\frac{3}{a}\left(\cs-w\right)\delta \label{eq:pert_cs_d}\\
\frac{V'}{k^2} &=& -\frac{1+3(\ca-w-\cs)}{a} \frac{V}{k^2} +\frac{1+w}{H a^2} \psi + \frac{\cs}{H a^2} \delta . \label{eq:pert_cs_v}
\eea
The spoiler here is the continued presence of $\ca$ which we know to diverge at
crossing. 

However, {\em by construction} the pressure perturbation looks perfectly
fine for precisely {\em the observer in the rest-frame}, 
as $\hdep = \cs \hder$ does {\em not}
diverge. Our prescription for the pressure perturbations has singled out
the one frame which we cannot use for fluids crossing the phantom divide.
The reason is that the gauge transformation
relating the pressure perturbations in the different gauges is: 
\be
\hdep = \dep + 3\HH \brho \ca \frac{V}{k^2} .
\label{eq:dep_gauge}
\ee
If $V$ does not vanish fast enough at the crossing then the pressure
perturbation has to diverge in at least one frame.

Clearly, the problem is to find a way of characterizing the pressure perturbations
in a physical way, in Ref.~\cite{ks1} is shown a clear example, called the Quintom 
model where it was found that the additional contributions to the pressure 
perturbation diverge but in such a way
that we end up with a finite result for $\dep$. In other words, even though the
propagation speed of sound waves remains finite and constant, the additional
internal and relative pressure perturbations lead to an apparent sound
speed which diverges. It is only the sum of all contributions to $\dep$
which remains finite.

\end{itemize}

At the end, models with purely
adiabatic perturbations cannot cross $w=-1$ without violating important
physical constraints (like causality or smallness of the perturbations),
it is possible to rectify the situation by allowing for non-adiabatic sources
of pressure perturbations. However, the parameterisation of $\dep$ in terms
of the rest-frame perturbations of the energy density cannot be used as this
frame becomes unphysical at $w=-1$. By parameterising $\dep$ instead in any
other frame the divergencies are avoided, see Ref.~\cite{vikmancross}.

The dark energy sector is still not well defined and several 
problems arise also at the perturbative level. 
The main problem here is that as far as we still believe the dark energy to be responsible 
for the acceleration we cannot demand their perturbations to be zero. 

\subsection{Equivalence between general fluid and scalar field}

Here we show that at the level of first-order
perturbation theory a scalar field behaves just like a non-adiabatic
fluid with $\cs=1$. To this end we decompose the scalar field into
a homogeneous mode $\p(t)$ and a perturbation $\delta\p(k,t)$. At the
background level we find

\bea
\brho &=& \frac{1}{2a^2}\dot{\p}^2+V\left( \p \right)\\
\bap&=& \frac{1}{2a^2}\dot{\p}^2-V\left( \p \right)
\eea
and the equation of conservation is just the equation of motion,
\be
\ddot{\p}+2 \HH \dot{\p}+a^2\frac{dV}{d\p}=0 .
\ee
The adiabatic sound speed is defined as:
\be
\ca=\frac{\dot \bap}{\dot\brho}=\frac{-\frac{3aH}{a^2}\dot{\p}^2 -2\frac{dV}{d\p}\dot{\p}}{-\frac{3aH}{a^2}\dot{\p}^2}=1+\frac{2a}{3H}\frac{\frac{dV}{d\p}}{\dot{\p}}
\ee
(and we remind the reader that $\dot{a}/a=\HH=a H$).
The perturbed energy momentum tensor is:
\bea
-\delta T_{0}^{0}&=&\delta\rho =
\frac{1}{a^2}\dot{\p}\dot{\delta\p}-\frac{1}{a^2}\dot{\p}^2\Psi+\frac{dV}{d\p}\delta\p\label{derho}\\
\delta T_{i}^{i} &=& \delta p = \frac{1}{a^2}\dot{\p}\dot{\delta\p}-\frac{1}{a^2}\dot{\p}^2\Psi-\frac{dV}{d\p}\delta\p\label{dep}\\
-ik\delta T_{0}^{i} &=& ik\delta T_{i}^{0} = \frac{k^2}{a^2}\dot{\p}\delta\p
=\brho V\label{detheta}
\eea
where we wrote $\Psi$ for the gravitational potential $\psi$ in order to avoid
confusions with the scalar field variables (only in this subsection).

In order to derive the rest frame sound speed $\cs$ of the scalar field
we use equation (\ref{eq:dp_rest})
\be
\delta p = \cs\delta\rho +
\frac{3aH}{k^2}\left(\cs-\ca\right)\brho V
\ee
and express everything in terms of scalar field quantities. We find
\bea
&&\frac{1}{a^2}\dot{\p}\dot{\delta\p}-\frac{1}{a^2}\dot{\p}^2\Psi-\frac{dV}{d\p}\delta\p=\nonumber\\
&&=\cs\left(\frac{1}{a^2}\dot{\p}\dot{\delta\p}-\frac{1}{a^2}\dot{\p}^2\Psi+\frac{dV}{d\p}\delta\p
+\frac{3aH}{a^2}\dot{\p}\delta \p\right)-\nonumber \\
&&-\frac{3aH}{a^2}\left(1+\frac{2a}{3H}\frac{1}{\dot{\p}}\frac{dV}{d\p}\right) \dot{\p}\delta\p
\eea
which after some algebraic manipulations turns into
\bea
&&\frac{1}{a^2}\dot{\p}\dot{\delta\p}-\frac{1}{a^2}\dot{\p}^2\Psi+\frac{dV}{d\p}\delta\p
+\frac{3aH}{a^2}\dot{\p}\delta \p=\nonumber\\
&&=\cs\left(\frac{1}{a^2}\dot{\p}\dot{\delta\p}-\frac{1}{a^2}\dot{\p}^2\Psi+\frac{dV}{d\p}\delta\p
+\frac{3aH}{a^2}\dot{\p}\delta \p\right) .
\eea
Therefore $\cs=1$.

Now let us derive the equation of motion for the scalar field perturbations
from the perturbation equation for a perfect fluid, Eq.~(\ref{d_pert}),
\be
\dot{\delta}=-\left(1+w\right)\left(\theta-3\dot\Psi\right)
-3aH\left(\frac{\delta p}{\brho} -w\delta \right)
\ee
which can be rewritten as
\be
\dot{\delta\rho}+3aH\left(\brho+\bap\right)\delta =
3\left(\brho+\bap\right)\dot{\Psi}-\rho V. \label{cont}
\ee
Expressing the time derivative $\delta\rho$ in terms of scalar field quantities,
\bea
\dot{\delta\rho}&=&-\frac{2aH}{a^2}\dot{\p}\dot{\delta\p}+\frac{1}{a^2}\ddot{\delta\p}\dot{\p}-\frac{2aH}{a^2}\dot{\p}\dot{\delta\p}-\frac{dV}{d\p}\dot{\delta\p}- \label{deltarhodot} \\
&&-\frac{1}{a^2}\dot{\p}^2\dot{\Psi}+\frac{6aH}{a^2}\dot{\p}^2\Psi+2\dot{\p}\frac{dV}{d\p}\Psi+\frac{dV}{d\p}\dot{\delta\p}+\frac{d^2V}{d\p^2}\dot{\p}\delta\p\nonumber
\eea
and doing likewise with the other terms,
\bea
3aH\left(\delta\rho +\delta p\right) &=& \frac{6aH}{a^2}\dot{\p}\dot{\delta\p}-\frac{6aH}{a^2}\dot{\p}^2\Psi\label{drho+dp} \\
3\left(\brho+\bap\right)\dot\Psi &=& \frac{3}{a^3}\dot{\p}^2\dot\Psi\label{rho+p} \\
\brho V &=&\frac{k^2}{a^2}\dot\p \delta\p\label{V}
\eea
we can insert all these expressions into Eq. (\ref{cont}) and obtain finally
\be
\ddot{\delta\p}+2aH\dot{\delta\p}+a^2\left( \frac{d^2V}{d\p^2}
+\frac{k^2}{a^2}\right)\delta\p=4\dot\p\dot\Psi -2a^2 \Psi\frac{dV}{d\p}\label{cons}
\ee
which is indeed the equation of motion for $\delta\p$ in the conformal
Newtonian gauge (see e.g. Ref.~\cite{hu_lect}).

\subsection{Comparing different models}\label{DE-vs-MG}

We are now in the position to compare different dark energy models also at 
perturbation level. In the previous section we assumed that 
dark energy has no anisotropic stress which implies that 
the metric perturbation variables are the same $\phi=\psi$, see Eq.~(\ref{ein-cond}). 
In general this is not true or better, there is no evidence to set the anisotropic 
stress equal to zero. This new quantity increases the degrees of freedom of dark 
energy since it depends on the intrinsic characteristics of the fluid itself and 
needs to be measured. It was shown in Ref.~\cite{ks2} that the anisotropic 
stress is a crucial quantity for the growth of perturbations moreover if we want 
to distinguish between dark energy and alternative gravity theories. 

We show here how the dark energy perturbations influence the dark matter 
and the metric perturbations providing an explicit example in which a 
general dark energy model can reproduce the metric perturbation of the 
DGP scenario; as a consequence the growth rate of structures is 
not sufficient to distinguish between dark energy and modifications of gravity.

As mentioned before, the physical properties of the fluid are given 
by the anisotropic stress $\sigma$ and the pressure perturbation 
$\delta p$ (in general both can be functions of time as well as of $k$). 
For the dark energy all these quantities
are {\em a priori} unknown functions and have to be measured. For 
the special case of dark energy due to a minimally
coupled scalar field we have a variable $w$ (corresponding to the choice
of the scalar field potential, and fixed by the expansion history of the
Universe), $c_{s,\de}^2=1$ and $\sigma=0$, see e.g.~Ref.~\cite{hu_lect,ks1}.

The perturbations in different fluids are linked via the perturbations
in the metric $\phi$ and $\psi$. Introducing the comoving density perturbation
$\Delta \equiv \delta + 3 H a V/k^2$, Eqs.~(\ref{ein-cona}) and (\ref{ein-cond}) 
in standard cosmology is given by
\bea
k^2\phi &=& -4\pi Ga^2 \sum_{i} \rho_i \Delta_i \label{eq:phi} \\
k^2\left( \phi -\psi\right) &=& 12\pi G a^2\sum_{i}\left(1+w_{i}\right)
\rho_{i}\sigma_{i} \label{eq:psi-phi}
\eea
where the sum runs over matter and dark energy in our case.

The quantity we are interested in is the growth factor $G\equiv \Delta_m/a$
which parameterises the growth of structure in dark matter. The growth
factor is normalized so that $G=1$ for $a\ll1$ (using that $\Delta_m \propto a$
during matter domination and on sub-horizon scales).
We assume that $G$ is an observable quantity (even though of course large 
scale structure surveys observe luminous baryonic matter, not dark matter, 
adding yet another layer of complications).

\begin{figure}
\begin{centering}
\epsfig{figure=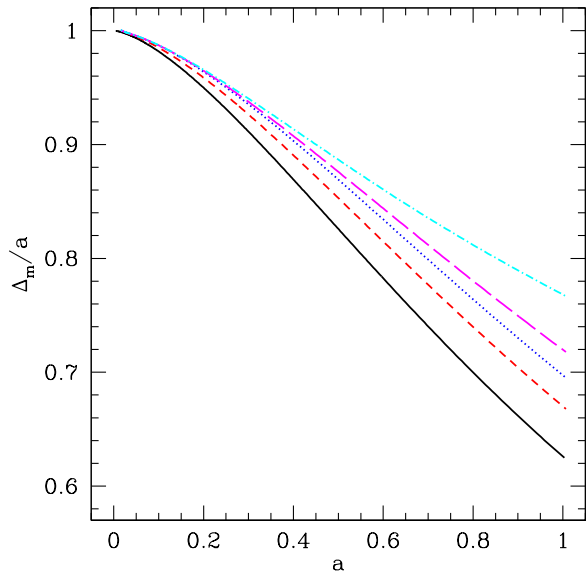,width=3.2in}
\caption{This figure shows how the growth of the matter perturbations depends
on the clustering properties of dark energy. From the top downward the
sound speed is $\cs=-2\times10^{-4}$ (cyan dash-dotted line),  $\cs=-10^{-4}$ 
(magenta long dashed line), $\cs=0$ (blue dotted line) and $\cs=1$ (red dashed line).
For comparison we also plot the growth factor of the DGP model (black solid line).}
\label{fig:depert}
\end{centering}
\end{figure}

The growth factor is {\em not} uniquely determined
by the expansion history of the Universe (and hence $w$ of dark energy). 
Although the main effect of the dark energy is to change $H$, leading to $G<1$ at late
times, there is an additional link through the gravitational potential $\psi$.
Different dark energy perturbations will lead to a different
evolution of $\psi$, which can modify the behavior of $G$. Conventionally 
dark energy perturbations are neglected because are considered to be 
unimportant, e.g. Ref.~\cite{g2}.
This is a good assumption for scalar field dark energy where the high sound
speed prevents clustering on basically all scales. However, a small sound
speed $c_{s,\de}^2\approx0$ is not excluded. Indeed, it could also vary in
time. We show in Fig.~(\ref{fig:depert}) how the growth factor of the dark matter
changes in response to large dark energy perturbations \footnote{We emphasize that
we discuss only how dark energy perturbations can modify the behavior
of dark matter, without taking into account limits from observations.}.

What happens is that, as we decrease the sound speed, dark energy is able
to cluster more and more. The increased dark energy perturbations lead to enhanced
metric perturbations. The dark matter in turn falls into the potential wells
created by dark energy, leading to an increase of the growth factor.
Although clearly $G$ is not uniquely determined by $w_\de$, we notice that
it always {\em increases} as we decrease $c_{s,\de}^2$ (at least as long as
the linearized theory is applicable).
Looking at the evolution equations
(\ref{eq:delta}) and (\ref{eq:v}) for $\sigma=0$ ($\Leftrightarrow \phi=\psi$)
we see that the response of the fluids to the metric perturbations is governed by 
the sign of $1+w$. Non-phantom dark energy clusters therefore 
in fundamentally the same way as the dark matter and 
can only {\em increase} the growth of matter relative to
the case of negligible dark energy perturbations.

So although dark energy perturbations can influence the growth factor of
dark matter, they only seem capable of enhancing it. In Fig.~(\ref{fig:depert})
is also shown the prediction for the growth factor in the DGP model from Ref.~\cite{KM},
and it is {\em smaller} than the one of a smooth dark energy component. Therefore, 
dark matter clusters weaker in DGP model than in any other model considered so far. 
Clearly, a sensitive experiment will be able to distinguish the different curves and 
rule out the wrong model. 

A natural question is: is it possible to have a dark energy model able 
to mimic DGP model also at perturbation level?

To this end we need to take a closer look at the DGP model.

An important aspect of DGP and other brane-world models
is that the dark matter does not see the higher-dimensional
aspects of the theory
as it is bound to the three-dimensional brane. Its evolution is then
{\em the same as in the standard model}. The modifications appear only
in the gravitational sector, represented by the metric perturbations.

The metric perturbation in DGP can be written as, see Refs.~\cite{KM,KK}
\bea
k^2 \phi &=& -4\pi G\,a^2\left(1-\frac{1}{3\beta}\right)
\rho_m \Delta_m  \label{eq:dgp_phi}\\
k^2 \psi &=& -4\pi G\,a^2\left(1+\frac{1}{3\beta}\right)
\rho_m \Delta_m \label{eq:dgp_psi}
\eea
where the parameter $\beta$ is defined as:
\be
\beta=1-2r_{c}H\left(1+\frac{\dot{H}}{3H^2}\right)
     = 1+2 r_c H w_\de.
\ee
Dark matter does not care if the metric perturbations
are generated (in addition to its own contribution) by
a modification of gravity or by an additional dark energy fluid. Its response
to them is identical. Or to put it differently, if dark energy and dark
matter together can create the $\phi$ and $\psi$ of
Eqs.~(\ref{eq:dgp_phi}) and (\ref{eq:dgp_psi}) then the growth factor (and
indeed all other cosmological observables) will be the same as in the 
DGP scenario.

\begin{figure}{}
\begin{centering}
\epsfig{figure=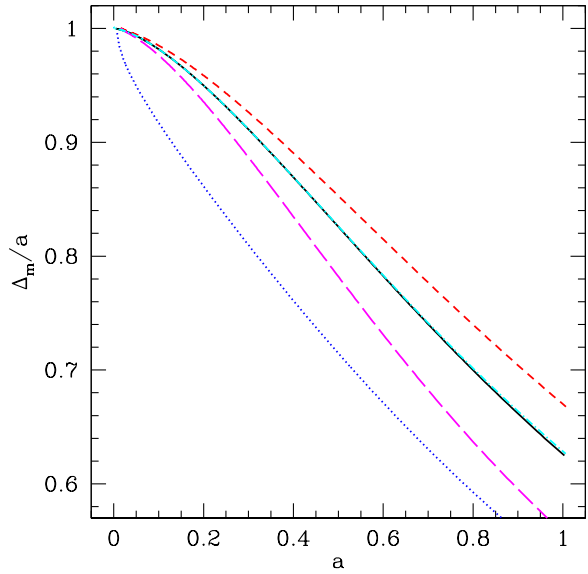,width=3.2in}
\caption{In this figure we show how the anisotropic stress of the
dark energy affects the growth of the dark matter perturbations.
The red dashed line corresponds to scalar field dark energy with
$\cs=1$ and $\sigma=0$. The dotted blue line shows how the dark matter
growth factor decreases for a constant $\sigma_\de=-0.1$. The
long-dashed magenta line uses the theoretical anisotropic stress
of Eq.~(\ref{eq:sigma_theo}) with $\cs=1$, which suppresses the growth
of the matter perturbations too much. Finally, the dash-dotted cyan
line (nearly on top of black solid DGP line) uses the same $\sigma_\de$
but sets the pressure perturbation of dark energy to 
$\delta p = (1+w)\rho \sigma$ in its rest frame.}
\label{fig:aniso}
\end{centering}
\end{figure}

We see immediately that in order to generate these metric perturbations
we will need to introduce an anisotropic stress since $\phi\neq\psi$. This
seems to be a very generic property of modified gravity  that
is also present in $f(R)$ models, see Ref.~\cite{fR_aniso} and has been noticed
before. We plot in Fig.~(\ref{fig:aniso}) again the growth factor for
scalar field dark energy and the DGP model, but now also a family of
dark energy models with non-vanishing anisotropic stress $\sigma$. We
notice that these models can easily suppress the growth of perturbations
in the dark matter for $\sigma<0$ and mimic the behavior of the DGP model.

Formally we can recover the DGP metric perturbations by choosing 
\be
\sigma_\de = \frac{2}{9\beta (1+w_\de)} \frac{\rho_m}{\rho_\de}
\Delta_m
\label{eq:sigma_theo}
\ee
for the anisotropic stress of the dark energy, if we can also generate dark
energy perturbations with
\be
\rho_\de \Delta_\de
= -\frac{1}{3\beta} \rho_m  \Delta_m .
\label{eq:de_theo}
\ee
We notice that these are very large dark energy perturbations.
Indeed, if we keep $\cs=1$ and set $\sigma$ to the expression
(\ref{eq:sigma_theo}) we suppress the growth of the matter perturbations
too much, see Fig.~(\ref{fig:aniso}). Since $\beta<0$
the large dark energy perturbations of Eq.~(\ref{eq:de_theo})
then increase the matter clustering back to the DGP value.

The required size of the dark energy perturbations in itself is no 
problem, as we can lower the sound speed and even make it negative. 
However, while for $\sigma=0$ we were not able to {\em decrease}
$\Delta_m$ with the help of dark energy perturbations, 
we find that with a large, negative anisotropic stress we
are unable to {\em increase} it. The required anisotropic stress is
far larger than the gravitational potential $\psi$,
and it starts to be the dominant
source of dark energy clustering in Eq.~(\ref{t_pert}). As it
enters with the opposite sign it now leads to anti-clustering
of the dark energy with respect to dark matter which
feels only $\psi$ (i.e. dark matter overdensities are dark 
energy voids). There is still enough freedom in the choice
of $\sigma$ to match the growth factor very precisely, but
if we could measure both $\phi$ and $\psi$ separately then
we could detect the differences between the two models.

Is it really not possible to match {\em both} $\psi$ and $\phi$
of the DGP model within a generalized fluid dark energy model?
Yes, it is: The metric perturbations have two degrees of freedom, and we do 
have two degrees of freedom of the dark energy to adjust, 
$\sigma$ and $\delta p$. Allowing free use of the pressure perturbations,
we can choose them for example to cancel the direct effect of
$\sigma$ onto the dark energy perturbation in Eq.~(\ref{t_pert})
by setting $\delta p = (1+w) \rho \sigma$. This reverses the
sign of $\Delta_\de$, and minor adjustments to the pressure
perturbations can then provide the required match to $\Delta_m$.
For the cyan dash-dotted curve in Fig.~\ref{fig:aniso} we
set $\delta p = (1+w) \rho \sigma+3 H a c_a^2 \rho V/k^2$, i.e.
we cancelled the contribution of the anisotropic stress in the
dark energy rest frame. This provides a very good solution to
Eqs.~(\ref{eq:sigma_theo}) and (\ref{eq:de_theo}) during matter
domination. It is easy to improve the solution to the point
where it is impossible to distinguish observationally between
the DGP scenario and this generalised dark energy model.

We can be even more general since the reader might have thought that the 
previous discussion worked only for the DGP model.

Let us assume that the (dark) matter is three-dimensional and conserved, and
that it does not have any direct interactions beyond gravity. We assume further
that it and the photons move on geodesics of the same (possibly effective) 
3+1 dimensional
space-time metric. In this case we can write the modified Einstein equations as
\be
X_{\mu\nu} = -8\pi G T_{\mu\nu}
\ee
where the matter energy momentum tensor still obeys $T_{\mu\, ;\nu}^\nu=0$. While
in GR this is a consequence of the Bianchi identity, this is now no longer the
case and so this is an additional condition on the behavior of the matter.

In this case, we can construct $Y_{\mu\nu} = X_{\mu\nu} - G_{\mu\nu}$,
so that $G_{\mu\nu}$ is the Einstein tensor of the 3+1 dimensional
space-time metric and we have that
\be
G_{\mu\nu} = -8\pi G T_{\mu\nu} - Y_{\mu\nu}.
\ee
Up to the prefactor we can consider $Y$ to be the energy momentum
tensor of a dark energy component. This component is also covariantly conserved
since $T$ is and since $G$ obeys the Bianchi identities. The equations governing
the matter are going to be exactly the same, by construction, so that the
effective dark energy described by $Y$ mimics the modified gravity model.

By looking at $Y$ we can then for example extract an effective anisotropic
stress and an effective pressure perturbation and build a dark energy model
which mimics the modified gravity model and leads to {\em exactly} the
same observational properties, see Refs.~\cite{ks2} and \cite{kas}.

This is both good and bad. It is bad since cosmology cannot directly distinguish
dark energy from modified gravity )although there could be clear hints, e.g.
a large anisotropic stress would favor modified gravity since in these
models it occurs generically while scalar fields have $\pi_i=0$, 
see for instance Ref.~\cite{sahnipert}). 
However, it is good since there is a clear target for future experiments: 
Their job is to measure the two additional functions
describing $Y$ as precisely as possible.

We will show in the next sections how baryon acoustic perturbations and 
weak lensing can help to constraint $H$, $\phi$ and $\psi$.

\section{A possible test}\label{tests}

We have shown in the previous section that the growth factor is not
sufficient to distinguish between modified gravity and generalised
dark energy, even if the expansion history (and so the effective
equation of state of the dark energy) has been fixed by observations.
Dark energy models can match the metric perturbations 
completely so that cosmological observations cannot distinguish between the 
two possibilities, as we have shown explicitly for the DGP case.

Although the construction of a matching dark energy model for the
DGP case may seem very fine tuned (and it is), we were more concerned with
the question to what degree this is possible at all. Just measuring
a growth factor that does not agree with scalar field dark energy
is not sufficient to rule out dark energy and General Relativity. 
But clearly, if the expansion history and the growth of matter perturbations
were to match those predicted from a physically motivated and 
self-consistent modified gravity model, a statistical analysis would
rule out a fine tuned dark energy model.

The big question now is: is there something that we can do? 
Among all the models we have at the moment to describe the acceleration of the Universe, 
can we rule out some of them from theoretical and/or observational considerations?
The answer is somewhat ambiguous. 
All the dark energy models that we can found in the literature 
are completely arbitrary (also those we reported in this review which are 
considered for faith as the most accepted ones) until fundamental physics 
is able to justify the form of the potential\footnote{If we are dealing 
with a scalar field being it associated to quintessence, K-essence etc...}.

As we saw, the $\Lambda$CDM is the simplest model to 
explain the acceleration and it is seems to be also the 
most accepted one because it is confirmed by all the observations 
we now possess. However, as we have also shown previously, it is always 
possible to have a scalar field which manifest an equation of state 
parameter close to $-1$ and we will never know the difference. 

However, this should not discourage the reader because we can 
always require some properties that our model has to respect\footnote{
However these are still open questions.}. 
We can consider it as a {\em test} that a particular model needs to pass 
in order to be considered as a good candidate.
We should always have in mind that our knowledge on dark energy 
is still poor and all the points listed below can be still debatable.

\begin{itemize}
\item {\bf If we are dealing with a real form of matter/energy}
\end{itemize}

We list here some properties that we think are fundamental for our 
model to be accepted.
Should our theory follow: 

\begin{enumerate}

\item {\bf Lorentz invariance?}: this means that the theory has to be 
not only covariant under transformations but also that the only 
absolute quantities possible in the theory are pure constants and they 
are the same in all the Universe.

\item {\bf Energy density?}: as mentioned in the previous sections, the energy 
density of a dark energy candidate should be in agreement with observations and 
the coincidence problem must be solved. In other words we should be able to solve 
the fine tuning problem. 

\item {\bf Close to cosmological constant?}: at the moment all the data we possess 
favour dark energy models whose equation of state parameters are close to 
$-1$. Probably we should think of it as a clear sign that regardless 
of what is the responsible for acceleration, a model should 
manifest a $w\sim -1$ (effective or not).
However, it is still not clear from theoretical basis if the equation of 
state parameter can be less than $-1$ (i.e. phantom). From the cosmological 
point of view there is no problem, however classical general relativity 
imposes energy conditions restricting the range of available values of 
$\left|w\right|\leq 1$, see Ref.~\cite{Carrollwless}.

\item {\bf Ghost?}: when a theory has a ghost it means that the kinetic term in the 
Lagrangian has a wrong sign; the main problem comes when we want to quantize the 
theory as it is not possible to define a ground state. So, in order for a theory to be 
considered as a good theory should be ghost-free.

\item {\bf Causality?}: each theory which is also Lorentz invariant should 
not send information with a speed larger than the speed of light: $c=1$.
In practice, we can say that the sound speed (speed with which perturbations, and 
hence information, propagate in the fluid) should not be bigger than unity: $\cs\leq 1$ 
(and possibly also positive in order to prevent perturbations to blow up).

\item {\bf Evolution of density perturbations?}: a dark energy model should lead to 
a proper evolution of the mass density field consistent with the observations: 
CMB, matter power spectrum, peculiar velocity and weak lensing power spectrum. 

\begin{itemize}
\item {\bf If we are dealing with a modification of gravity}
\end{itemize}

An obvious requirement is that the theory should be well defined and it has to follow 
the same criteria listed above; moreover, should our model also follow:

\item {\bf Proper evolution of different epochs?}: a modified gravity model should 
account for a correct evolution of the scale factor in the matter dominated era 
(as required from structure formations).

\item {\bf Local gravity test?}: we are in the position to accept that General Relativity works 
extremely well in our local Universe, hence a particular model of modified gravity 
should pass the solar system test, i.e.  $f\left(R\right)\rightarrow R-2\Lambda$.

\end{enumerate}

According to the requirements listed above we can examine the models we presented 
in this review and see whether or not they pass the tests; in Tab.~1 are 
presented  a subset of {\em dark energy} candidate, when the candidate satisfies the 
corresponding requirement the symbol \chk is used whereas the symbol $\times$ is used instead. 
If the symbol \chk is followed by the symbol ! this means that the model partially satisfies 
the criteria (or it is still not clear whether or not it is applicable).
Clearly a candidate can be ruled out if contains at least one $\times$.

\begin{table}
\begin{centering}
\begin{tabular}{|c|c|c|c|c|c|c|c|c|}
\hline 
Candidate & \multicolumn{8}{|c|}{tests}\tabularnewline
\hline
 & 1 & 2 & 3 & 4 & 5 & 6 & 7 & 8  \tabularnewline 
\hline
$\Lambda$ & \chk &  $\times$ & \chk & \chk & \chk & \chk & \chk & \chk  \tabularnewline 
\hline
Quintessence & \chk & $\times$ & \chk  & \chk & \chk & \chk & \chk & \chk \tabularnewline 
\hline 
k-essence & \chk & \chk ! & \chk & \chk & $\times$ & \chk & \chk & \chk  \tabularnewline 
\hline 
Phantom & \chk\ & $\times$ & \chk  & $\times$ & \chk & \chk & \chk & \chk \tabularnewline 
\hline 
Chaplygin & \chk & \chk ! & \chk & \chk & $\times$ & \chk ! & \chk & \chk \tabularnewline
\hline 
Coupled DM-DE & \chk & \chk ! & \chk &  \chk & \chk & \chk ! & \chk & \chk  \tabularnewline
\hline 
$f\left(R\right)$ &\chk  &  & \chk ! & \chk & \chk & \chk ! & \chk & $\times$  \tabularnewline 
\hline 
DGP & \chk &  & $\times$ & $\times$ & $\times$ & \chk ! & \chk & $\times$  \tabularnewline 
\hline 
\end{tabular}
\par\end{centering}
\label{tab:tests}
\caption{The symbol \chk is used when the models satisfies the corresponding 
requirement; the symbol $\times$ is used when it does not satisfy the criteria. 
The symbol \chk ! means that the model partially satisfies the criteria (or it is still not clear 
whether or not it is applicable). The empty boxes mean that it is not required to 
the model to satisfy the corresponding criteria.}
\end{table}

%\begin{table}[ph]
%\tbl{Tests.}
%{\begin{tabular}{@{}ccccccccccc@{}}\toprule
%\hline 
%Candidate & \multicolumn{10}{c}{tests}\\
%\hline
% & 1 & 2 & 3 & 4 & 5 & 6 & 7 & 8 & 9 & 10 \\ 
%\hline
%$\Lambda$ & \chk &  $\times$ & \chk & \chk & \chk & \chk & \chk & \chk & \chk & \chk \\ 
%\hline
%Quintessence & \chk & $\times$ & \chk  & \chk & \chk & \chk & \chk & \chk & \chk & \chk \\ 
%\hline 
%k-essence & \chk & \chk ! & \chk & $\times$ & \chk & \chk & \chk & \chk & \chk  & \chk \\ 
%\hline 
%Phantom & \chk\ & $\times$ & \chk  & $\times$ & $\times$ & $\times$ & \chk & \chk & $\times$ & \chk \\ 
%\hline 
%Chaplygin & \chk & \chk ! & \chk & \chk & $\times^{\cite{urakawa}}$ & $\times$ & \chk & \chk & $\times$ & \chk \\ 
%\hline 
%Coupled DM-DE & \chk & \chk ! & \chk &  \chk & \chk & \chk ! & \chk & \chk & \chk! & \chk \\ 
%\hline 
%$f\left(R\right)$ & \chk & \chk & \chk ! & 4 & 5 & 6 & 7 & 8 & 9 & 10 \\ 
%\hline 
%DGP & 1 & 2 & 3 & 4 & 5 & 6 & 7 & 8 & 9 & 10 \\ 
%\hline 
%\end{tabular}
%\par\end{centering}
%\caption{Tests.}
%\label{tab:tests}}
%\end{table}

\newpage

\section{The importance of being phenomenological}\label{phenomenological}

In this section we introduce again some concepts that we already 
treated before and we apologize if sometimes we look repetitive 
but we believe that recalling concepts may set the discussion clearer.

An alternative approach to understand the observed late-time accelerated 
expansion of the Universe is to construct general parameterisations
of dark energy, in the hope that measuring these parameters will give us some
insight into the mechanism underlying the dark energy phenomenon.

A successful example for such a phenomenological parameterisation
in the dark energy context is the equation
of state parameter of the dark energy component, $w\equiv p/\rho$. If
we can consider the Universe as evolving like a homogeneous and isotropic 
FLRW Universe, and if dark energy is not coupled to anything except through gravity, 
then $w(z)$ completely specifies its evolution: as it was mentioned before, 
dark energy (or anything else) is described by the homogeneous energy 
density $\rho_\dde$ and the isotropic pressure $p_\dde$, corresponding 
to the $T_0^0$ and $T_i^i$ elements respectively
in the energy momentum tensor in the rest-frame of the dark energy. Any other non-zero
components would require us to go beyond the FLRW description of the Universe.

As it was mentioned before, the equation of state parameter can be measured from SNIa. 
However, to include other observables like CMB, Weak Lensing and Matter power spectrum (we will 
come back on these topics) we need to improve our description of the Universe, by
using perturbation theory. If we work in the Newtonian gauge, we add two gravitational
potentials $\phi$ and $\psi$ to the metric, see Section \ref{pert-section}. 
They can be considered as being similar to $H$ in that they enter the 
description of the space-time (but they
are functions of scale as well as time). Also the energy momentum tensor becomes
more general, and $\rho$ is complemented by perturbations $\delta\rho$ as well
as a velocity $V_i$. The pressure $p$ now can also have perturbations $\delta p$
and there can further be an anisotropic stress $\sigma$.

The reason why we grouped the new parameters in this way is to emphasize their
role: at the background level, the evolution of the Universe is described by $H$,
which is linked to $\rho$ by the Einstein equations, and $p$ controls the evolution
of $\rho$ but is {\em a priori} a free quantity describing the physical properties of the 
fluid. Now in addition there are $\phi$ and $\psi$ describing the Universe, and
they are linked to $\delta\rho_i$ and $V_i$ of the fluids through the Einstein 
equations. $\delta p_i$ and
$\sigma$ in turn describe the fluids. Actually, there is a simplification: the
total anisotropic stress $\sigma$ directly controls the difference between the potentials,
$\phi-\psi$.

The gravitational potential is given by Eq.~(\ref{eq:phi}). 
All the fluids with non-zero perturbations will contribute to it. If we assume for instance 
that dark energy does exist we cannot demand that its perturbations are too small. 
In addition it might be that we are dealing with some modified gravity model or 
something more complicated, see for instance Eq.~(\ref{eq:dgp_phi}); 
in order to account for all these effects we parameterise the Poisson's equation as:
\be
k^{2}\phi = -4\pi G a^{2} Q\left(k,a\right)\bar{\rho}_{m}\Delta_{m}\label{eq:phi-Q}
\ee
where $\Delta = \delta+3aHV/k^2$ is always the comoving density contrast.

Here $Q(k,a)$ is a phenomenological quantity that, in General Relativity, is due to 
the contributions of the non-matter fluids (and in this case depends on 
their $\delta p$ and $\sigma$). But it is more general, as it can describe a 
change of the gravitational constant $G$ due to a modification of gravity. 
It could even be apparent: If there is 
non-clustering early quintessence contributing to the expansion rate after 
last scattering then we added its contribution to the total energy density during 
that period wrongly to the dark matter, through the definition of $\Omega_{m}$. 
In this case we will observe less clustering than expected, and we need to be able 
to model this aspect. This is the role of $Q(k,a)$.

For a dark energy model we have to admit an arbitrary anisotropic stress $\sigma$ 
(see Eq.~(\ref{eq:psi-phi})) and we use it to parameterise $\psi$:
\be
\psi = \left[1+\eta\left(k,a\right)\right]\phi \label{eq:psi-eta}.
\ee
There is no sign for a non-vanishing anisotropic stress beyond that generated by the 
free-streaming of photons and neutrinos.
However, it is expected to be non-zero in the case of topological defects, 
see Ref. \cite{KuDu} or very generically it is expected to be non-zero 
for modified gravity models (see Eq.~(\ref{eq:dgp_psi})).
With $Q$ and $\eta$ (usually called in the literature gravitational slip) then we 
can recover all the possible models.

However, it is not always possible to have an analytic expression for the parameters above 
mostly because the equation of perturbations are too complicated to be solved. 
Of course, it is always possible to solve 
them numerically, However, analytical results allow a much better insight and 
also an easy way to see how the behavior of some observables changes as a 
function of the parameters. 
For instance, for all the class of dark energy models the parameter 
$Q\left(k,t\right)$ can be thought as:
\be
Q\left(k,t\right)= 1+\frac{\bar{\rho}_{DE}\Delta_{DE}}{\bar{\rho}_{m}\Delta_{m}}.
\label{eq:q-ratio}
\ee
In Ref.~\cite{sk} can be found an analytical expression for $Q\left(k,t\right)$:
\be
Q = 1+\frac{1-\Omega_{m}}{\Omega_{m}}\left(1+w\right)\frac{a^{-3w}}{1-3w+\frac{2k^2\hat{c}_{s}^{2}a}{3H_{0}^{2}\Omega_{m}}}
\ee
here $\eta\left(k,t\right) = 0$ because it was assumed zero anisotropic stress for dark energy.
This expression is evaluated under the assumption that both $\hat{c}_{s}^2$ and $w$ are constant.
This latter assumption is usually violated, but the results should nonetheless allow insight
into the behavior of the perturbations. For models where the quantities
vary only slowly with time, it is expected that the results will still hold in an averaged
sense, due to the indirect nature of most observations. The big advantage of
making these assumptions is that it allows us to solve the perturbations
analytically under the additional condition of matter domination, leading to 
surprisingly simple results. 
In addition, the resulting expression for $Q$ is surprisingly accurate even for late times. 
The reason is that both fluids, dark energy and dark matter, respond similarly to the change in the expansion 
rate so that most of the deviations cancel, see Eq.~(\ref{eq:q-ratio}).

We would like now to connect the parameterisation introduced before to the 
observables. A very promising experiment regards weak lensing (we will come back later 
in more details), the distortion of foreground galaxies will help us to constrain 
cosmological parameters, like dark energy parameter of equation of state, with 
extraordinary accuracy.

The weak lensing effect is proportional to the lensing potential $\Phi = \phi+\psi$, 
it is then useful to consider a new parameterisation:
\be
k^{2}\Phi=Q(2+\eta)\frac{3H_{0}^{2}\Omega_{m}}{2a}\Delta_{m}=2\Sigma\frac{3H_{0}^{2}\Omega_{m}}{2a}\Delta_{m}.\label{eq:lensde}
\ee
Correspondingly, the total lensing effect is obtained by multiplying
the usual equations by a factor $\Sigma=(1+\eta/2)Q$.

In linear perturbation theory all $k$ modes evolve independently,
so that we can decompose the dark matter density contrast as: 
\begin{equation}
\Delta_{m}=aG\left(a,k\right)\Delta_{m}\left(k\right).
\label{eq:deltamatter1}
\end{equation}
Here $\Delta_{m}\left(k\right)=\Delta_{m}\left(a=1,k\right)$ determines
the matter power spectrum today, $P(k)=|\Delta_m(k)|^2$.
$G\left(a,k\right)$ is the growth factor which is scale-independent in the
usual $\Lambda$CDM cosmology. However, the contribution from the dark energy perturbations
induces a scale dependence, as we showed in the previous section.

However, this is not the only way the evolution of dark matter perturbations
is affected by the evolution of dark energy perturbations. To see
this we can write the weak lensing potential as: 
\begin{equation}
k^{2}\Phi=-3H\left(a\right)^2 a^{3}\Sigma\left(a,k\right)\Omega_{m}\left(a\right)G\left(a,k\right)\Delta_{m}\left(k\right)
\end{equation}
where we used Eq. (\ref{eq:deltamatter1}).

Hence, the lensing potential contains three conceptually different contributions
from the dark energy perturbations: 
\begin{itemize}
\item The direct contribution of the perturbations to the gravitational
potential through the factor $\Sigma$. 
\item The impact of the dark energy perturbations on the growth
rate of the dark matter perturbations, affecting the time dependence
of $\Delta_{m}$, through $G\left(a,k\right)$. 
\item A change in the shape of the matter power spectrum $P(k)$, corresponding
to the dark energy induced $k$ dependence of $\Delta_{m}$. 
\end{itemize}

In the standard $\Lambda$CDM model of cosmology, the dark matter
perturbations on sub-horizon scales grow linearly with the scale factor
$a$ during matter domination. During radiation domination they grow
logarithmically, and also at late times, when the dark energy begins
to dominate, their growth is slowed. The growth factor $G\equiv\Delta_{m}/a$
is therefore expected to be constant at early times (but after matter-radiation
equality) and to decrease at late times. In addition to this effect
which is due to the expansion rate of the Universe, there is also
the possibility that fluctuations in the dark energy can change the
gravitational potentials and so affect the dark matter clustering.

In $\Lambda$CDM $G$ can be approximated very well through
\be
G(a)=\exp\left\{ \int_{0}^{a}d\ln a\left(\Omega_{m}(a)^{\gamma}-1\right)\right\}
\ee
where $\gamma\approx0.545$. There are two ways that the growth rate
can be changed with respect to $\Lambda$CDM: Firstly, a general $w$
of the dark energy will lead to a different expansion rate, and so
to a different Hubble drag. Secondly, if in the Poisson equation (\ref{eq:phi-Q})
we have $Q\neq1$ then this will also affect the growth rate of the
dark matter, as will $\eta\neq0$. We therefore expect that $\gamma$
is a function of $w$, $\eta$ and $Q$.

For standard quintessence models we have $\eta=0$ and $Q\approx1$
on small scales as the scalar field does not cluster due to a sound
speed $c_{s}^{2}=1$. In this case $\gamma$ is only a weak function
of $w$. A constant $\gamma$ turns out to be an excellent approximation 
for some coupled dark energy models as well, see Ref. \cite{amque}, 
and for some modified gravity models, see Ref. \cite{linder1}. 
A similar change in $\gamma$ can also be obtained in models where the effective 
dark energy clusters, see Refs. \cite{ks2,KoiMo}.

However, we would like of course to be more general; 
on sub-horizon scales it can be assumed that $k^{2}\psi$ dominates over
$\phi'$ as the source of matter clustering  in the equation of perturbations, 
also because the potentials are normally slowly
varying. The term with $\psi$ is now multiplied by $(1+\eta)Q$ and
following the discussion in Ref. \cite{LiCa} we have: 
\begin{equation}
A=\frac{(1+\eta)Q-1}{1-\Omega_{m}(a)}
\end{equation}
and the asymptotic growth index becomes: 
\begin{equation}
\gamma_{\infty}=\frac{3(1-w_{\infty}-A(Q,\eta))}{5-6w_{\infty}}.
\end{equation}
Although these relations are useful to gain physical insight and for
a rough idea of what to expect, for an actual data analysis one would
specify $Q$ and $\eta$ and then integrate the perturbation equations.

Then, the set of parameters $\{\gamma,\Sigma\}$ is completely general 
and it can be used also for modified gravity models as we can reconstruct a 
set of effective parameters also for this class of models.

Therefore, there is a clear target for future experiments: together 
with $w$, {\em only two extra quantities} ($\delta p$ and $\sigma$ or 
equivalently $\gamma$ and $\Sigma$) are needed to be 
measured in order to span the complete  model space for both modified gravity and 
dark energy models. 
We can find first observational tests and future constraints on the 
sound speed and the anisotropic stress in Refs.~\cite{mota}, \cite{xia}, 
\cite{csHI}, \cite{ballesteros} and \cite{ska}.

\section{The new observational tests}\label{newtests}

\subsection{Baryon acoustic oscillations: BAO}

Let us now consider in more details the standard rulers introduced in 
the section \ref{sec-standard-ruler}. 
If we observe a transverse comoving scale $\lambda_{1}$ which subtends 
an angle $\theta$, then the angular diameter distance is 
\be
D_{1}=\lambda_{1}/\left(1+z\right)\theta \label{d1-bao}
\ee
where the subscript indicates a given cosmology. In a different cosmology, the angular 
diameter distance will be given by
\be
D_{2}=\lambda_{2}/\left(1+z\right)\theta \label{d2-bao},
\ee
in brief we can say that the scale has to change in order to keep the same subtended 
angle at the same redshift. 
Comparing Eq.~(\ref{d1-bao}) and Eq.~(\ref{d2-bao}) it can be seen that the ration $D/\lambda$ 
is constant for any given angle.
We can write everything in terms of wavenumber; therefore if we refer with the subscript 
$r$ to the reference cosmology, then we have that for any other cosmology the 
transverse wavenumber is given by:
\be
k_{\perp}=k_{r,\perp}\frac{D_{r}}{D}.
\ee
The same discussion can be done with the comoving scale extending a long the line of sight from 
two different redshifts, say $z_1$ and $z_2$. The scale is given by:
\be
\lambda = \frac{\rmd z}{H(z)}
\ee
where $\rmd z =z_{2}-z_{1}$. In order for this scale to be seen at the same 
$\rmd z$ the product $\lambda H(z)$ has to remain constant when changing the 
cosmology. As before, we can have a relation for the radial mode:
\be
k_{\parallel}=k_{r,\parallel}\frac{H(z)}{H_{r}(z)}.
\ee
The relations above can be applied to any mode. Moreover, every mode can be written in term 
of a reference mode $k_{r}$ with a dependence on the cosmological parameters: 
\be
k=\left(k_{\parallel}^{2}+k_{\perp}^{2}\right)^{1/2}=Rk_{r}
\ee
where
\be
R=\frac{\sqrt{H^2D^2\mu_{r}^2-H_{r}^2D_{r}^2\left(\mu_{r}^2-1\right)}}{H_{r}D}
\ee
and $\mu={\bf{k \cdot x}}/k$ is the direction cosine along the line of sight, 
here ${\bf x}$ is the unit vector parallel to the line of sight. 

We showed that the wavenumber changes with the cosmology, this implies that if 
the matter power spectrum is isotropic in the reference cosmology it 
will be anisotropic for any other cosmology since $k_{\parallel}$ and 
$k_{\perp}$ scale differently: this is called the Alcock-Paczynski effect, 
see Ref. \cite{alc-pac}.

Moreover, the power spectrum is proportional to the volume $V$ in which we 
measure the perturbations, hence if cosmology changes then we need to take 
into account also the change of the volume. 
Let us assume that we measure the power spectrum within a solid angle $\theta^2$ 
and a shell of width $\rmd z$, then the volume is:
\be
V = \theta^2r^2\rmd r = \theta^2 \frac{r^2(z)}{H(z)}\rmd z= \frac{D^2(z)}{H(z)}\rmd z
\ee
where $\rmd r=\rmd z/H(z)$ and $D^2(z)=\theta^2r^2(z)$. Following the same 
argument used before, we can express the volume in terms of the reference volume, it reads:
\be
V = V_{r}\frac{H_{r}(z)D^2(z)}{H(z)D_{r}^2(z)}.
\ee
The matter power spectrum is simply a product of the spectrum of primordial fluctuations 
and the modifications of those fluctuations in the later epoch. Linear perturbation theory 
fixes the matter power spectrum in comoving coordinates and changes only the amplitude as 
the structures evolve (once matter-radiation equality is fixed). The growth 
factor $G\left(z\right)$ rescales the amplitude of the fixed matter 
power spectrum to account for the growth of structures 
from the recombination to a redshift $z$. Then the matter power spectrum 
can be further simplified as: 
\be
P(z,k)=G^2(z)P(k)
\ee 
where $G(z)= \delta_{m}(z)/\delta_{m}(0)$ is called growth 
factor\footnote{This is certainly true if dark energy 
is accounted for only at background level. The growth factor can also be 
a function of space if dark energy can cluster as pointed out before.}, where 
$\delta$ is the energy density 
contrast of the matter field.

However, the observable in the Universe is the galaxy over-density which 
is assumed to trace the underlying matter distribution through a function called bias factor. 
In principle this quantity could be arbitrary; it could depend either on scale and time. 
Usually it is assumed that the bias on large scale is independent on scale hence, in the 
matter power spectrum, this term appears as a multiplicative factor which modulate the 
overall amplitude of the galaxy power spectrum, $b^2(z)$. 
It is worth noticing that there are works which pointed out that 
this assumption is too strong and it may be wrong, see Ref.~\cite{durrerbias}.

Then, an observer can only measure the galaxy power spectrum in 
redshift space, which is distorted compared to the power spectrum in the real space, 
where the theoretical predictions were made. 
Redshift distortions are angle dependent distortion on the power spectrum caused by the 
peculiar velocities of galaxies. In linear theory these will affect structures along the 
line of sight; i.e. for a structure which is isotropic in the real space, an observer will 
measure more power in the radial direction than in the transverse direction. On large scales, 
these distortions follow a simple form and it appears as a multiplicative 
angle dependent factor in the power spectrum, 
$\left(1+\beta\mu^2\right)^2$, see Ref.~\cite{kaiser}. The faction $\beta$ 
is defined by $\beta = f/b$ where $f$ is the growth rate of matter perturbations 
defined as:
\be
f=\frac{\dot{\delta}}{H\delta}=\frac{\dot{G}}{HG}-1.
\ee
Putting all the effects together we finally have the observed galaxy power spectrum:
\be
P_{obs}(z,k_{r}) =  \frac{D_{r}^{2}(z)H(z)}{D^{2}(z)H_{r}(z)}G^{2}(z)b^2(z)\left(1+\beta\mu^{2}\right)^{2}P_{0r}(k) +P_{shot}(z)
\ee
where  we added the shot noise $P_{s}(z)$ which depends only on redshift and 
it is a Poissonian-like noise coming directly from the number of galaxy 
counted in the survey volume.

Another layer of complication is the non linear evolution of the matter density fields. 
As time goes by matter density fields start to interact with each other breaking the linearity, 
linear theory does not describe correctly the evolution of structures if observations are made 
in the local Universe. As a consequence, this will increase the power spectrum at small scales 
(large wavenumber) starting to become an important term; non linear effects 
will automatically smear out the acoustic peaks and all the information is lost. 
Therefore, in order to observe more peaks, telescopes need to measure galaxies power 
spectrum at high redshift where the non linear effects are still small.

What does it mean to measure the BAO scale in a survey covering different redshifts?
Let suppose we have a survey from redshift $z_{1}$ up to $z_{2}$ divided in different 
bins of width $\Delta z$. Let us also assume we measure the correlation function 
as galaxy pairs in different bins. 
Then, our estimate of the correlation function will be the average of the 
correlation functions in different bins. 
BAO on the power spectrum correspond to a bump in the correlation function.

Measuring the BAO in the galaxy power spectrum will help us to say more about the 
dark energy parameters, such as: its energy density $\Omega_{DE}$ and the 
parameter of equation of state $w$, as it will measure the Hubble parameter 
and angular diameter distance. Moreover, the galaxy power spectrum 
depends on the growth factor which happens to be an important quantity nowadays. 
A test of the growth factor would be important both as a consistency check 
for the standard cosmological model because $G$ is also determined by the 
Hubble parameter and as a constraint on non-standard models like e.g. modify gravity, 
see section \ref{DE-vs-MG}. In fact, models that modify the Poisson 
equation will also generically modify the perturbation equation for the matter 
density contrast $\delta$.

\begin{figure}
\begin{centering}
\epsfig{figure=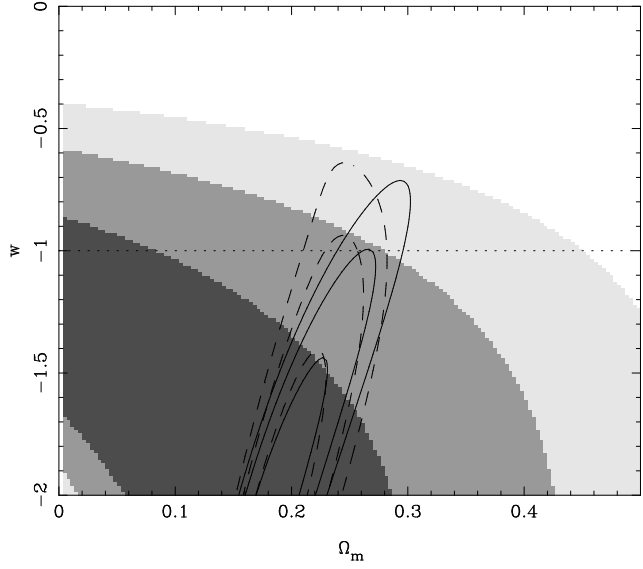,
  width=3.3in}
\caption{
Likelihood surface in the $\left(\Omega_{m},w_{DE}\right)$ plane assuming a flat space with 
constant dark energy equation of state, for spectra in the $2$dF and SDSS redshift survey. The shaded 
regions show the likelihood for the redshift survey alone. The solid contours are calculated by modelling 
the CMB sound horizon scale and the dashed contours by including the CMB peak position measurement. From Ref.~\cite{w-percival-bao}
}
\label{fig:bao-constraints}
\end{centering}
\end{figure}

We report here the constraints on the cosmological parameters from the latest experiments. 
In Fig. (\ref{fig:bao-constraints}) the confidence region for the parameters 
$\Omega_{m}$ and $w_{DE}$ is plotted. 
It is important to notice that the CMB information is important 
to reduce the errors especially on the matter density parameter but the errors on the 
equation of state parameter $w_{DE}$ are basically unaltered.

The reason is that at moment we have survey only at low redshift ($z\leq 0.5$) whereas 
the CMB peaks give a measurement at redshift $z\sim 1100$ but some parameter, 
like $w_{DE}$, at that redshift are very diluted.

\begin{figure}
\begin{centering}
\epsfig{figure=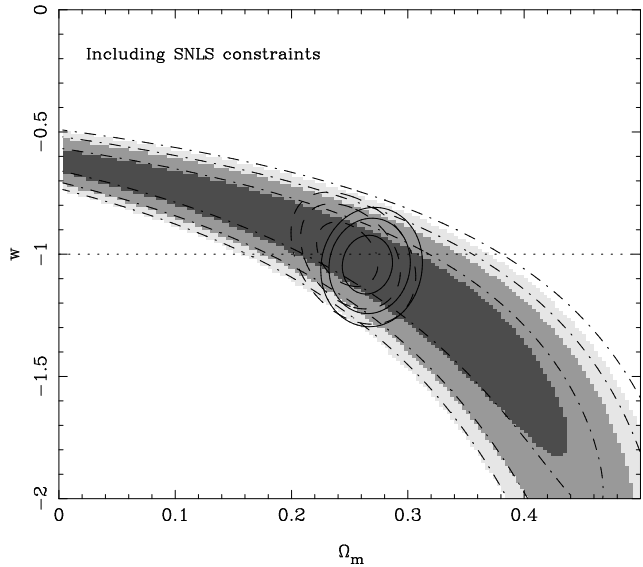,
  width=3.3in}
\caption{As Fig.~(\ref{fig:bao-constraints}), but now additionally using the
    SNIa in the Likelihood calculation. The shaded region, dashed and solid contours were
    calculated using the BAO based measurements described in the
    caption to Fig.~(\ref{fig:bao-constraints}). The dot-dashed contours show the
    likelihood surface calculated from just the SNLS
    data.
}
\label{fig:bao-constraints1}
\end{centering}
\end{figure}

In Fig.~(\ref{fig:bao-constraints1}) the constraints for $\Omega_{m},w_{DE}$ are shown
including the set of supernovae given in Ref.~{\cite{SNLS}}.

\subsection{Weak Lensing: WL}

The idea of focusing on WL comes from two reasons: first, contrary to eg supernovae or
baryon-oscillation tests, WL makes use of both background and linear
perturbation dynamics, allowing to break degeneracies that arise at
the background level (this is particularly important for testing modified
gravity); second, several groups are planning or proposing large WL
experiments in the next decade (e.g.~Refs.~\cite{DUNE,JDEM,SNAP}) that
will reach the sensitivity to test cosmology at unprecedented depth
and it is therefore important to optimize the science return of these
proposals. We find it inappropriate to write the whole calculation for the 
lensing effect, so here we report the main features. Excellent reviews on 
lensing can be found in Refs. \cite{bart-schn,schneider,bhuter}.

Light rays are deflected when they propagate through an inhomogeneous gravitational field. 
It was Einstein theory which elevated the deflection of light by masses from a hypothesis 
to a firm prediction. Assuming light behaves like a stream of particles, its deflection can 
be calculated within Newton's theory of gravitation, but General Relativity predicts that the 
effect is two times larger. The first experiment was in the 1919 when the deflection of a 
light ray, coming from a star, due to the gravitational field of the Sun was measured. 
The confirmation of the larger value was the most important step towards accepting 
General Relativity as the correct theory of gravity.

Cosmic bodies more distant, more massive, or more compact than the Sun can bend light rays 
from a single source sufficiently strongly so that we can also have 
multiple images
Tidal gravitation fields lead to differential deflection of light bundles. 
The size and shape of the sources are therefore changed. Since photons are neither emitted nor 
absorbed in the process of gravitational light deflection, the intrinsic surface brightness of 
lensed sources remains unchanged. Changing the size of the cross section of a light 
bundle therefore changes the flux. Since astronomical sources like galaxies are 
not circular, this deformation is generally very difficult to identify in 
individual images. In some cases, however, the distortion 
is so strong that it can be recognized (this is the case of Einstein rings and 
arcs in galaxy clusters). 
Such strong effects are not very common in nature and most of the time we are dealing with very 
weak distortions which, in turn, are difficult to see. Although weak distortions in individual 
images cannot be detected, the net distortion averaged over an ensemble 
of images can still be detected. 
This is called weak gravitational lensing. 

There are two reason why gravitational lensing has become a powerful tool in modern cosmology:
\begin{itemize}
\item The deflection angle of a light ray is determined by the gravitational 
fields ($\psi+\phi$) 
of the matter distribution along its path. According to General Relativity, 
the gravitational fields are determined by the energy momentum tensor of 
matter distribution. So information about the matter density distribution 
can be obtained. Moreover, the light distortion is sensible to the overall mass 
distribution and not only to the luminous matter, hence weak lensing is 
not affected by biasing uncertainty (as it was for the BAO).

\item Once the deflection angle is given, gravitational lensing can be easily reproduced. 
Since most of the lens system involves sources at moderate and high redshift, lensing can 
probe the geometry of the Universe.

\end{itemize}

The two main distortion effects here that can be used in weak lensing to probe the statistical 
properties of the matter distribution between the observer and an ensemble of distant 
sources are: 
{\em shear} and {\em magnification}.

The images are distorted in shape and size. The shape distortion is due to the tidal component 
of the gravitational field, described by the shear, whereas the magnification is given by both 
isotropic focusing caused by the local matter density and anisotropic focusing 
due to the shear, hence lensing changes the apparent brightness of a source.

For cosmological use, the weak lensing equation takes a simple form because of 
different assumption that can be made:

\begin{itemize}
\item Density perturbations are well localized in an homogeneous and isotropic background 
(each perturbation can be surrounded by a spatially flat neighborhood).

\item The Newtonian potential of the perturbations is small and typical velocities are much 
smaller than the speed of light.
\end{itemize}

Light rays are deflected by gravitational potential due to a mass distribution, then 
an effective surface mass density can be defined (and it will define the impact parameter). 
The power spectrum of the effective surface mass density is closely related to the power 
spectrum of the matter fluctuations (the central physical object in cosmology). 
Any two point statistics of cosmic magnification and cosmic shear can 
be then expressed in terms of the effective convergence power spectrum.

Briefly we can say that the convergence $\kappa$ is related to the first derivative of the 
deflection angle $\alpha$ which is proportional to the first derivative of the 
gravitational potential. 
The effective convergence $\kappa_{eff}$ then involves the Laplacian of the potential 
which is the Poisson's equation. 
The effective convergence along a light ray is therefore an integral over the density 
contrast along the light path, weighted by a combination of comoving angular 
diameter distances (sources and lens at different redshift). Of course we are 
interested in the statistical properties of the effective convergence, 
especially the power spectrum. According to the definition above the convergence 
power spectrum $P_{\kappa}(\ell)$ is a weighted integral of the matter power spectrum; 
then any weak lensing experiment will give information on the mass distribution 
(more correctly the gravitational potential) in the Universe.

The convergence weak lensing power spectrum (which in linear regime is identical to the 
ellipticity power spectrum) is a linear function of the matter power spectrum convoluted 
with the lensing properties of space. 
In $\Lambda$CDM cosmology it can be written as, see Ref.~\cite{hujain}:
\be
P_{ij}(\ell)=H_{0}^{3}\int_{0}^{\infty}\frac{dz}{E(z)}W_{i}(z)W_{j}(z)P_{nl}\left[P_{l}\left(\frac{H_{0}\ell}{r(z)},z\right)\right]\label{eq:conv-ps}
\ee
where $\ell$ denotes the angular multiple, $P_{nl}[P_{l}(k,z)]$ is the non-linear 
matter power spectrum at redshift $z$ obtained correcting the linear power 
spectrum $P_{l}(k,z)$ and the weight function $W_i$ is the efficiency for 
lensing a population of source galaxy and it is determined 
by the distribution function of source and lens galaxy.

However, there are some difficulties. Besides the problem of the linear correction, 
there are several source of systematics that can affect the outcome of the experiment. 
A good example of a work dealing with many of these details and how to control their 
effects is reported in Ref.~\cite{bernestein}.

The statistical uncertainty in measuring the shear power spectrum is, see Ref.~\cite{kaiser92}:
\be
\Delta P_{ij} = \frac{\langle\gamma_{int}^2\rangle}{n_{eff}}\delta_{ij}
\ee
where $\gamma_{int}$ is the rms intrinsic shear (arising from the intrinsic ellipticity of 
the galaxies and the measurement noise), $n_{eff}$ is the number of 
galaxy per steradians belonging to the $i$-th bin.

WL test will be the new frontier of the experiments and powerful probes on the evolution of 
the dark energy parameter. As it can be noticed the convergence power spectrum depends 
on the matter power spectrum. As often pointed out, the growth of the matter perturbations 
depends strongly on the behavior of the dark energy and its perturbations. If we 
refer to the parameter introduced in the section \ref{phenomenological}, 
we immediately realize that constraints on the set of parameters 
$\{w,\gamma,\Sigma\}$ can be inferred or equivalently on the intrinsic 
characteristic of dark energy, i.e. $\{w,\delta p,\sigma\}$.

This is not the whole story, we have to remember the set of parameters 
$\{w,\gamma,\Sigma\}$ 
is a generic way to express our ignorance. We should have clear in mind that 
these parameters do not necessarily refer to a {\em real} dark energy fluid but 
they can be also effective parameters.
Having an good measurement of this set of parameters will enable us to say something more on 
the nature of our Universe.

\begin{figure}
\begin{centering}
~

\vspace{0.2in}
\epsfig{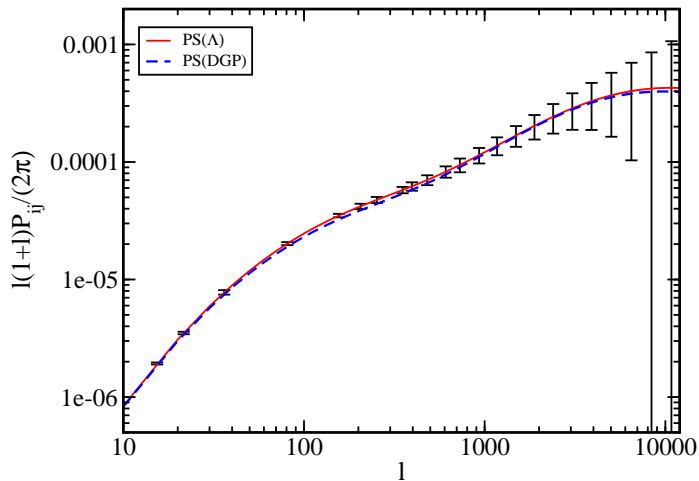}
\caption{The non linear power spectrum for $\Lambda$CDM
and DGP model for one single bin. The red (central) solid
line and the blue dashed line show the convergence power spectrum
for the $\Lambda$CDM and the DGP model respectively. The error-bars
represent the noise errors on the $\Lambda$CDM convergence power
spectrum in logarithmically spaced bins.From Ref.~\cite{aks}.
}
\label{fig:ps-wl}
\end{centering}
\end{figure}

In Fig.~(\ref{fig:ps-wl})  is reported the weak lensing spectrum for 
$\Lambda$CDM with the noise due to the intrinsic ellipticity 
and for comparison the DGP spectrum for a survey such as DUNE/Euclid, assuming a 
fraction of sky $f_{sky}=1/2$ and an effective number density of $n_{eff}=35$ 
galaxy per sq. arcmin. 
It can be seen that the DGP spectrum is outside the noise at low $\ell$'s.

\begin{figure}
\begin{centering}
\epsfig{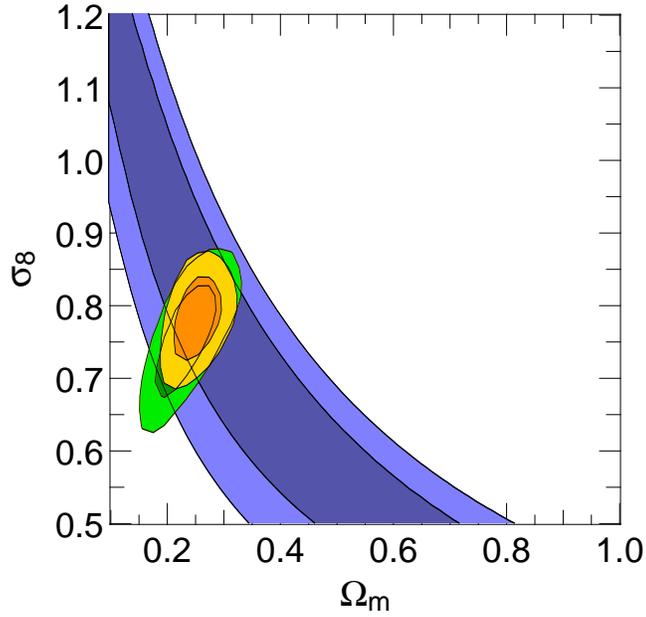}
\caption{
Constrains on  $\left(\Omega_{m},\sigma_8\right)$ from weak lensing observations CFHTLS and the CMB 
observation WMAP3 adopting $\Lambda$CDM. }
\label{fig:wl-constraints}
\end{centering}
\end{figure}

So far, the existing cosmic shear estimation have provided interesting and complementary 
constraints on $\Omega_{m}$ and on the spectrum amplitude $\sigma_8$ but not directly on the dark energy 
parameters. These are shown in Fig.~(\ref{fig:wl-constraints}), taken from Ref.~\cite{futal}.

An increasing number of papers are giving forecasts on cosmological and on dark energy parameters, 
putting in evidence the outstanding capability of such experiments to constrain parameters with 
an extremely high precision, and possibly capable of ruling out different models, 
see Refs.~\cite{aks,heavens-wl,saponea,joucooholz,holl-sapone,zhan-tyson}.

\begin{figure}
\begin{centering}
\includegraphics[angle=-90,width=2.4in]{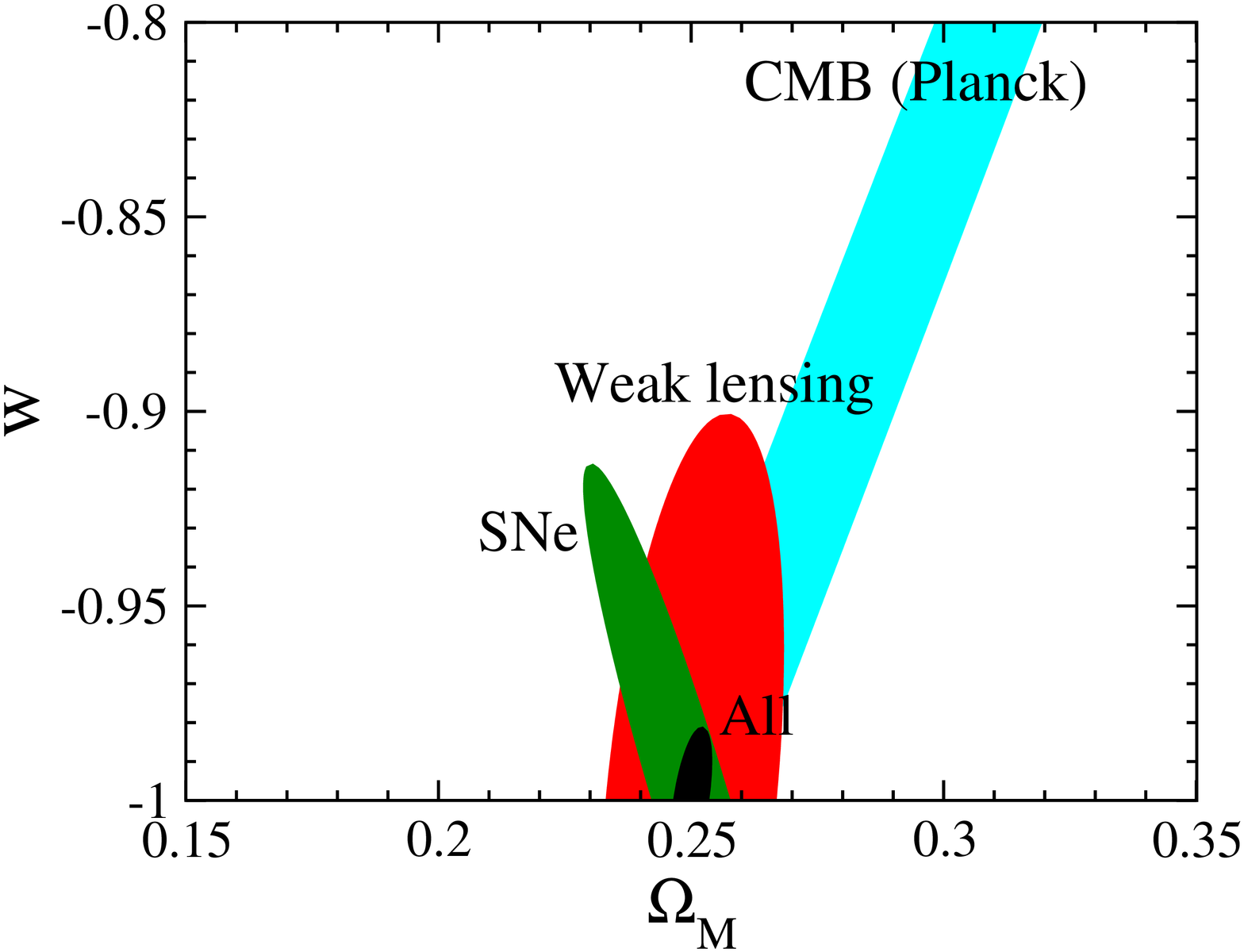}
\hspace{0.1in}
\includegraphics[angle=-90,width=2.4in]{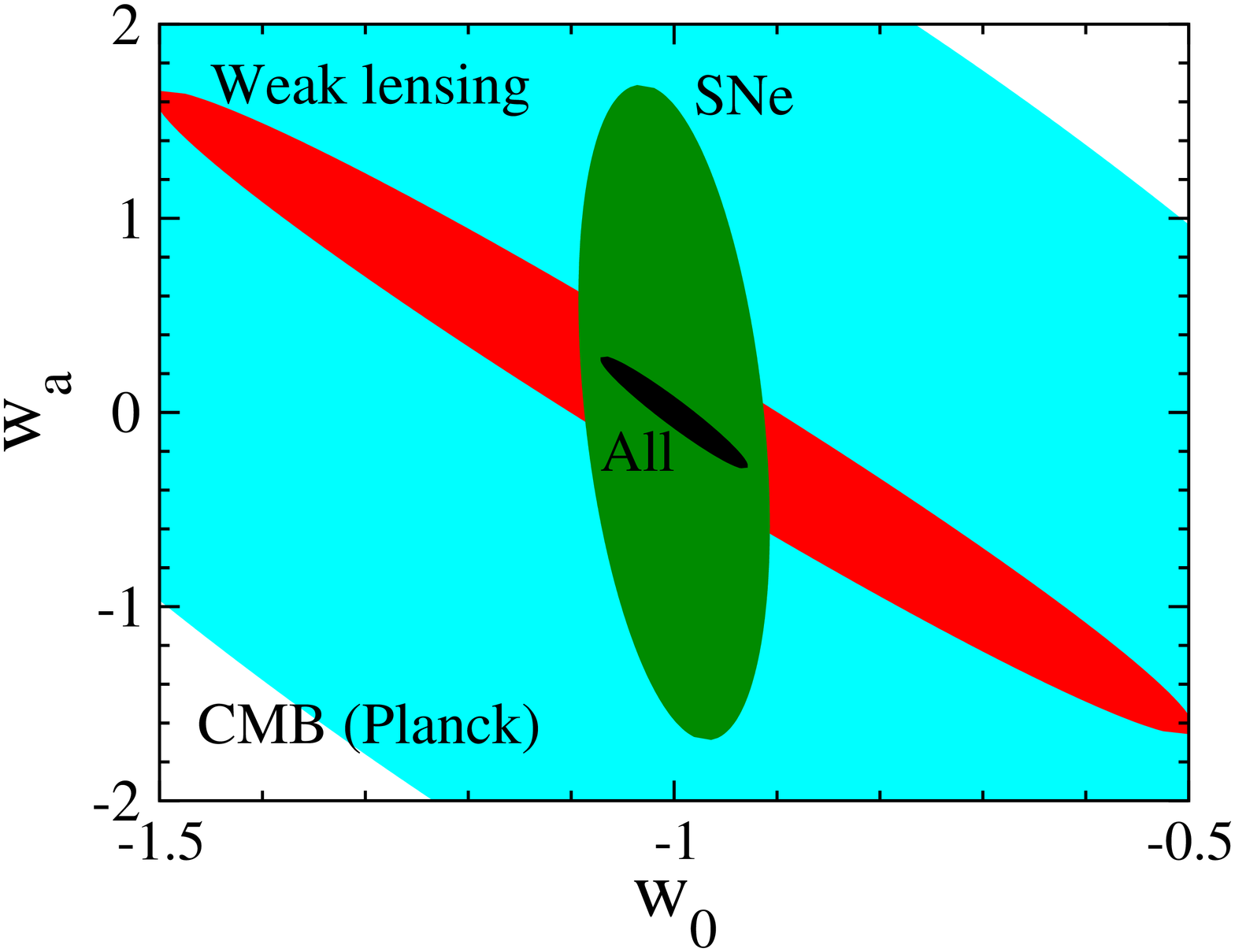}\\
\caption{Confidence region at 68\% for the dark energy parameters for SNAP
experiment, which combines 2000 SNe to $z=1.7$ and a weak lensing survey of 1000 sq. deg.  
{\it Left panel:} Constraints in the $\Omega_{m}$-$w$ plane, assuming constant $w$. 
{\it Right panel:} Constraints in the $w_0$-$w_a$ plane for time-varying
dark energy equation of state, marginalized over $\Omega_{m}$ for a flat
Universe. }
\label{fig:future_constraints}
\end{centering}
\end{figure}

In Fig.~(\ref{fig:future_constraints}) are shown the statistical constraints that the proposed 
SNAP mission could achieve, combining SN, WL observation and results from CMB Planck mission.

\subsection{Other probes}

Besides the methods discussed before i.e. SN Ia, BAO, CMB and WL, a number of other methods 
have been proposed, giving the possibility to cross check the same parameters. 
We report here the most promising ones: 
\begin{itemize}
\item Gamma ray bursts: GRB are powerful explosions, probably generated by the collapse 
of a rapidly-rotating, high-mass star into a black hole. They have a huge luminosity, which can be seen 
at very large distances. They could become the competitor of SN Ia as they can cover different redshifts 
(up to $z=8.3$, see Ref.~\cite{tanvir}) hence testing the behavior of the dark energy in different epochs. 
Although the physics of the GRB is still uncertain, several papers have been 
published in order to find a correlation between the luminosity of GRB and other observables so that they 
could be used as standard candles. So far the constraints on dark energy parameter are 
very weak. 
\item Integrated Sachs-Wolfe effect: when photon enters a potential well it gains 
energy while climbing out of a potential loses energy. During matter domination 
era the gravitational potentials are constant, hence a 
photon falling down and climbing up will not manifest a change in the energy. 
If matter is not the dominant component, the gravitational potentials are 
expected to evolve. As a consequence, photons traveling through a 
shallowing (or deepening) potential well will  manifest a blueshift (or redshift). 
This is known as ISW effect. We have shown in the previous sections that the 
dark energy affects the evolution of the gravitational 
potentials either at background level and at the perturbative level. 
At the present the ISW effect has been measured 
up to redshift $z\approx 1.5$, see Refs~\cite{giannantonio,shirleyetal}.
\item Age tests: the Hubble parameter is defined as $H=-\frac{1}{1+z}\frac{\rmd z}{\rmd t}$; 
if we know the age difference $\Delta t$ of two galaxies separated by the redshift $\Delta z$ then it is 
possible to measure the Hubble parameter $H(z)$ directly, 
see Ref.~\cite{dunlop,lima,simon,dantas, deepak1,deepak2}. 
The idea of using stellar age seems very interesting since it will use a complete different method. Here 
we are not interested in geometry but on cosmic chronology. Of course the difficult part of this method 
is that we have to make sure that the clocks we are using are reliable.
\item Gas fraction: in standard cosmology is expected that all clusters 
contain the fixed ratio of baryons to total matter $\Omega_{b}/\Omega_{m}={\rm const}$. 
Now, the X-ray thermal bremsstrahlung luminosity that comes from 
those baryons present in galaxies and clusters of galaxies is proportional to the volume 
of the emitting region which is proportional to the 
total gas mass. It follows that the gas mass is proportional to the luminosity. 
An observer will measure the flux which is a function of the luminosity and 
of the luminosity distance which depends on the cosmology. 
The constraints obtained using these methods, see Refs. \cite{allen,ettori,rapetti}, 
are still not accurate because of the different assumptions made, like spherical model, 
ideal gas equation of state, isothermal distribution, etc.
\end{itemize}

\section{Conclusion}
There is now strong evidence that the Universe is accelerating. 
However the main question is: is it really accelerating or it is just an observed 
effect probably due to other unknown physics? And, if it is really accelerating, is this acceleration 
due to a strange, small component called dark energy? 
Finding out the {\em real} nature of the dark energy is one of the most exciting and challenging problems 
facing physicists. 
In this review we have not investigated all the dark energy models
because there have been more than $1000$ papers with the 
dark energy in the title written since $1998$. 
In this review we were more interested in the possible explanation of the acceleration of the Universe 
focusing of course on dark energy but also highlighting alternative solutions. 

After more than $10$ years, dark energy still remains an open question. 
What we have done is firstly to add a new fluid component in the Universe, concentrating on the 
possibility that the dark energy may be dynamical; furthermore we also allowed the possibility that the 
Einstein equations themselves require modification. Although there is not yet reason to do so, we think 
it was appropriate to consider alternative theories which in principle are plausible.

We paid special attention to the observational tests because the list of planned 
and undergoing observational projects related to dark energy is impressive, 
see Ref.~\cite{turnerhuterer}.
The amount of data we will possess in the future will be enormous and we will 
be able to constraint other parameters apart from the {\em classical} 
cosmological ones.
This will help us to say a bit more about the candidate responsible for 
the expansion of the Universe. It might be that we need to go 
beyond the cosmological constant $\Lambda$ or it is also possible that the 
experiments will keep confirming $\Lambda$ as the best candidate. In the last case, 
we will learn more about gravity or we will be able to map the total 
matter content or we will be able to justify all the assumptions we have been making so far. 
For instance, it has been shown via weak lensing and galaxy velocity that  
General Relativity is confirmed also at large scales, see Ref. \cite{reyes-seljak}.

Again, this review should not be seen as a list of all the models on dark energy and the possible 
suggestions to explain several problems. At the moment, it is impossible to say anything! 
In other words, the term {\em Dark Energy} is the name that we give to something that we still do not 
understand to explain the observed late time acceleration of the Universe. 

Moreover, this review was not meant to discourage the reader because she (or he) 
might have thought that a lot was already done in order to understand 
the Universe where we live. On the contrary, this review shows that there is still a lot to do.

In the end, studying and understanding the dark energy problem will not be a 
waste of time and money but it will most likely be an access to {\em a complete new world}.

\section*{Acknowledgments}

I would like to thank in first place Ruth Durrer and K.~K.~Phua for giving me the 
opportunity to write this review for the International Journal of Modern Physics A.

I am very grateful to Martin Kunz who spent a lot of time reading this work. 
I would like to thank to Luca Amendola, Elisabetta Majerotto, Elisa Fenu, Riccardo 
Sturani, Juan Garcia-Bellido, Robert Crittenden, Lukas Hollenstein and Bj\"orn 
Malte Sch\"afer for interesting discussions and fruitful collaborations. 

I would like also to thank Willow Prair Vanderbosch for suggestions on the reading. 

I thank  S.~Perlmutter, W~Percival,  M.~Tegmark, Y.~Mellier, D.~Jain,  Z.~K.~Guo, S.~M.~Carroll 
for giving the permission to incorportate their figures on this review. 

This work is supported by the Spanish ${\rm MICINN}$ under the project 
${\rm AYA}2009-13936-{\rm C}06-06$ and the EU FP6 Marie Curie Research and Training Network 
"UniverseNet" (${\rm MRTN-CT}-2006-035863$).

%%%%%%%%%%%%%%%%%%%%%%%%%%%%%%%%%%%%%%%%%%%%%%%%%%%%%%%%%%%%%%%%%%%%%%%%%%%%%%


\begin{thebibliography}{0}
\bibitem{sn1} A.~G.~Riess {\it et al.}, {\it Astronomical J. {\bf 116}, 1009} (1998).
\bibitem{sn2} S.~Perlmutter {\it et al.}, {\it Astrophys. J. {\bf 517}, 565} (1999).
\bibitem{cmb-5y} E.~Komatsu {\it et al.}, {\it Astrophys. J. Suppl. {\bf 180}, 330} (2009).

%%%%%%%%%%%%%%%%%%%%%%% Physics of BAO %%%%%%%%%%%%%%%%%%%%%
\bibitem{bond-efst1} J.~R.~Bond and G.~Efstathiou, {\it Astrophys. J. {\bf 285}, L45} (1984).
\bibitem{bond-efst2} J.~R.~Bond and G.~Efstathiou, {\it Mon. Not. Roy. Astron. Soc. {\bf 226}, 655} (1987).
\bibitem{peebles-bao} P.~J.~E.~Peebles and J.~T.~Yu, {\it Atrophys. J. {\bf 162}, 815} (1970).
\bibitem{silk-bao} J.~Silk, {\it Astrophys. J. {\bf 151}, 459} (1968).

%%%%%%%%%%%%%%%%%%%%%%%%%% BAO in Galaxy %%%%%%%%%%%%%%%%%
\bibitem{meikins} A.~Meikins , M.~White and J.~A.~Peacock, {\it  Mon. Not. Roy. Astron. Soc. {\bf 304}, 851} (1999). 
\bibitem{springel} V.~Springel {\it et al.}, {\it Nature {\bf 435}, 629} (2005).
\bibitem{seo-eisenstein-bao} H.~J.~Seo and D.~J.~Eisenstein, {\it Atrophysic. J. {\bf 633}, 575} (2005).
\bibitem{white-bao} M.~White, {\it Astroparticlephys. {\bf 24}, 334} (2005).
\bibitem{eisensteinetal} D.~J.~Eisenstein, H.~J.~Seo and M.~White, {\it Astrophysic. J. {\bf 664}, 660} (2007).
\bibitem{blake-glaze} C.~Blake and K.~Glazebrook, {\it Astrophysic. J. {\bf 594}, 665} (2003).
\bibitem{seo-eisenstein-bao-ruler} H.~J.~Seo and D.~J.~Eisenstein, {\it Atrophysic. J. {\bf 598}, 720} (2003).
\bibitem{coleetal} S.~Cole {\it et al.}, {\it Mon. Not. Roy. Astron. Soc. {\bf 362}, 505} (2005).
\bibitem{sdss_bao}D.~Eisenstein {\it et al. {[}SDSS collaboration], Astrophys. J. {\bf 633}, 560} (2005).
\bibitem{huetsi06} G.~Huetsi, {\it Astron. \& Astrophys., {\bf 449}. 891} (2006).

\bibitem{liddle-lyth} E.~R.~Liddle and D.~H.~Lyth, {\it Cosmological 
inflation and large scale structure}, Cambridge University Press (2000).
\bibitem{Hubble} E.~Hubble, {\it Proc. Nat. Acad. Sci. {\bf 15}, 168} (1929).
\bibitem{HST} W.~L.~Freedman {\it et al.},{\it  Astrophys. J. {\bf 553}, 47} (2001).
\bibitem{Etherington} J.~M.~H.~Etherington, {\it Phil. Mag. {\bf 15}, 761} (1933).
\bibitem{BassKunz} B.~A.~Basset and M.~Kunz, {\it Phys.Rev. D {\bf 69}, 101305} (2004).
\bibitem{sahni} V.~Sahni, {\it Lect. NotesPhys. {\bf 653}, 141} (2004).
%\bibitem{sn-low} M.~Hamuy, {\it Astronom. J. {\bf 112}, 2391} (1996).
\bibitem{SNLS} P.~Astier {\it et al.}, {\it Astrophys. J. {\bf 447}, 31} (2006).
\bibitem{HST1} A.~G.~Riess {\it et al.}, {\it Astrophys. J. {\bf 659}, 98} (2007).
\bibitem{ESSENCE} W.~M.~Wood-Vasey {\it et al.}, {\it Astrophys. J. {\bf 666}, 716} (2007).
\bibitem{Carretta} E.~Carretta {\it et al.}, {\it Astrophys. J. {\bf 533}, 215} (2000).
\bibitem{Jimenez} R.~Jimenez {\it et al.}, {\it Mon. Not. Roy. Astron. Soc. {\bf 282}, 926} (1996).
\bibitem{Hansen} B.~M.~S.~Hansen {\it et al.}, {\it Astrophys. J {\bf 574}, 155} (2002).

\bibitem{camb} A.~Lewis, A.~Challinor and A.~Lasenby, {\it Astrophys. J. {\bf 538}, 473} (2000).
\bibitem{SDSSgal} M.~Tegmark {\it et al. {[}SDSS collaboration], Phys. Rev. D {\bf 74}, 123507} (2006).
\bibitem{lss} M.~Tegmark {\it et al.}, {\it Phys. Rev. D {\bf 69}, 103501}  (2004);  
M. Tegmark et al.,{\it  Astrophys. J. {\bf 606}, 702} (2004).

%%%%%% From here there is cosmological constant interpretations
\bibitem{kachru} S.~Kachru {\it et al., Phys. Rev. D {\bf 68}, 046005} (2003).
\bibitem{susskind} L.~Sussking, {\it arXiv:hep-th/0302219}.
\bibitem{Weinberg} S.~Weinberg, {\it Rev. Mod. Phys. {\bf 61}, 1} (1989).
\bibitem{garriga} J.~Garriga and A.~Vilenkin, {\it Phys. Rev. D {\bf 61}, 083502} (2000).
\bibitem{hawking} S.~W.~Hawking, {\it Phys. Lett. B {\bf 134}, 403} (1984).
\bibitem{tye} S~H.~H.~Tye and I.~Wasserman, {\it Phys. Rev. Lett. {\bf 86}, 1682} (2001).
\bibitem{yoko} J.~Yokoyama, {\it Phys. Rev. Lett. {\bf 88}, 151302} (2002).
\bibitem{muko} S.~Mukohyama and L.~Randall, {\it Phys. Rev. Lett. {\bf 92}, 211302} (2004).
\bibitem{kane} G.~L.~Kane M.~J.~Perry and A.~N.~Zytkow, {\it Phys. Lett B {\bf 609}, 7} (2005).
\bibitem{dolgov} A.~D.~Dolgov and F.~R.~Urban, {\it Phys. Rev. D {\bf 77}, 083503} (2008).
\bibitem{albrechtstein} A.~J.~Albrecht and P.~J.~Steinhardtm, {\it Phys. Rev. Lett. {\bf 48}, 1220} (1982).
\bibitem{brown} J.~B.~Brown and C.~Teitelboim, {\it Nucl. Phys. B {\it 195}, 177} (1987).
\bibitem{brown1} J.~B.~Brown and C.~Teitelboim, {\it Nucl. Phys. B {\it 297}, 787} (1988).
\bibitem{abbott} L.~F.~Abbott, {\it Phys. Lett. B {\bf 150}, 427} (1985).
\bibitem{bousso} R.~Bousso and J.~Polchinski, {\it JHEP {\bf 0006}, 006} (2000).
\bibitem{denef} F.~Denef and M.~R.~Douglas, {\it JHEP {\bf 0405}, 072} (2004).
%%%%%%%%%%%%%%%%%%%%%%%%%%%%%%%%%
\bibitem{martel} H.~Martel, P.~R.~Shapiro and S.~Weinberg, {\it Astrophys. J. {\bf 492}, 29} (1998);
\bibitem{bludman} S.~A.~Bludman, {\it Nucl. Phys. A {\bf 663}, 865} (2000);
\bibitem{hogan} C.~J.~Hogan, {\it Rev.  Mod.  Phys.  {\bf 72}, 1149} (2000);
\bibitem{Anth2} J.~Garriga, M.~Livio and A.~Vilenkin,{\it  Phys. Rev. D {\bf 61}, 023503} (2000).
\bibitem{Anth4} J.~Garriga, A.~Linde and A.~Vilenkin, {\it Phys. Rev. D {\bf 69}, 063521} (2004).
\bibitem{Gurzadyan:2000ku} V.~G.~Gurzadyan and S.~S.~Xue, {\it Mod. Phys. Lett. A {\bf 18}, 561} (2003).
\bibitem{Carroll:2001zy} S.~M.~Carroll and L.~Mersini-Houghton, {\it Phys. Rev D {\bf 64}, 124008} (2001).
\bibitem{Kane:2003qh} G.~L.~Kane, M.~J.~Perry and A.~N.~Zytkow, {\it Phys. Lett. B {\bf 609}, 7} (2005).


%%%%%%%%%%%%%%% Quintessence %%%%%%%%%%%%%%%%%%%
\bibitem{carroll} S.~M.~Carroll, {\it Living Rev. Rel. {\bf 4}, 1} (2001). 
\bibitem{caldwell} R.~R.~Caldwell, R.~Dave and P.~J.~Steinhardt, {\it Phys. Rev. Lett. {\bf 80}, 1582} (1998).
\bibitem{kolbturner} E.~Kolb and M.~Turner, {\it The early Universe} (Addison-Wesley, Redwood City, CA, 1990).
\bibitem{bassetwands} B.~A.~Bassett, S.~Tsujikawa and D.~Wands, {\it Rev. Mod. Phys. {\bf 78}, 537} (2006).
\bibitem{ratrapeebles} B.~Ratra and P.~J.~E.~Peebles, {\it Phys. Rev. D {\bf 37} 3406} (1988).
\bibitem{carroll1} S.~M.~Carroll, {\it Phys. Rev. Lett. {\bf 81} 3067} (1998).

%%%%%%%%%%% Quintessense models from particle physics %%%%%%%%%%%%%%
\bibitem{binetruy} P.~Binetruy, {\it Phys. Rev. D {\bf 60} 063502} (1999).
\bibitem{frieman} J.~A.~Frieman {\it et al.}, {\it Phys. Rev. Lett. {\bf 75} 2077} (1995).
\bibitem{copeland1} E.~J.~Copeland, N.~J.~Numes and F.~Rosati, {\it Phys. Rev. D {\bf 62} 123503} (2000).
\bibitem{kolda} C.~F.~Kolda and D.~H.~Lyth, {\it Phys. Lett. B {\bf 458} 3681} (1996).
\bibitem{nomura} Y.~Nomura, T.~Watari and T.~Yanagida, {\it Phys. Lett. B {\bf 484} 103} (2000).
\bibitem{choi} K.~Choi, {\it Phys. Rev. D {\bf 62} 043509}  (2000).
\bibitem{kim} J.~E.~Kim and H.~P.~Nilles, {\it Phys. Lett. B {\bf 553} 1} (2003).
\bibitem{gasperni} M.~Gasperini, F.~Piazza and G.~Veneziano, {\it Phys. Rev. D {\bf 65} 023508} (2002).
\bibitem{Townsend} P.~K.~Townsend, {\it JHEP {\bf 0111}, 042} (2001).
\bibitem{Townsend1} P.~K.~Townsend and M.~N.~R.~Wohlfarth, {\it Phys. Rev. Lett.  {\bf 91}, 061302} (2003).
\bibitem{Cornalba} L.~Cornalba and M.~S.~Costa, {\it Phys. Rev. D {\bf 66}, 066001} (2002).
\bibitem{hellerman01} S.~Hellerman, N.~Kaloper and L.~Susskind, {\it JHEP {\bf 0106} 003} (2001). 
\bibitem{fischler} W. Fischler, A. Kashani-Poor, R. McNees and S. Paban, {\it JHEP {\bf 0107} 003} (2001).  
\bibitem{Hull} C.~M.~Hull, {\it JHEP {\bf 0111}, 012} (2001).
\bibitem{Burgess1} C.~P.~Burgess {\it et al.}, {\it JHEP {\bf 0308}, 056} (2003).
\bibitem{Burgess2} C.~P.~Burgess, {\it Annals Phys.  {\bf 313} 283} (2004).
\bibitem{Matias} J.~Matias and C.~P.~Burgess, {\it JHEP {\bf 0509}, 052} (2005).
\bibitem{Tolley} A.~J.~Tolley {\it et al.}, arXiv:hep-th/0512218.
\bibitem{QCD} A.~Masiero, M.~Pietroni, and F.~Rosati,  {\it Phys. Rev. D {\bf 61}, 023504} (2000).
\bibitem{ohta1} N.~Ohta, {\it Phys. Rev. Lett. {\bf 91}, 061303} (2003).
\bibitem{ohta2} Z.~K.~Guo, N.~Ohta and Y.~Z.~Zhang, {\it Phys.\ Rev.\  D {\bf 72}, 023504} (2005).
\bibitem{ohta3} Z.~K.~Guo, N.~Ohta and Y.~Z.~Zhang, {\it Mod.\ Phys.\ Lett.\  A {\bf 22}, 883} (2007).
\bibitem{brax1} P.~Brax, and J.~Martin, Phys. Lett. B{\bf 468}, 40-45 (1999);
\bibitem{brax2} P.~Brax and J.~Martin, {\it Phys. Rev. D {\bf 61}, 103502} (2000).
\bibitem{kallosh1} R.~Kallosh {\it et al.}, {\it Phys. Rev. D {\bf 65}, 105016} (2002);
\bibitem{kallosh2} R.~Kallosh {\it et al.}, {\it Phys. Rev. D {\bf 66}, 123503} (2002).
\bibitem{Chacko} Z.~Chacko, L.~J.~Hall and Y.~Nomura, {\it JCAP {\bf 0410}, 011} (2004).
\bibitem{Hall} L.~J.~Hall, Y.~Nomura and S.~J.~Oliver, {\it Phys. Rev. Lett.\  {\bf 95}, 141302} (2005).
\bibitem{Barbieri} R.~Barbieri {\it et al.}, {\it Phys. Lett. B {\bf 625}, 189} (2005).
\bibitem{huterer-peiris} D.~Huterer and H.~V.~Peiris, {\it Phys. Rev. D {\bf 75}, 083503} (2007).
\bibitem{linder-quint} E.~Linder, {\it Den. Rel. Grav. {\bf 40}, 329} (2008).
\bibitem{amendola-wetterich} L.~Amendola, M.~Balbi and C.~Wetterich, {\it Phys. Rev. D {\bf78}, 023015} (2008).
\bibitem{jeromemartin} J. Martin, {\it Mod. Phys. Lett. A {\bf 23}, 1252} (2008).
\bibitem{iwazaki} A. Iwazaki, {\it arXiv:1006.4890}.
\bibitem{chibasiino} T. Chiba, M. Siino and M. Yamaguchi, {\it Phys. Rev. D{\bf 81}, 083530} (2010).

%%%%%%%%%%%%%%% K-essence %%%%%%%%%%%%%%
\bibitem{armendariz1} C.~Armend\'eriz-Picon, V.~Mukhanov and P.~J.~Steinhardt, {\it Phys. Lett. B {\bf 458}, 209} (1999). 
\bibitem{garrida} J.~Garrida and V.~Mukhaniv, {\it Phys. Lett. B {\bf 458}, 219} (1999).
\bibitem{armendariz2} C.~Armend\'eriz-Picon, V.~Mukhanov and P.~J.~Steinhardt, {\it Phys. Rev. Lett. {\bf 85}, 4438} (2000).
\bibitem{armendariz3} C.~Armend\'eriz-Picon, V.~Mukhanov and P.~J.~Steinhardt, {\it Phys. Rev. D {\bf 63}, 103510} (2001).
\bibitem{camille} C.~Bonvin, C.~Caprini and R.~Durrer, {\it Phys. Rev. Lett. {\bf 97}, 081303} (2006).

\bibitem{barger-marfatia} V.~D.~Barger and D.~Marfatia, {\it Phys. Lett. B {\bf 498}, 67} (2001).
\bibitem{lizang-K} M.~Li and X.~Zhang, {\it Phys. Lett. B {\bf 573}, 20} (2003).
\bibitem{copelandtrodden} M.~Malquarti {\it et al.}, {\it Phys. Rev. D {\bf 67}, 123503} (2003).
\bibitem{copelandliddle-K} M.~Malquarti, E.~J.~Copeland and A.~R.~Liddle, {\it Phys. Rev. D {\bf 68}, 023512} (2003).
\bibitem{chimentolazkoz} L.~P.~Chimento and R.~Lazkoz, {\it Phys. Rev. D {\bf 71}, 023505} (2005).
\bibitem{lazkoz-k} R.~Lazkoz, {\it Int. J. Mod. Phys. D {\bf 14}, 635} (2005).
\bibitem{aguirregabiria} J.~M.~Aguirregabiria, L.~P.~Chimento and R.~Lazkoz, {\it Phys. Lett. B {\bf 631}, 93} (2005).
\bibitem{daskephart} R.~Das, T.~W.~Kephart and R.~J.~Scherrer, {\it Phys. Rev. D {\bf 74}, 103515} (2006).
\bibitem{quercellini-bruni} C.~Quercellini, M.~Bruni, A.~Balbi {\it Class. Quant. Grav. {\bf 24}, 5413} (2007).
\bibitem{bosemaj} N.~Bose and A.~S.~Majumdar, {\it Phys. Rev. D {\bf 79}, 103517} (2009).
\bibitem{chibascherrer} T.~Chiba, S.~Dutta and R.~J.~Scherrer, {\it Phys. Rev. D {\bf 80}, 043517} (2009).


%%%%%%%%%%%%%%%% Phantom %%%%%%%%%%%%%%%%%%
\bibitem{Carrollwless} S.~M.~Carroll, M.~Hoffman and M.~Trodden, {\it Phys. Rev. D {\bf 68}, 023509} (2003).
\bibitem{saridakis} J.~F.~Cai {\it et al.}, {\it arXiv:0909.2776}.
\bibitem{Cline} J.~M.~Cline, S.~Jeon and G.~D.~Moore, {\it Phys. Rev. D?{\bf 70}, 043543} (2004).




\bibitem{nojiriodint-phontom} S.~Nojiri and S.~D.~Odintsov, {\it Phys. Lett. B {\bf 562}, 147} (2003).
\bibitem{nojiriodint-phontom1} S.~Nojiri and S.~D.~Odintsov, {\it Phys. Lett. B {\bf 565}, 1} (2003).
\bibitem{diaz1} P.~F.~Gonzalez-Diaz, {\it Phys. Rev. D {\bf 68}, 021303} (2003).
\bibitem{diaz2} P.~F.~Gonzalez-Diaz, {\it Phys. Lett. B {\bf 586}, 1} (2006).
\bibitem{diaz3} P.~F.~Gonzalez-Diaz, {\it Phys. Rev. D {\bf 69}, 063522} (2004).
\bibitem{phantomcosmo} M.~P.~Dabrowski, T.~Stachowiak and M.~Szydlowski,{\it Phys. Rev. D {\bf 68}, 103519} (2003).
\bibitem{chimentolazkoz-phantom} L.~P.~Chimento and R.~Lazkoz, {\it Phys. Rev. Lett.  {\bf 91}, 211301} (2003).
\bibitem{stefancic} H.~Stefancic, {\it Phys. Lett. B {\bf 586}, 5} (2004).
\bibitem{stabiliti-phantom} Z.~K.~Guo, Y.~S.~Piao and Y.~Z.~Zhang, {\it Phys. Lett. B {\bf 594}, 247} (2004).
\bibitem{urenalopez} L.~A.~Urena-Lopez, {\it JCAP {\bf 0509}, 013} (2005).
\bibitem{faraoni} V.~Faraoni, {\it Class. Quant. Grav.  {\bf 22}, 3235} (2005).
\bibitem{sur-das} S.~Sur and S.~Das, {\it JCAP {\bf 0901}, 007} (2009). 
\bibitem{dutta-scherrer-phantom} S.~Dutta, E.~N.~Saridakis, R.~J.~Scherrer, {\it Phys. Rev. D {\bf 79}, 103005} (2009).



%%%%%%%%%%%%%%%%% Chaplygin Gas%%%%%%%%%%%%%%%%%%%%%%
\bibitem{AFBC} L.~Amendola {\it et al., JCAP {\bf 0307}, 005} (2003).
\bibitem{sandvik} H.~Sandvik {\it et al., Phys. Rev. D {\bf 69}, 123524} (2004).
\bibitem{fabris1} J.~C.~Fabris, S.~V.~B.~Goncalves and P.~E.~de Souza, {\it Gen. Rel. Grav.  {\bf 34}, 53} (2002).
\bibitem{cartfin} D.~Carturan and F.~Finelli, {\it Phys. Rev. D {\bf 68}, 103501} (2003).
\bibitem{devjain} A.~Dev, D.~Jain and J.~S.~Alcaniz, {\it Phys. Rev. D {\bf 67}, 023515} (2003).
\bibitem{bento} M.~C.~Bento, O.~Bertolami and A.~A.~Sen, {\it Phys. Rev. D {\bf 67}, 063003} (2003).
\bibitem{gorikame} V.~Gorini, A.~Kamenshchik and U.~Moschella, {\it Phys. Rev. D {\bf 67}, 063509} (2003).
\bibitem{makler} M.~Makler, S.~Quinet de Oliveira and I.~Waga, {\it Phys. Lett. B {\bf 555}, 1} (2003)
\bibitem{szy} M.~Szydlowski and W.~Czaja, {\it Phys. Rev. D {\bf 69}, 023506} (2004).
\bibitem{perrotta} F.~Perrotta, S.~Matarrese and M.~Torki, {\it Phys. Rev. D {\bf 70}, 121304} (2004).
\bibitem{zhu} Z.~H.~Zhu, {\it Astron. \& Astrophys. {\bf 423}, 421} (2004).
\bibitem{chimento}L.~P.~Chimento and R.~Lazkoz, {\it Phys. Lett. B {\bf 615}, 146} (2005).
\bibitem{gong} Y.~G.~Gong, {\it JCAP {\bf 0503}, 007} (2005).
\bibitem{bilic} N.~Bilic, G.~B.~Tupper and R.~D.~Viollier, {\it JCAP {\bf 0510}, 003} (2005).
\bibitem{biesiada} M.~Biesiada, W.~Godlowski and M.~Szydlowski, {\it Astrophys. J.  {\bf 622}, 28} (2005).
\bibitem{zimd} W.~Zimdahl and J.~C.~Fabris, {\it Class. Quant. Grav.  {\bf 22}, 4311} (2005).
\bibitem{sen} A.~A.~Sen and R.~J.~Scherrer, {\it Phys. Rev. D {\bf 72}, 063511} (2005).
\bibitem{fabris2} J.~C.~Fabris {\it et al.}, {\it Phys. Rev. D {\bf 78}, 103523} (2008).
\bibitem{pun} C.~S.~J.~Pun {\it et al.}, {\it Phys.~Rev.~D {\bf 77}, 063528} (2008).
\bibitem{jlu} J.~Lu, {\it Phys. Let B {\bf 680}, 404} (2009).

%%%%%%%%%%%%%%%%% COUPLED DARK ENERGY %%%%%%%%%%%%%%%%%%%%%
\bibitem{amendola-coupled} L.~Amendola, {\it Phys. Rev. D {\bf 62}, 043511} (2000).
\bibitem{majsapam} E.~Majerotto, D.~Sapone and L.~Amendola, {\it astro-ph/0410543}.
\bibitem{fuzfa} A.~Fuzfa and J.~M.~Alimi, {\it Phys.~Rev.~D {\bf 75},123007} (2007).
\bibitem{dalal} N.~Dalal {\it et al.}, {\it Phys.~Rev.~Lett. {\bf 87}, 141302} (2001).
\bibitem{zimdhal} W~Zimdahl, D.~Pav\'on and L.~P.~Chimento, {\it Phys.~Lett.~B {\bf 521}, 133} (2001).
\bibitem{campoherrera} S.~del Campo {\it et al.}, {\it Phys.~Rev.~D {\bf 74}, 023501} (2006).
\bibitem{caldera} G.~Caldera-Cabral, R.~Maartens and L.~A.~Urena-Lopez, {\it Phys.~Rev.~D {\bf 79} 063518} (2009).
\bibitem{guota} Z.-K.~Guo, N.~Ohta and S.~Tsujikawa, {\it Phys.~Rev.~D {\bf 76}, 023508} (2007).
%%%%%%%%%%%%%%%%%%%% Modified gravity %%%%%%%%%%%%%%%%%%
\bibitem{starobinsky} A.~A.~Starobinsky, {\it Phys.~Lett.~B {\bf 91}, 99} (1980).
\bibitem{capoz} S.~Capozziello, S.~Carloni and A.~Troisi,{\it arXiv:astro-ph/0303041}; S.~Capozziello {\it et al.},{\it  Int.~ J.~ Mod.~ Phys.~ D {\bf 12}, 1969} (2003).
\bibitem{carroll-d} S.~M.~Carroll {\it et al.}, {\it  Phys. Rev. D  {\bf 70}, 043528} (2004).
\bibitem{chiba} T.~Chiba, {\it Phys. Rev. Lett. B {\bf 575}, 1} (2003).
\bibitem{olmo1} G.~J.~Olmo, {\it Phys. Rev. Lett. {\bf 95}, 261102} (2005).
\bibitem{navarro} I.~Navarro and K.~Van Acoleyen, {\it JCAP {\bf 0702}, 022} (2007).
\bibitem{chibaerick} T.~Chiba, T.~L.~Smith and A.~L.~Erickcek, {\it Phys. Rev. D {\bf 75}, 124014} (2007).
\bibitem{nojiriodint} S.~Nojiri and S.~D.~Odintsov, {\it Phys. Rev. D {\bf 68}, 123512} (2003).
\bibitem{amenpolar} L.~Amendola, D.~Polarski and S.~Tsujikawa, {\it Phys.~Rev.~Lett.~{\bf 98},131302} (2007).
\bibitem{amenpolar1} L.~Amendola, D.~Polarski and S.~Tsujikawa, {\it Int.~J.~Mod.Phys.~D {\bf 16},155} (2007).
\bibitem{tsujuddin} S.~Tsujikawa, K.~Uddin and R.~Tavakol, {\it Phys. Rev. D {\bf 77}, 043007} (2007).
\bibitem{sotiriou} T.~P.~Sotiriou and V. Faraoni, {\it Rev.~Mod.~Phys. {\bf 82}, 451} (2010).
\bibitem{felicetsujikawa} A.~De Felice and S.~Tsujikawa, arXiv:1002.4928.
\bibitem{dgp} G.~R.~Dvali, G.~Gabadadze and M.~Porrati, {\it Phys. Lett. B {\bf 484}, 112} (2000).
\bibitem{martmajer} R.~Maartens and E.~Majerotto, {\it Phys. Rev. D {\bf 74}, 023004} (2006).
\bibitem{kkoyama} K.~Koyama, {\it Phys. Rev. D {\bf 74}, 123511} (2005).
\bibitem{gorbunov} D.~Gorbunov, K.~Koyama and S.~Sibiryakov, {\it Phys. Rev. D {\bf 73}, 044016} (2006).
\bibitem{dgpr} G.~Dvali {\it et al.}, {\it Phys. Rev. D {\bf 75}, 124013} (2007).

\bibitem{fanget} W.~Fang {\it et al.}, {\it Phys. Rev. D {\bf 78}, 103509} (2008). 
\bibitem{lombriser} L.~Lombriser {\it et al.}, {\it Pys. Rev. D {80}, 063536} (2009).
\bibitem{cascadingdgp1} C.~de~Rham {\it et al.}, {\it JCAP {\bf 0802}, 011} (2008).
\bibitem{cascadingdgp2} O.~Corradini, K.~Koyama and G.~Tasinato, {\it Phys. Rev. D {\bf 77}, 084006} (2008)
\bibitem{garciahoeg1} J.~Garcia-Bellido and T.~Haugboelle, {\it JCAP {\bf 0804}, 003} (2008).
\bibitem{garciahoeg2} J.~Garcia-Bellido and T.~Haugboelle, {\it JCAP {\bf 0809}, 016} (2008).
\bibitem{cascadingcosmology} N.~Agarwal {\it et al.}, {\it Phys. Rev. D{\bf 81}, 084020} (2010).
\bibitem{nicolis} A.~Nicolis, R.~Rattazzi and E.~Trincherini, {\it Phys. Rev. D {\bf 79}, 064036} (2009).
\bibitem{silvakoyama} F.~P.~Silva and K.~Koyama, {\it Phys. Rev. D {80}, 121301} (2009).
\bibitem{wands} D.~Wands, {\it ERE2005 proceeding, the XXVIII Spanish Relativity Meeting, Oviedo, Spain}, arXiv:gr-qc/0601078.

\bibitem{buchert} T.~Buchert, {\it Gen. Rel. Grav. {\bf 40}, 467} (2008).
\bibitem{rasanen} S.~R${\rm \ddot{a}}$s${\rm \ddot{a}}$nen, {\it arXiv:0811.2364}.
\bibitem{alnes} H.~Alnes, M.~Amarzguioui and O.~Gron, {\it Phys. Rev. D {\bf 73}, 083519} (2006).
\bibitem{enqvist} K.~Enqvist, {\it arXiv:0709.2044}.



%%%%%%%%%%%%%%%%%%%%%%%%% DE vs MG %%%%%%%%%%%%%%%%%%%
\bibitem{cmb} D.~N.~Spergel {\it et al.}, {\it ApJS {\bf 148}, 175} (2003).
\bibitem{CD} C.~Deffayet, {\it Phys. Lett. B {\bf 502}, 199} (2001).
\bibitem{BDEL} P.~Binetruy {\it et al.}, {\it Phys. Lett. B {\bf 477} 285} (2000).
\bibitem{RM} R.~Maartens, {\it astro-ph/0602415}.
%\bibitem{bdk} C.~Bonvin, R.~Durrer and M. Kunz, {\it Phys. Rev. Lett. {\bf 96}, 191302} (2006).
\bibitem{vikmancross} E.~A.~Lim. I.~Sawicki and A.~Vikman, {\it JCAP {\bf 1005}, 012} (2010).
\bibitem{bedo} R.~Bean and O.~Dor\'e, {\it Phys. Rev. D {\bf 69}, 083503} (2004).

%%%%%%%%%%%%%%%%%Perturbations%%%%%%%%%%%%%%%
\bibitem{hu_lect} W.~Hu, {\it astro-ph/0402060}.
\bibitem{mabe} C.~P.~Ma and E.~Bertschinger, {\it Astrophys. J. {\bf 455}, 7} (1995).
\bibitem{cora} P.~S.~Corasaniti {\it et al.}, {\it Phys. Rev. D {\bf 70} 083006} (2004).
\bibitem{wlim} D.~N.~Spergel {\it et al.}, {\it Astrophys. J. Suppl. {\bf 170}, 377} (2007).
\bibitem{phantom} U.~Alam {\it et al.}, {\it Mon. Not. Roy. Astron. Soc. {\bf 354} 275} (2004).
\bibitem{caldwell1} R.~R.~Caldwell, {\it Phys. Lett. B {\bf 545}, 23} (2002).
\bibitem{crossing} R.~R.~Caldwell and M.~Doran, {\it Phys. Rev. D {\bf 72}, 043527} (2005).
\bibitem{cht} S.~M.~Carroll, M.~Hoffmann and M.~Trodden, {\it Phys. Rev. D {\bf 68}, 023509} (2003).
\bibitem{Cline:2003gs} J.~M.~Cline, S.~Jeon and G.~D.~Moore, {\it Phys. Rev. D {\bf 70}, 043543} (2004).
\bibitem{para} L.~Perivolaropoulos, {\it JCAP {\bf 0510}, 001} (2005).
\bibitem{nojiri} S.~Nojiri and S.~D.~Odintsov, {\it Gen. Rel. Grav. {\bf 38}, 1285} (2006).
\bibitem{vli} M.~Li, B.~Feng and X.~Zhang, {\it JCAP {\bf 0512}, 002} (2005).
\bibitem{vik1} A.~Vikman, {\it Phys. Rev. D {\bf 71}, 023515} (2005).
\bibitem{ah} A.~Adams {\it et al.}, {\it JHEP {\bf 0610}, 014} (2006).
\bibitem{bcd} C.~Bonvin, C.~Caprini and R.~Durrer, {\it Phys. Rev. Lett. {\bf 97}, 081303} (2006).
\bibitem{vik2} A.~Anisimov, E.~Babichev and A.~Vikman, {\it JCAP {\bf 0506}, 006} (2005).
\bibitem{kosh} I.~Ya.~Aref'eva, A.~S.~Koshelev and S.~Yu.~Vernov, {\it Phys. Rev. D {\bf 72}, 064017} (2005).
\bibitem{ks1} M.~Kunz and D.~Sapone, {\it Phys. Rev. D {\bf 74}, 123503} (2006).
\bibitem{ks2} M.~Kunz and D.~Sapone, {\it Phys. Rev. Lett \textbf{98}, 121301} (2007). 
\bibitem{kas} M.~Kunz, L.~Amendola and D.~Sapone, {\it arXiv:0806.1323}.
\bibitem{KS} H.~Kodama and M.~Sasaki, {\it Progr. Theor. Phys. Suppl. {\bf 78}, 1} (1984).
\bibitem{KM} K.~Koyama and R.~Maartens, {\it JCAP \bf{0601}, 016} (2006).
\bibitem{KK} K.~Koyama, {\it JCAP \bf{0603}, 017} (2006).
\bibitem{g2} W.~J.~Percival, {\it Astron. \& Astrophys. {\bf 443}, 819} (2005).
\bibitem{fR_aniso} V.~F.~Mukhanov, H.~A.~Feldman and R.~H.~Brandenberger, Phys. Rep. {\bf 215}, 206 (1992).
\bibitem{sahnipert} U.~Alam, V.~Sahni, A.~A.~Starobinsky, {\it Astrophys.~J.~{\bf 704}, 1086} (2009).

%%%%%%%%%%%%%% TESTS %%%%%%%%%%%%%%%
%\bibitem{urakawa} Y. Urakawa and T. Kobayashi, {\it arXiv:0907.1191}.
%%%%%%%%%%% Observational tests %%%%%%%%%%%%%
\bibitem{aks} L.~Amendola, M.~Kunz and D.~Sapone, {\it JCAP  \bf{0804}, 013} (2008).
\bibitem{DUNE} A.~R\'{e}fr\'{e}gier {\it et al}, {\it astro-ph/0610062} (2006). 
\bibitem{JDEM} A.~Crotts {\it et al}, {\it astro-ph/0507043} (2005). 
\bibitem{SNAP} J.~Albert {\it et al}, {\it astro-ph/0507460} (2005). 
%\bibitem{doran} R.~Caldwell {\it et al.}, {\it Astrophys. J. \bf{591}, 75} (2003). 
%\bibitem{DTV} L.~Amendola and D.~Tocchini-Valentini, {\it Phys. Rev. D \bf{66}, 043528} (2002). 
\bibitem{chevpol} A.~Chevallier and D.~Polarski, {\it IJMPD \bf{10}, 213} (2001). 
%\bibitem{BaCoKu} B.~A.~Bassett, P.~S.~Corasaniti and M.~Kunz, {\it Astrophys. J. \bf{617}, L1} (2004). 
\bibitem{KuDu} M.~Kunz and R.~Durrer, {\it Phys. Rev. D \bf{55} 4516} (1997). 
\bibitem{sk} D.~Sapone and M.~Kunz, {\it Phys. Rev. D {\bf 80}, 083519}, (2009).
%\bibitem{lahav} O.~Lahav {\it et al.}, {\it MNRAS \bf{251}, 128} (1991).
\bibitem{linder2003} E.~V.~Linder, {\it Phys. Rev. Lett. {\bf 90}, 091301} (2003).
%\bibitem{wang} L.~Wang and P.~J.~Steinhardt, {\it Astrophys. J. \bf{508}, 483} (1998). 
\bibitem{AQG} L.~Amendola, C.~Quercellini and L.~Giallongo, {\it MNRAS {\bf 357}, 429} (2005).
\bibitem{linder1} E.~V.~Linder, {\it Phys. Rev. D {\bf 72} 043529} (2005).
\bibitem{linder2} D.~Huterer and E.~V.~Linder, {\it Phys. Rev. D {\bf 75}, 023519} (2007). 
\bibitem{alc-pac} C.~Alcock and B.~Paczynski, {\it Nature {\bf 281}, 358} (1979).
\bibitem{durrerbias} R.~Durrer {\it et al.}, {\it Astrophys. J. {\bf 585}, L1-L4} (2002).
\bibitem{kaiser} N.~Kaiser, {\it Mon. Not. Roy. Astron. Soc. {\bf 227}, 1} (1987).
%\bibitem{Kunz} M.~Kunz, {\it astro-ph/0702615} (2007). 
\bibitem{bart-schn} M.~Bartelmann and P.~Schneider, {\it Phys. Rept. {\bf 340}, 291} (2001).
\bibitem{schneider} P.~Schneider, {\it Gravitational lensing: Strong, Weak and Micro.},  Saas-Fe Advanced Courses, Volume 33. ISBN 978-3-540-30309-1, Springer-Varlag Heidelberg, 269 (2006)
\bibitem{bhuter} B.~Huterer, {\it arXiv:1001.1758} (2010).
\bibitem{amque} L.~Amendola and C.~Quercellini, {\it Phys. Rev. Lett. \textbf{92}, 181102} (2004).
\bibitem{KoiMo} T.~Koivisto and F.~Mota, Phys. {\it Rev. D {\bf 73}, 083502} (2006) 
\bibitem{LiCa} E.~V.~Linder and R.~N.~Cahn, {\it Astropart. Phys. {\bf 28} 481} (2007). 
\bibitem{mota} D.~F.~Mota {\it et al.}, {\it Not. R. Astron. Soc. {\bf 382}, 793} (2007).
\bibitem{xia} J.~Q.~Xia {\it et al.}, {\it  Int.~J.~Mod.~Phys. D {\bf 17}, 1229} (2008).
\bibitem{csHI} A.~Torres-Rodriguez, C.~M.~Cress and K.~Moodley,{\it MNRAS {\bf 388}, 669} (2008).
\bibitem{ballesteros} G.~Ballesteros and J.~Lesgourgues, {\it arXiv:1004.5509}.
\bibitem{ska} D.~Sapone, M.~Kunz and L.~Amendola, {\it arXiv:1007.2188}.


%%%%%%%%%%%%%%%%%%%%%%%% BAO RESULTS %%%%%%%%%%%%%%%%%%%%%
\bibitem{w-percival-bao} W.~J.~Percival {\it et al.}, {\it  Mon. Not. Roy. Astron. Soc. {\bf 381}, 1053} (2007).

%%%%%%%%%%%%%%%%%%%%%%%%%%%%%% WL RESULTS %%%%%%%%%%%%%
\bibitem{futal} L.~Fu {\it et al.}, {\it Astron. Astrophys. {\bf 479}, 9} (2008)
%\bibitem{hoekstra} H.~Hoekstra and B.~Jain, {\it Ann. Rev. Nucl. Part. Sci. {\bf 58}, 99} (2008). 
\bibitem{bernestein} G.~M.~Bernstein, {\it Astrophys. J. {\bf 695}, 652} (2009).

%%%%%%%%% Other papers on WL
\bibitem{hujain} W.~Hu and B.~Jain, {\it Phys. Rev. D {\bf 70}, 043009} (2004).
\bibitem{kaiser92} N.~Kaiser, {\it Astrophys. J. {\bf 388}, 272} (1992).
\bibitem{heavens-wl} A.~Heavens, {\it Nucl.Phys.Proc.Suppl. {\bf 194} 76} (2009). 
\bibitem{saponea} D.~Sapone and L.~Amendola, {\it arXiv:0704.2421}.
\bibitem{joucooholz} S.~Joudaki, A.~Cooray, D.~E.~Holz, {\it Phys.Rev.D {\bf 80} 023003} (2009).
\bibitem{holl-sapone} L.~Hollenstein {\it et al.}, {\it JCAP {\bf 0904} 012} (2009).
\bibitem{zhan-tyson} H.~Zhan, L.~Knox and J.~A.~Tyson, {\it Astrophys.J. {\bf 690} 923} (2009).


%%%%%%%%%%%%%%%%%%%%%%%OTHER PROBES %%%%%%%%%%%%%%%%%555
\bibitem{tanvir} N.~R.~Tanvir {\it et al.}, {\it arXiv:0906.1577}.
\bibitem{giannantonio} T.~Giannantonio {\it et al.}, {\it Phys. Rev. D {\bf 77}, 043519} (2008).
\bibitem{shirleyetal} H.~Shirley {\it et al}, {\it Phys. Rev. D {\bf 78}, 043519} (2008).
\bibitem{dunlop} J.~Dunlop {\it et al.}, {\it Nature {\bf 381}, 581} (1996). 
\bibitem{lima} J.~A.~S.~Lima and J.~S.~Alcaniz, {\it Mon. Not. Roy. Astron. Soc. {\bf 317}, 581} (2000).
\bibitem{simon} J.~Simon, L.~Verde and R.~Jimenz, {\it Phys. Rev. D {\bf 71}. 123001} (2005).
\bibitem{dantas} M.~A.~Dantas and J.~S.~Alcaniz, {\it Phys.Lett. B {\bf 679}, 423} (2009).
\bibitem{deepak1} L.~Samushia {\it et al.}, {\it arXiv:0906.2734}.
\bibitem{deepak2} D.~Jain, {\it arXiv:0910.4825}.
\bibitem{allen} A.~W~Allen {\it et al.}, {\it Mon. Not. Roy. Astron. Soc. {\bf 383}, 879} (2008).
\bibitem{ettori} S.~Ettori {\it et al.}, {\it arXiv:0904.2740}.
\bibitem{rapetti}D.~Rapetti {\it et al.}, {\it arXiv:0812.2259}.
\bibitem{turnerhuterer} J.~Frieman, M.~Turner, D.~Huterer, {\it Ann. Rev. Astron. Astrophys. {\bf 46}, 385} (2008).

% CONCLUSIONS %%%%%%%%%%
\bibitem{reyes-seljak} R.~Reyes {\it et. al}, {\it NATURE {\bf 464} 256} (2010).

\end{thebibliography}
\end{document}